\documentclass{aa}
\usepackage{amsmath} % or simply amstext
\usepackage{graphicx}
\usepackage{natbib}
\bibpunct{(}{)}{;}{a}{}{,} % to follow the A&A style
\usepackage{amsmath}	% Advanced maths commands
\usepackage{amssymb}	% Extra maths symbols 

\makeatother

\titlerunning {Luminosity function of LOFAR radio-selected quasars} 
\authorrunning{Retana-Montenegro \&  R\"{o}ttgering}

\begin{document}

\title{}

\title{The optical luminosity function of LOFAR radio-selected quasars at
$1.4\leq z\leq5.0$ in the NDWFS-Bo\"otes field }

\author{E. Retana-Montenegro\inst{1} \and  H. J. A. R\"{o}ttgering\inst{1}
}

\offprints{E. Retana-Montenegro}

\institute{Leiden Observatory, Leiden University, P.O. Box 9513, 2300 RA, Leiden,
The Netherlands\\
\email{edwinretana@gmail.com}\\
}

\date{Received September xx, xxxx; accepted March xx, xxxx}

\abstract{We present an estimate of the optical luminosity function (OLF) of
LOFAR radio-selected quasars (RSQs) at $1.4<z<5.0$ in the $9.3\:\textrm{deg}^{2}$
NOAO Deep Wide-field survey (NDWFS) of the Bo\"otes field. The selection
was based on optical and mid-infrared photometry used to train three
different machine learning (ML) algorithms (Random forest, SVM, Bootstrap
aggregation). Objects taken as quasars by the ML algorithms are required
to be detected at $\geq5\sigma$ significance in deep radio maps to
be classified as candidate quasars. The optical imaging came from
the Sloan Digital Sky Survey and the Pan-STARRS1 $3\pi$ survey; mid-infrared
photometry was taken from the Spitzer Deep, Wide-Field Survey; and
radio data was obtained from deep LOFAR imaging of the NDWFS-Bo\"otes
field. The requirement of a $5\sigma$ LOFAR detection allowed us
to reduce the stellar contamination in our sample by two orders of
magnitude. The sample comprises 130 objects, including both photometrically
selected candidate quasars (47) and spectroscopically confirmed quasars
(83). The spectral energy distributions calculated using deep photometry
available for the NDWFS-Bo\"otes field confirm the validity of the
photometrically selected quasars using the ML algorithms as robust
candidate quasars. The depth of our LOFAR observations allowed us
to detect the radio-emission of quasars that would be otherwise classified
as radio-quiet. Around $65\%$ of the quasars in the sample are fainter
than $M_{\textrm{1450}}=-24.0$, a regime where the OLF of quasars
selected through their radio emission, has not been investigated in
detail. It has been demonstrated that in cases where mid-infrared
wedge-based AGN selection is not possible due to a lack of appropriate
data, the selection of quasars using ML algorithms trained with optical
and infrared photometry in combination with LOFAR data provides an
excellent approach for obtaining samples of quasars. The OLF of RSQs
can be described by pure luminosity evolution at $z<2.4$, and a combined
luminosity and density evolution at $z>2.4$. The faint-end slope,
$\alpha$, becomes steeper with increasing redshift. This trend is
consistent with previous studies of faint quasars $(M_{\textrm{1450}}\leq-22.0)$.
We demonstrate that RSQs show an evolution that is  very similar to
that exhibited by faint quasars. By comparing the spatial density
of RSQs with that of the total (radio-detected plus radio-undetected)
faint quasar population at similar redshifts, we find that RSQs may
compose up to $\sim20\%$ of the whole faint quasar population. This
fraction, within uncertainties, is constant with redshift. Finally,
we discuss how the compactness of the RSQs radio-morphologies and
their steep spectral indices could provide valuable insights into
how quasar and radio activity are triggered in these systems.}

\keywords{quasars: general \textendash{} quasars: supermassive black holes
\textendash{} Radio continuum: galaxies \textendash{} galaxies: high-redshift }
\maketitle

\section{Introduction}

A good determination of the quasar luminosity function (QLF) is important
for gaining an understanding of several aspects of the cosmological
evolution of supermassive black holes (SMBHs). These aspects include:
i) the build-up of black hole (BH) demography; ii) the integrated
UV contribution from quasars to the ionization of the intergalactic
medium; iii) the accretion history of BHs across cosmic time; iv)
triggering and fueling quasar mechanisms and their co-evolution with
host galaxies. 

\noindent The cosmological evolution of quasars has been studied in
detail over a wide range of optical luminosities at $z<3$ (e.g.,
\citealt{2006AJ....131.2766R,2009MNRAS.399.1755C,2013ApJ...773...14R}).
These studies show that the comoving space density of quasars evolves
strongly, with lower luminosity quasars peaking in their space density
at lower redshift than higher luminosity quasars. This result is interpreted
as a downsizing evolutionary scenario for SMBHs in which very massive
BHs were already in place at very early times, whereas less massive
BHs evolve predominantly at lower redshifts. These results provide
valuable benchmarks to BH formation models (e.g., \citealt{2005ApJ...633..624V,2008MNRAS.391..481S}). 

\noindent At $z>3$, the shallow flux limits of the majority of current
optical quasar surveys restricts our understanding of BH growth to
the brightest optical objects. Faint quasar and AGN surveys at optical
\citep{2018AJ....155..131M,2018AJ....155..110Y,2015AAA...578A..83G,2015ApJ...813...53M,2011ApJ...728L..26G,2019ApJ...884...19G}
and X-ray \citep{2005A&A...441..417H,2005ApJ...624..630S,2014MNRAS.445.3557V,2015MNRAS.453.1946G}
wavelengths, respectively, have shed some light on the evolution of
these objects. However, the BH downsizing behavior in the early universe
is still not very well understood and the role of faint quasars in
the cosmic reionization of hydrogen remains poorly constrained. While
studies considering only the brightest quasars found that their contribution
to cosmic reionization is not significant (e.g., \citealt{2012ApJ...746..125H}),
other authors which take into account faint quasars claim that potentially
they can produce the high emissivity rate required to ionize the intergalactic
medium (e.g., \citealt{2011ApJ...728L..26G,2015AAA...578A..83G,2019ApJ...884...19G}).
For a good picture of the quasar phenomena at high-z, it is important
to study significant numbers of these low luminosity objects at high
redshifts.

\noindent The selection of quasars at $z>2.2$ is challenging, particularly
for low luminosity quasars with optical magnitudes close to the detecti
limit. In this regime, the photometric errors broaden the stellar
locus, and color distributions of quasars and stars are hard to distinguish
using color selection. One approach to circumvent this difficulty,
is to build quasar samples using optical and infrared color selection
combined with a radio detection. The spectral energy distribution
of most stars does not extent to radio wavelengths, which implies
that the number of stars with radio detections is very small. For
example, \citet{2009ApJ...701..535K} estimated that there are approximately
two radio-loud stars per every million of stars in the magnitude range
$15\leq i\leq19$. The main advantage of searching for quasars using
a radio selection over typical color selection is that stellar contamination
is reduced very significantly due to the small numbers of radio stars.
The application of this technique has therefore led to the discovery
of quasars outside the typical color boxes used to select them (see
\citealp{2002A&A...391..509H,2009AJ....138.1925M,2011ApJ...736...57Z,2015ApJ...804..118B}).

One caveat of using radio selection is that the quasars selected may
not be representative of the entire quasar demographics. In fact,
the majority of studies of the luminosity function of radio-selected
quasars (RSQs) \citep{1996Natur.384..439S,2003ApJ...591...43V,2009AJ....138.1925M,2006MNRAS.370.1034C,2015MNRAS.449.2818T}
include only radio-loud quasars (RLQs) with luminosities $L_{1.4GHz}\gtrsim1\times10^{26}\:\textrm{W/Hz}$
that are selected using shallow all-sky radio and optical surveys
such as FIRST \citep{1995ApJ...450..559B} and SDSS \citep{2000AJ....120.1579Y},
respectively. Interestingly, the works presented by \citet{2009AJ....138.1925M}
and \citet{2015MNRAS.449.2818T} found that the luminosity function
of RSQs shows a flattening of the bright-end that is similar to the
whole quasar population at $3.5\leq z\leq4.4$ \citep{2006AJ....131.2766R}.
Moreover, the analysis by \citet{2005MNRAS.357.1267C} suggests a
decrement of the space density of faint RSQs from $z\simeq1.8$ to
$z=2.2$ by a factor of 2. An issue is the origin of the radio emission
in radio-quiet quasars (RQQs). It is still a matter of debate whether
it is linked to star-forming activity occurring in the host galaxy
\citep{2011ApJ...739L..29K,2011ApJ...740...20P,2013ApJ...768...37C,2013MNRAS.436.3759B}
or non-thermal processes near the SMBH \citep{2010A&A...510A..42P,2016A&A...589L...2H}.

The number of known quasars has increased dramatically in the course
of the last two decades (e.g., \citet{2009MNRAS.392...19C,2018A&A...613A..51P}).
This number will continue increasing with forthcoming facilities such
as the Large Synoptic Survey Telescope (LSST, \citealt{2019ApJ...873..111I}),
WFIRST \citep{2015arXiv150303757S}, DESI \citep{2016arXiv161100036D},
and Euclid \citep{2011arXiv1110.3193L}, expected to deliver millions
of quasars (e.g. \citealt{2014IAUS..304...11I}). One of the major
challenges is the identification of quasars without spectroscopic
observations, which are costly in terms of telescope time for such
large samples. Several machine learning (ML) techniques have been
proposed to photometrically create large samples of quasars, including
artificial neural networks \citep{2010A&A...523A..14Y,2015MNRAS.449.2818T},
random forest \citep{2017ApJ...851...13S}, support vector machine
\citep{2008MNRAS.386.1417G,2012MNRAS.425.2599P,2019MNRAS.485.4539J},
extreme deconvolution \citep{2011ApJ...729..141B,2015ApJS..221...27M},
bayesian selection \citep{2011ApJ...743..125K,2015ApJ...811...95P,2017AJ....154..269Y},
and bootstrap aggregation \citep{2018ApJ...859...20T}. 

Many of these quasars will be detected by the next generation of radio
surveys \citep{2017NatAs...1..671N}. Particularly, low-frequency
radio telescopes such as the Low Frequency Array (LOFAR, \citealt{2013A&A...556A...2V})
open a new observational spectral window to study the evolution of
quasar activity. The LOFAR Surveys Key Science Project (LSKSP, \citealt{2011JApA...32..557R})
aims to map the entire Northern Sky down to $\lesssim100\,\mu\textrm{Jy}$,
while for extragalatic fields, greater than a few square degrees in
size and with extensive multi-wavelength data, the target rms noise
is of a few tens of $\mu\,\textrm{Jy}$. In this paper, we investigate
the evolution of the luminosity function of RSQs. For this purpose,
we take advantage of the deep optical, infrared and LOFAR data available
for the NOAO Deep Wide-field survey (NDWFS) Bo\"otes field \citep{1999ASPC..191..111J}.

This paper is organized as follows. In Section \ref{sec:Section2},
we present the surveys used in this work. In Section \ref{sec:classification},
we briefly discuss the classification algorithms employed to compile
our quasar sample. We explain the method utilized to compute the photometric
redshifts for our photometricallyselected candidate quasars in Section
\ref{sec:photo_zs}. In Section \ref{subsec:lofar_irac}, we compare
the LOFAR and wedge-based mid-infrared selection for objects classified
as quasars by the ML classification algorithms. In Section \ref{sec:Section5},
we present the optical luminosity function of our RSQs and compare
our results with previous works from the literature in Sections \ref{sec:qlf-models}
and \ref{sec:comparison_previouswork}. The comoving spatial density
of RSQs is studied in Section \ref{sec:density_evolution}. Section
\ref{sec:Section7} discusses how the compactness of the RSQs radio-morphologies
and their steep spectral indices could provide insights into the way
quasar and radio activities are triggered. In Section \ref{sec:RSQs_quasar_evolution},
we discuss the possible location of RSQs in different spectroscopic
parameter spaces. Finally, we summarize our conclusions in Section
\ref{sec:conclusions}. For the purposes of simplicity, we henceforth
refer to photometrically selected candidate quasars as photometric
quasars and to spectroscopically confirmed quasars as spectroscopic
quasars. Also, we refer to published samples of quasars with $M_{\textrm{1450}}\leq-22.0$
\citep{2008ApJ...675...49S,2011ApJ...728L..26G,2015ApJ...813...53M,2018AJ....155..110Y,2018AJ....155..131M,2018PASJ...70S..34A}
as faint quasars. Through this paper, we use a $\Lambda$ cosmology
with the matter density $\Omega_{m}=0.30$, the cosmological constant
$\Omega_{\Lambda}=0.70$, and the Hubble constant $H_{0}=70\,\textrm{km}\,\textrm{s}^{-1}\,\textrm{Mpc}^{-1}$.
We assume a definition of the form $S_{\nu}\propto\nu^{-\alpha}$,
where $S_{\nu}$ is the source flux, $\nu$ the observing frequency,
and $\alpha$ the spectral index. To estimate radio luminosities,
we adopt an radio spectral index of $\alpha=0.7$. The optical luminosities
are calculated using a power-law continuum index of $\epsilon_{\textrm{opt}}=0.5$.
All the magnitudes are expressed in the AB magnitude system \citep{1983ApJ...266..713O}. 

\section{Data \label{sec:Section2}}

In this section, we introduce the datasets that will be utilized for
the selection of quasars, and for the estimation of photometric redshifts
for objects without spectroscopy.

\subsection{NOAO Deep Wide-field survey \label{sec:bootes_data}}

\noindent The NOAO Deep Wide-field survey (NDWFS) is a deep imaging
survey that covers approximately two $9.3\;\textrm{deg}^{2}$ fields
\citep{1999ASPC..191..111J}. One of these regions, the Bo\"otes
field has a large wealth of data available at a range of observing
windows including:\emph{ }X-ray (\emph{Chandra}; \citealt{2005ApJS..161....9K}),
UV-optical (\emph{$NUV$,$U_{\textrm{spec}}$,}$B_{W}$,$R$,$I$\emph{,$Z_{\textrm{Subaru}}$
}bands; \citealt{1999ASPC..191..111J,2005ApJ...619L...1M,2007ApJS..169...21C,2013ApJ...774...28B}),
infrared ($Y$,$J,H,K,K_{s}$\emph{ }bands\emph{, Spitzer}; \citealt{2009ApJ...701..428A,2010AAS...21547001J})\emph{,
}and radio (150-1400MHz; \citealt{2002AJ....123.1784D,2013AA...549A..55W,2016MNRAS.460.2385W,2018arXiv180704878R}).
We used the Spitzer/IRAC-$3.6\,\mu\textrm{m}$ matched photometry
catalog presented by \citealt{2009ApJ...701..428A}. This catalog
contains 677522 sources detected at $5\sigma$ limiting magnitudes
measured in a $4^{\prime\prime}$ diameter (aperture-corrected) of
22.56, 22.08, 20.24, and 20.19 at $3.6$, $4.5$, $5.8$, and $8.0\,\mu\textrm{m}$,
respectively. The $3.6\,\mu\textrm{m}$ and $4.5\,\mu\textrm{m}$
magnitudes were converted to AB units using the relations: $\textrm{\ensuremath{\left[3.6\,\mu\textrm{m}\right]}}_{AB}=\textrm{\ensuremath{\left[3.6\,\mu\textrm{m}\right]}}_{Vega}+2.788$
and $\textrm{\ensuremath{\left[4.5\,\mu\textrm{m}\right]}}_{AB}=\textrm{\ensuremath{\left[4.5\,\mu\textrm{m}\right]}}_{Vega}+3.255$\footnote{http://irsa.ipac.caltech.edu/data/COSMOS/gator\_docs/ scosmos\_irac\_colDescriptions.html }.
To select our RSQs, we used the deep 150MHz LOFAR observations of
the Bo\"{o}tes field presented by \citet{2018arXiv180704878R}. The
image obtained covers more than $20\;\textrm{deg}^{2}$ and was based
on 55 hours of observation. The central rms noise of the mosaic is
$55\,\mu\textrm{Jy}$ with an angular resolution of $3.98^{''}\times6.45^{''}$.
The final radio catalog contains 10091 sources detected above a $5\sigma$
peak flux density threshold. There are 170 extended sources in the
catalog, whose components were merged according to a visual inspection.
This reduces the possibility of missing sources without detected cores
in the LOFAR mosaic. Here, we focus only in the $9.3\;\textrm{deg}^{2}$
covered by the optical and infrared data. A total of 5646 LOFAR sources
are found in the Spitzer/IRAC-$3.6\,\mu\textrm{m}$ matched catalog
using a matching radius of $2^{\prime\prime}.$ 

\noindent 
\begin{figure}[tp]
\centering{}\includegraphics[bb=20bp 8bp 800bp 610bp,clip,scale=0.32]{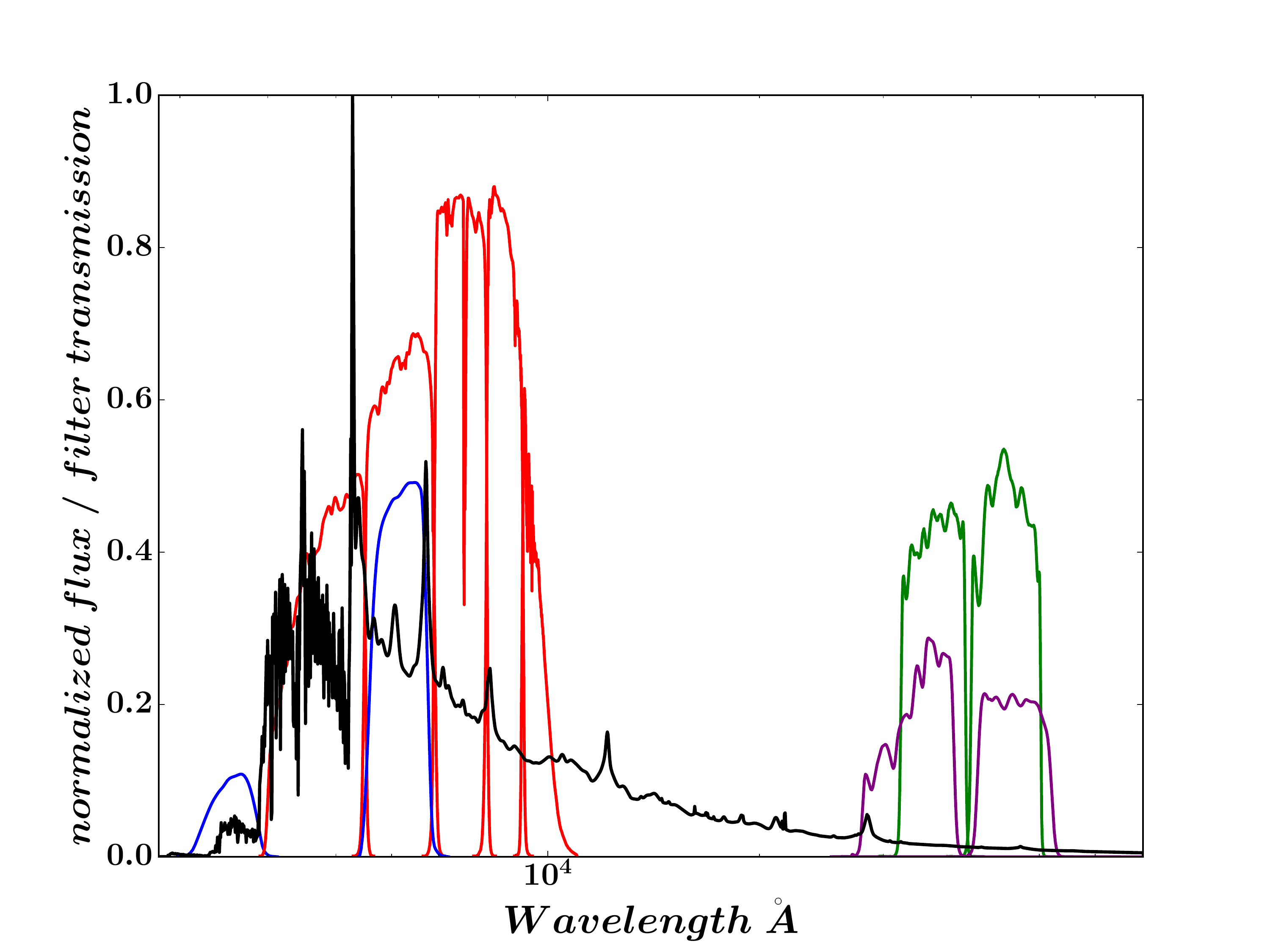}\centering\caption{\label{fig:ndwfs_filters} Transmission curves of the filters used
in this work. \emph{Blue lines}: SDSS-u and SDSS-r filters; \emph{red
lines}: the Pan-STARRS1 filter set: $g_{\textrm{PS}},r_{\textrm{PS}},i_{\textrm{PS}},z_{\textrm{PS}},y_{\textrm{PS}}$;
\emph{green lines: Spitzer-IRAC }$[3.6\,\mu\textrm{m}]$ and $[4.5\,\mu\textrm{m}]$
bands; \emph{purple lines: }WISE W1 and W2 bands; and the solid black
line shows a simulated quasar spectrum from our library at $z=3.4$
(See Section \ref{sec:simuqso}). }
\end{figure}

\subsection{SDSS, Pan-STARRS1, WISE, and Spitzer surveys \label{sec:survey_data}}

The Sloan Digital Sky Survey (SDSS, \citealt{2000AJ....120.1579Y})
is a multi-filter imaging and spectroscopic survey conducted with
the 2.5m wide-field Sloan telescope \citep{2006AJ....131.2332G} located
at the Apache Point observatory in New Mexico, USA. The SDSS-DR14
\citep{2018ApJS..235...42A} provides photometry for $14955\,\,\textrm{deg}^{2}$
in five broad-band optical filters ($u,g,r,i,z;$ \citealt{1996AJ....111.1748F}).
The magnitude limits (95\% completeness for point sources) in the
five filters are $u=22.0$, $g=22.2$, $r=22.2$, $i=21.3$, and $z=20.5$
mag, respectively. 

We also use optical and near-infrared imaging from the 1.8m Pan-STARRS1
telescope \citep{2004SPIE.5489..667H} located on the summit of Haleakala
on the Hawaiian island of Maui, which provides five band photometry
($g_{\textrm{PS}},r_{\textrm{PS}},i_{\textrm{PS}},z_{\textrm{PS}},y_{\textrm{PS}}$).
The Pan-STARRS1 first and second data releases \citep{2016arXiv161205560C}
are dedicated to the $3\pi$ survey, which observed, for almost four
years the sky north of $-30^{\circ}$ declination, reaching $5\sigma$
limiting magnitudes in the $g_{\textrm{PS}},r_{\textrm{PS}},i_{\textrm{PS}},z_{\textrm{PS}},y_{\textrm{PS}}$
bands of 23.3, 23.2, 23.1, 22.3, 21.3, respectively. Pan-STARRS1 provides
deeper imaging in overlapping optical bands (except the $\textrm{SDSS}-u$
band) and has the near-IR filter $y_{\textrm{PS}}$. SDSS has the
$u$ band covering wavelengths between 3000-4000 \AA, which
contains the Lyman alpha emission at redshifts $1.3\lesssim z\lesssim2.2$.
This makes the $\textrm{SDSS}-u$ band important for the selection
of $z\lesssim2.2$ quasars. For these reasons, we combined the $\textrm{SDSS}-u$
band with the Pan-STARRS1 filter set ($g_{\textrm{PS}},r_{\textrm{PS}},i_{\textrm{PS}},z_{\textrm{PS}},y_{\textrm{PS}}$)
to have wavelength coverage from 3000 \AA to 10800 \AA
(see Figure \ref{fig:ndwfs_filters}). 

As a first step to obtaining mid-infrared photometry for the spectroscopic
quasars, we combined the observations from all deep Spitzer-IRAC surveys
including: XFLS \citep{2005ApJS..161...41L}, SERVS \citep{2012PASP..124.1135M},
SWIRE \citep{2003PASP..115..897L}, S-COSMOS \citep{2007ApJS..172...86S},
SDWFS \citep{2009ApJ...701..428A}, SHELA \citep{2016ApJS..224...28P},
and SpIES \citep{2016ApJS..225....1T}. We followed the same procedure
described by \citet{2015ApJS..219...39R} to combine all the Spitzer-IRAC
observations. The final catalog contains over 6.2 million Spitzer-IRAC
sources. In cases where an IRAC source has been observed multiple
times, we used only the deepest IRAC observation. 

The mid-infrared photometry for spectroscopic quasars outside the
footprint of Spitzer-IRAC surveys comes from observations by NASA's
Wide Infrared Survey Explorer (WISE, \citealt{2010AJ....140.1868W}).
WISE mapped the sky at $3.4$, $4.6$, $12$ and $22\:\mu\textrm{m}$
(known as W1, W2, W3, and W4). After the cryogenic fuel of the satellite
was exhausted in 2010, WISE continued its observations as part of
the post-cryogenic NEOWISE mission phase using only its two shortest
bands (W1 and W2). We used the SDSS/unWISE forced photometry catalog
by \citet{2017AJ....154..161M}. This catalog provides forced photometry
of custom WISE coadds, at the positions of over 400 million SDSS primary
sources. This approach provides WISE flux measurements for sources
blended in WISE coadds, but resolved in SDSS images and non-detected
objects below the ``official'' WISE magnitude limits (i.e., ALLWISE;
\citealt{2013yCat.2328....0C}). We only used the W1 and W2 bands
as the other two bands are shallower and, thus, have lower detection
rates. 

We retrieved the SDSS, Pan-STARRS1, and WISE photometry from the SDSS
database via CasJobs\footnote{http://skyserver.sdss.org/CasJobs/}
and the Mikulski Archive for Space Telescopes (MAST) with PS1 CasJobs\footnote{https://mastweb.stsci.edu/ps1casjobs/}.
We made sure that the objects in our samples had clean photometry
by excluding sources with the SDSS bad photometry flags described
in \citet{2015ApJS..219...39R}. However, we opted to keep objects
with the flag \texttt{BLENDED, as} high-z quasars could be flagged
as \texttt{BLENDED }in some instances, despite being isolated objects
(e.g., \citealt{2009AJ....138.1925M}). Only \texttt{PRIMARY} sources
were selected from the SDSS data. The flags that describe the quality
of the Pan-STARRS1 sources are taken from Table 2 in \citet{2016arXiv161205244M}.
For SDSS and Pan-STARRS1, we use \texttt{PSF} magnitudes, and adopt
the \texttt{w1mag} and\texttt{ w2mag} columns from the unWISE catalog
as the WISE measurements for the W1 and W2 bands, respectively. These
WISE magnitudes are converted to AB units using the relations: $\textrm{W1}_{AB}=\textrm{W1}_{Vega}+2.699$
and $\textrm{W2}_{AB}=\textrm{W2}_{Vega}+3.339$\footnote{http://wise2.ipac.caltech.edu/docs/release/allsky/expsup/}.
We considered only WISE sources that met the following criteria: \texttt{w1\_prochi2$\leq2.0$}
\&\& \texttt{w2\_prochi2$\leq2.0$} (to avoid sources with low-quality
profile fittings) and \texttt{w1\_profracflux$\leq0.1$ \&\&} \texttt{w1\_profracflux$\leq0.1$}
(to exclude sources with fluxes severely affected by bright neighbors).
The SDSS \texttt{cMODELMAG}\footnote{https://www.sdss.org/dr14/algorithms/magnitudes/}
magnitudes were also retrieved to investigate the separation of point
and extended sources (Section \ref{subsec:final_qso_sample}). The
SDSS magnitudes in the $u$ filter, originally in inverse hyperbolic
sine magnitudes \citep{1999AJ....118.1406L}, were converted to the
AB system using $u_{AB}=u_{SDSS}-0.04$ \citep{1996AJ....111.1748F}.
The WISE-W1 and WISE-W2 photometry was converted to the IRAC $3.6\,\mu\textrm{m}$
and $4.5\,\mu\textrm{m}$ bands, respectively, using the transformations
derived by \citet{2015ApJS..219...39R}. We crossmatched the WISE
and Spitzer-IRAC catalogs using a radius of $2^{\prime\prime}$. If
the crossmatch was positive, we kept only the IRAC measurement. The
SDSS, Pan-STARRS1 and IRAC magnitudes were corrected for Galactic
extinction using the prescription by \citet{2011ApJ...737..103S}.
Figure \ref{fig:ndwfs_filters} shows the transmission curves of the
filters utilized in this work.

\subsection{Spectroscopic quasars with optical and mid-infrared photometry \label{sec:qso_sdss_wise}}

To efficiently discover new quasars using ML techniques, requires
the compilation of a large and representative sample of spectroscopic
quasars (e.g., \citet{2015ApJS..219...39R,2018A&A...611A..97P,2019MNRAS.485.4539J}).
For this purpose, we used the Million Quasars (Milliquas) catalog
v6.2 2019\footnote{http://quasars.org/} by \citet{2015PASA...32...10F}.
This catalog contains more than 600000 type-I quasars and active galactic
nuclei (AGN) from the literature, and is updated on a regular basis.
The majority of quasars included in the Milliquas catalog were discovered
as part of SDSS/BOSS \citep{2010AJ....139.2360S,2018A&A...613A..51P},
LAMOST \citep{2019ApJS..240....6Y}, ELQS \citep{2017ApJ...851...13S},
2QZ \citep{2004MNRAS.349.1397C}, 2SLAQ \citep{2009MNRAS.392...19C},
and many other surveys (e.g., \citealt{2006AJ....132..231P,2009ApJ...696.1195T,2012ApJS..200....8K,2012MNRAS.424.2876M,2013ApJ...768..105M}).
We only considered quasars with magnitudes measured for each band
to maximize the use of the multi-dimensional color information available.

\section{Classification \label{sec:classification}}

In this section, we explain how the training and target (objects to
be classified) samples were compiled and the different algorithms
used for the classification of quasars in the NDWFS-Bo\"otes field.
We also assess the performance of the classification algorithms by
calculating their efficiency and completeness. 

\subsection{Training sample \label{subsec:training_sample}}

\noindent A critical success factor for any ML technique to classify
astronomical sources is the use of an appropriate training sample
to identify new objects in the target sample. The training sample
must have measurements in the relevant filters to identify the characteristic
spectral features (e.g., colors) of the sources of interest (e.g.,
quasars) in order to map their parameter space. At the same time,
the training sample has to be representative of the target data. This
means not only including a significant number of the sources of interest,
but also the other types of astronomical objects expected to be part
of the target sample (i.e. stars and galaxies). In particular for
quasars, the training samples require several thousands of these objects
to robustly extract their color trends as a function of redshift \citep{2010A&A...523A..14Y,2015ApJS..219...39R,2019A&A...624A..13N,2019A&A...621A..26P}.
Unfortunately, there are only 2042 quasars with $0.126\leq z\leq6.12$
in the NDWFS-Bo\"otes field, with only 1259 of these quasars having
redshifts larger than 1.4 \citep{2015PASA...32...10F}. The Bo\"{o}tes
quasars in this catalog are drawn mainly from the AGN and Galaxy Evolution
Survey (AGES, \citealt{2012ApJS..200....8K}), but other quasar surveys
\citep{2006AJ....132..823C,2006ApJ...652..157M,2011ApJ...728L..26G,2018A&A...613A..51P}
have been used as well. These objects are included in the Milliquas
catalog \citep{2015PASA...32...10F}. However, this sample is too
small and sparse, to properly map the parameter space of quasars in
the NDWFS-Bo\"otes field. Instead of relying only on the NDWFS-Bo\"otes
quasar sample to identify new quasars, we created a training sample
using as starting point the Milliquas catalog presented in Section
\ref{sec:qso_sdss_wise}.

We restricted the redshift range of our analysis to $1.4<z<5.0$ for
the following reasons. First, the host galaxies of some $z<1$ quasars
can be detected in the NDWFS images. This implies that the contribution
of the host galaxy component to the overall quasar spectra has to
be considered for these sources. This makes it difficult to separate
low-z galaxies and quasars using morphological classification based
on standard photometric criteria. Secondly, low-z contaminants, such
as star-forming, blue, and emission-line galaxies, can mimic the colors
of high-z quasars due to their $4000$ \AA breaks \citep{1993ASPC...43..189S,2002AJ....123.2945R}.
Therefore, we set the lower redhift limit to $z=1.4$. This choice
is a compromise between reducing contamination by galaxies and excluding
good candidate quasars from the sample, but ensures that the degree
of contamination due to galaxies across the redshift interval considered
is as low as possible. Thirdly, at redshifts $z>5$ the number of
known quasars is significantly low compared with lower redshifts.
Thus, we set $z=5.0$ as the upper redshift limit for our analysis.
In the training sample, we included spectroscopic quasars with redshifts
slightly larger than the redshift limits of our analysis. This helps
to reduce the possibility of quasars located at the edges of the redshift
intervals not being identified by the classification algorithms. Therefore,
we included spectroscopic quasars with $1.2\leq z\leq5.5$ in our
training sample.

The first step to create our training sample is to crossmatch the
entire Milliquas catalog with the SDSS, Pan-STARRS1, WISE, and IRAC
surveys as described in Section \ref{sec:survey_data}. Only objects
with a spectroscopic confirmation as quasars are considered. We made
sure that the quasars in the training sample have clean photometry
by excluding quasars with SDSS, IRAC, and WISE bad photometry flags
(see Section \ref{sec:survey_data}). We limited the spectroscopic
quasar sample to magnitudes $16.0\leq i_{\textrm{PS}}\leq23.0$ as
this range contains 99.9\% of all optical/mid-infrared quasars in
the training sample. In total, our sample contains 328956 spectroscopic
quasars with $1.2\leq z\leq5.5$. These quasars have clear photometry
and are detected in the optical and near-infrared ($\,u\,,g_{\textrm{PS}}\,,r_{\textrm{PS}}\,,i_{\textrm{PS}}\,,z_{\textrm{PS}}\,,y_{\textrm{PS}}\,$),
and mid-infrared ($3.6\,\mu\textrm{m},\;4.5\,\mu\textrm{m}$) bands
considered in our analysis. These objects are assigned the label ``quasar''
in our training sample. This label assignment could be seen as trivial,
but it is fundamental because the algorithms introduced in Section
\ref{subsec:classification_algorithms} require labels to categorize
new unlabeled data in the target sample.

The rest of the training sample that comprises non-quasar objects
was compiled as follows. First, we needed to consider that the target
sample does not only contain $1.2\leq z\leq5.5$ quasars, but ones
with redshifts lower than 1.2 as well. Thus, to ensure that the training
sample was representative, we included quasars with redshifts $0\leq z<1.2$
from the Milliquas catalog. There is a total of 71830 quasars that
were selected as described in Section \ref{sec:survey_data}, and
were assigned the ``non-quasar'' label. Secondly, stars and galaxies
were expected to be part of the target sample. We included them in
the training sample by randomly drawing objects with clean photometry
(see Section \ref{sec:survey_data}) from the SDSS database with CasJobs,
and excluding sources located in the NDWFS-Bo\"otes region. These
objects have $m_{AB}<15$ in all the bands to avoid saturated pixels,
and must have been classified spectroscopically as stars or galaxies
by the SDSS pipeline. If the source was matched within a radius of
$2^{\prime\prime}$ to a known spectroscopic quasar it is excluded.
We did not apply any morphological criteria to the sources added to
the training sample. The drawing process was repeated until a total
of 1098858 ``non-quasar'' objects are selected. This size was chosen
to have, in combination with the spectroscopic quasar sample, a total
of a million and a half of objects in the training sample. The inclusion
of galaxies, stars, and $z<1.2$ quasars in the training sample is
important because it helps the classification algorithms to delimit
the color space of ``quasars'' and ``non-quasars''. The details
of the training sample are summarized in Table \ref{tab:properties_samples}.

\begin{table*}
\noindent \begin{centering}
\caption{Properties of the training and target samples. \label{tab:properties_samples}}
\begin{tabular}{ccc}
\hline 
{\tiny{}Sample} & {\tiny{} Number of objects } & {\tiny{}$i_{\textrm{PS}}$ magnitude}\tabularnewline
 &  & {\tiny{}{[}AB{]}}\tabularnewline
\hline 
{\tiny{}Training sample} & {\tiny{}$1.5\times10^{6}$} & {\tiny{}$15\leq i_{\textrm{PS}}\leq24$ }\tabularnewline
{\tiny{}$1.2\leq z\leq5.5$ quasars} & {\tiny{}328956} & {\tiny{}$16\leq i_{\textrm{PS}}\leq23$ }\tabularnewline
{\tiny{}$0<z<1.2$ quasars} & {\tiny{}71830} & {\tiny{}$16\leq i_{\textrm{PS}}\leq23$ }\tabularnewline
{\tiny{}Target sample} & {\tiny{}287218} & {\tiny{}$15\leq i_{\textrm{PS}}\leq24$ }\tabularnewline
\hline 
\end{tabular}
\par\end{centering}
$\;$
\end{table*}

\subsection{Target sample \label{subsec:test_sample}}

The target sample is the Spitzer/IRAC-$3.6\,\mu\textrm{m}$ matched
catalog presented in Section \ref{sec:bootes_data}. As mentioned
in Section \ref{subsec:training_sample}, there is deep optical photometry
available for the NDWFS-Bo\"otes field in the $U_{\textrm{spec}},B_{w},R,I,Z_{\textrm{Subaru}}$
bands, however, the number of quasars with photometry in these bands
is small. Fortunately, the entire NDWFS-Bo\"otes field is located
inside the SDSS/Pan-STARRS1 footprint, and thus photometry from these
surveys is available for the target sample. We obtained SDSS and Pan-STARRS1
photometry for all the objects in the NDWFS-Bo\"otes catalog following
the same procedure as for the training set. We removed sources with
$m_{AB}\geq15$ in all the bands (SDSS, Pan-STARRS1, IRAC) regardless
of their flags to avoid saturated pixels. We kept only sources with
the SDSS/Pan-STARRS1 good photometry flags. We had also considered
using the IRAC flags (SExtractor flags indicating possible blending
issues in the source extraction) but found that around $60\%$ of
the $z>1.4$ spectroscopic quasars in Bo\"otes are flagged. Considering
that the removal of such a significant fraction of quasars from the
analysis could affect our conclusions, we decided not to remove flagged
IRAC sources from the target sample at this point. In Sections \ref{subsec:3.3 classification_results}
and \ref{sec:redshift_estimation}, we confirmed that including sources
marked by the IRAC flags does not result in a deterioration of the
quality of the classification, and determination of the photometric
redshifts. Finally, the details of the target sample are summarized
in Table \ref{tab:properties_samples}.

\subsection{Classification algorithms \label{subsec:classification_algorithms} }

In supervised ML, classification algorithms rely on labeled input
data (training sample) to produce an inferred function, which can
be used to categorize new unlabelled data (target sample). If there
is a strong correlation between the input data and the labels a robust
inferred function can be obtained. This usually results in a better
performance of the ML classification algorithms. In this work, our
aim is to identify new quasars in the NDWFS-Bo\"otes field in the
target sample using supervised ML classification algorithms. For quasars,
the obvious choice is to use their colors for classification purposes
(e.g., \citealt{2002AJ....123.2945R,2009ApJS..180...67R,2015ApJS..219...39R,2018ApJ...859...20T}).
Therefore, we used the color indices ($u-g_{\textrm{PS}}$, $u-r_{\textrm{PS}}$,
$g_{\textrm{PS}}-r_{\textrm{PS}}$, $r_{\textrm{PS}}-i_{\textrm{PS}}$,
$i_{\textrm{PS}}-z_{\textrm{PS}}$, $z_{\textrm{PS}}-y_{\textrm{PS}}$,
$y_{\textrm{PS}}-[3.6\,\mu\textrm{m}]$, $[3.6\,\mu\textrm{m}]-[4.5\,\mu\textrm{m}]$)
of the training and target samples as inputs to the classification
algorithms. The algorithms used in our analysis (Random Forest, Support
vector machine, and Bootstrap aggregation) were selected because of
their extensive use in previous studies of quasars (e.g., \citealt{2008MNRAS.386.1417G,2012MNRAS.425.2599P,2015A&A...584A..44C,2018ApJ...859...20T,2019MNRAS.485.4539J}).
Each one of these algorithms is briefly explained below. They are
also part of the open-source scikit-learn\footnote{https://scikit-learn.org }
Python library.

\subsubsection{Random forest \label{subsec:random_forest}}

A random forest (RF, \citealt{Breiman2001}) ensemble is composed
of random decision trees, with each one created from a random subset
of the data. The outputs of the decision trees are combined to make
a consensus prediction. The final RF classification of an unlabeled
instance is determined using the majority vote of all decision trees.

\subsubsection{Support vector machines \label{subsec:svc}}

The support vector machines (SVM, \citealt{Cortes1995}) is a discriminative
classifier for two-group problems. The basic idea is to find an optimal
hyperplane in an N-dimensional space (N is the number
of features) that distinctly categorizes the data points. In two dimensions,
the hyperplane is a line dividing the parameter space in two parts
wherein each group is located on either side. For unlabeled instances,
the SVM classifier outputs an optimal hyperplane which is used to
classify them. 

\subsubsection{Bootstrap aggregation on K-nearest neighbors \label{subsec:bagging}}

Bootstrap Aggregation \citep{Breiman1996}, also called bagging, is
a method for generating multiple versions from a training set, by
sampling the original sample uniformly and with replacement. Subsequently,
each one of these subsets is used to train the K-Nearest Neighbors
algorithm (KNN, \citealp{doi:10.1080/00031305.1992.10475879}). In
the KNN algorithm, an unlabelled object is classified by a simple
majority vote of its neighbors, with the object being assigned the
label most common among its $k$ nearest neighbors ($k$ is a positive
integer). In the case $k=1$, the label assigned is of that of the
single nearest neighbor. For each bagging subset, we use a value of
$k=50$. Finally, the results of the bagging subsets are aggregated
by averaging to obtain a final classification.

\subsubsection{Performance \label{subsec:performance_classi}}

We assessed the performance of the classification algorithms with
the quasar training sample presented in Section \ref{subsec:training_sample},
by calculating their completeness and efficiency.

\noindent The completeness $C$ \citep{2000A&A...359....9H,10.3389/fspas.2018.00005}
is defined as the ratio between the number of quasars correctly identified
as quasars, and the total number of quasars in the sample:
\noindent \begin{center}
\begin{equation}
C=\dfrac{\textrm{Number of identified quasars }}{\textrm{Total number of quasars }}\times100.\label{eq:winhorst_size_dist-2}
\end{equation}
\par\end{center}

\noindent The efficiency $E$ is defined as the ratio between the
number of quasars correctly identified as quasars, and the total number
of objects identified as quasars:
\noindent \begin{center}
\begin{equation}
E=\dfrac{\textrm{Number of identified quasars }}{\textrm{Total no. of objects identified as quasars }}\times100.\label{eq:winhorst_size_dist-2-1}
\end{equation}
\par\end{center}

\noindent As is common practice in supervised ML, the training sample
described in Section \ref{subsec:training_sample} is separated into
validation and test samples for cross-validation (CV) purposes, in
order to calculate the performance on observations of the classification
algorithms. The validation set is used as the training sample, while
the CV test sample, which contains spectroscopic quasars, is employed
to report the completeness and efficiency of the classification algorithms.
As the CV test set, we chose a random 25\% subsample of the full training
set. The remaining 75\% of the training data was the validation sample,
and is used to train the algorithms.

\noindent In Table \ref{tab:classification_performance}, we present
the completeness and efficiency for all the classification algorithms.
While the differences are small, performance of the SVM algorithm
is the worst, while RF has the highest completeness and efficiency
values.

\begin{table*}
\noindent \begin{centering}
\caption{Performance of the classification algorithms for the quasar training
sample. The statistics are indicated for each subsample.\label{tab:classification_performance}}
\begin{tabular}{ccc}
\hline 
{\tiny{}Algorithm} & {\tiny{} Completeness} & {\tiny{}Efficiency}\tabularnewline
 & {\tiny{}\%} & {\tiny{}\%}\tabularnewline
\hline 
{\tiny{}Random Forest} & {\tiny{}93.09} & {\tiny{}90.50 }\tabularnewline
{\tiny{}Support Vector Machines} & {\tiny{}89.36} & {\tiny{}81.63 }\tabularnewline
{\tiny{}Bootstrap Aggregation on K-Nearest Neighbors} & {\tiny{}93.36} & {\tiny{}85.97 }\tabularnewline
\hline 
\end{tabular}
\par\end{centering}
$\;$

\emph{Notes:} {\tiny{}All experiments used the same training and test
samples.}{\tiny \par}
\end{table*}

\subsubsection{Classification results \label{subsec:3.3 classification_results}}

In this section, we discuss the results of the application of the
classification algorithms to our target sample. To identify the maximum
number of quasars, we combined the results of the three classification
algorithms. While this step increases the sample completeness, it
also increases the amount of contamination on the sample despite each
algorithm having low degree of contamination (see Table \ref{tab:classification_performance}).
In Sections \ref{subsec:radio_matching} and \ref{subsec:final_qso_sample},
we took additional steps to eliminate contaminants from our quasar
sample. At this point, the classification algorithms identified 39160
objects as quasar candidates. Of these, 36699 lack spectroscopic observations,
and 2470 sources had been classified spectroscopically. By crossmatching
our sample with the Milliquas catalog, we found that 1374 are already
known quasars. From these quasars, 1042 have redshifts larger than
$z>1.4$. This corresponds to a completeness of 87\% for the sample
of Bo\"otes spectroscopic quasars. Additionally, we checked the AGES
catalog \citep{2012ApJS..200....8K} and NED\footnote{https://ned.ipac.caltech.edu/}
database to find that 1096 objects had been classified spectroscopically
as either galaxies or galaxies. The confirmed stars and galaxies were
removed from the sample.

\subsubsection{Radio data \label{subsec:radio_matching}}

The contamination by the stars and galaxies in quasar samples is inevitable
as they occupy similar regions in the color space used to train the
classification algorithms. As has already been discussed, an efficient
way to eliminate stellar sources from quasar samples is to include
information from radio surveys \citep{2002AJ....123.2945R,10.3389/fspas.2018.00005}.
For this purpose, we used the LOFAR observations of the NDWFS-Bo\"otes
field by \citet{2018arXiv180704878R} introduced in Section \ref{sec:bootes_data}.
We crossmatched our quasar sample and the LOFAR catalog using a radius
of $2^{\prime\prime}$ to identify the spectroscopic quasars detected
in our LOFAR mosaic. We also inspected their stacked RGB (R=$B_{W}$,
G=$R$, B=$I$) images with radio contours overlaid to ensure that
the match between optical and radio counterparts is correct and to
eliminate likely contaminants due to image artifacts or spurious matches
still present in our sample. We identified 83 spectroscopic quasars
with $5\sigma$ LOFAR detections. We identified the photometric quasars
in Section \ref{subsec:final_qso_sample} calculated their photometric
redshifts in Section \ref{sec:photo_zs}, and used a morphology cut
in the optical as a function of redshift to eliminate extended sources.

\noindent 
\begin{figure*}[tp]
\begin{centering}
\includegraphics[clip,scale=0.39]{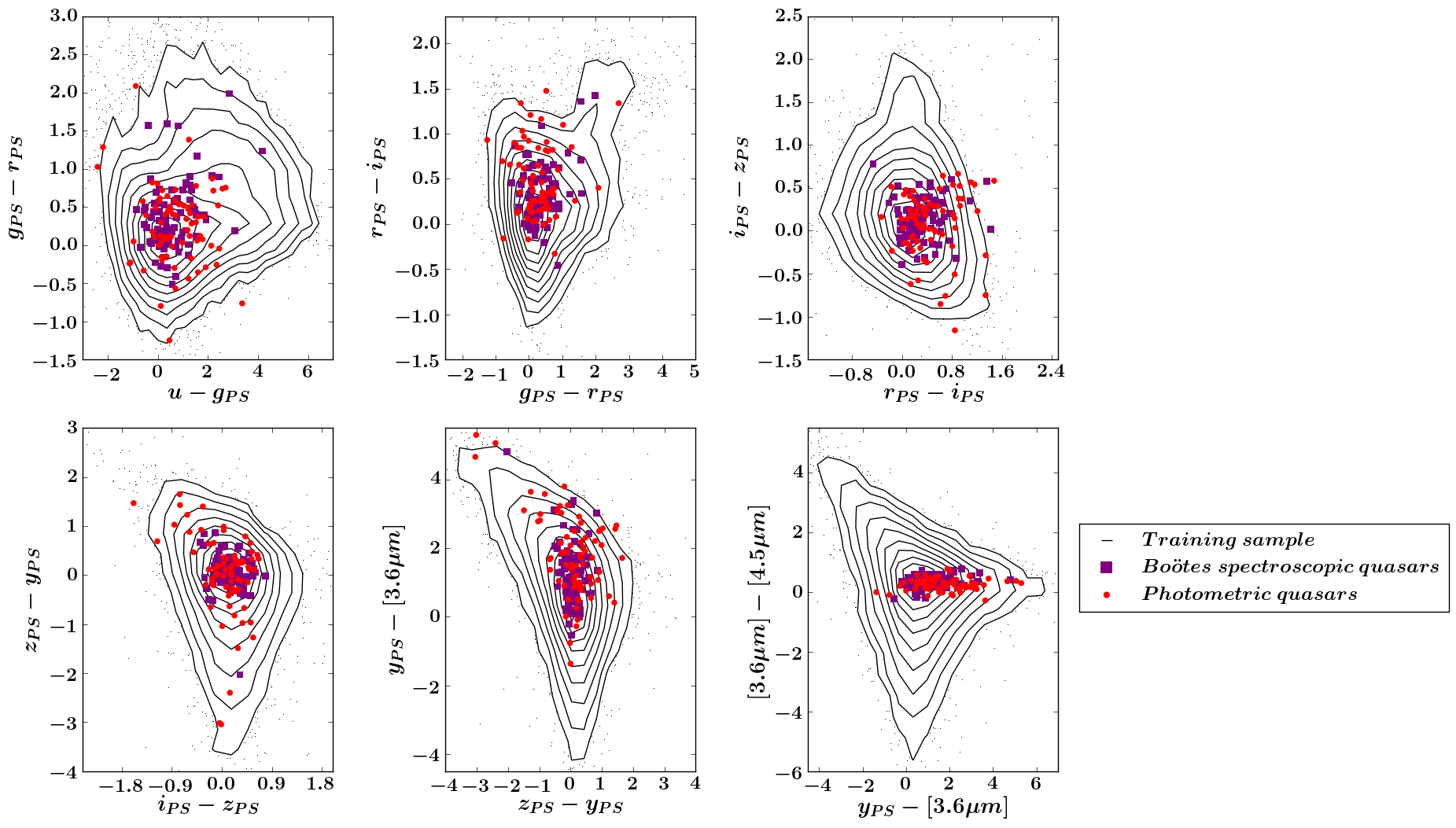}
\par\end{centering}
\centering{}\centering\caption{\label{fig:colors_sdss} Comparison between of the colors of the training
and NDWFS-Bo\"otes (photometric and spectroscopic quasars) samples.
Black contours and points denote the spectroscopic quasars with $z>1.4$
of the training sample, while purple squares indicate all the spectroscopic
quasars in Bo\"otes with $z>1.4$. Photometric quasars in our sample
are denoted by red circles. The training sample is employed to identify
quasars in the target sample (see Section \ref{sec:classification}),
and to determine their photometric redshifts with the NW kernel regression
method (see Section \ref{sec:nadaraya_watson}).}
\end{figure*}

\section{Photometric redshifts \label{sec:photo_zs}}

\subsection{Nadaraya-Watson kernel regression \label{sec:nadaraya_watson}}

Our sample contains only 83 spectroscopic quasars. For the photometric
quasars, we estimated their photometric redshifts $z_{\textrm{phot}}$
using the Nadaraya-Watson (NW) kernel regression estimator \citep{doi:10.1137/1109020,10.2307/25049340}.
The NW estimator is part of the family of kernel regression methods,
in which the expectation of the variable $Y$ relative to a variable
$X$ does not depend on all the $X$-values, as in traditional regression
methods, but on sets of locally weighted values. A bandwidth scale
parameter determines the amount of local averaging that is performed
to obtain the estimate of $Y$. The NW estimator has been widely applied
for nonparametric classification and regression (e.g., \citealt{li2011nonparametric,campbell2012econometrics,qiu2013introduction})
and to derive photometric redshifts \citep{2007MNRAS.382.1601W,2018ApJ...859...20T}.

\noindent The NW estimate $\hat{y}$ is a weighted average of the
observed variable $y_{i}$ calculated utilizing nearby points around
the test point $x_{0}$. The estimate is calculated using the following
equation:
\noindent \begin{center}
\begin{equation}
\hat{y}\left(x_{0}\right)={\textstyle \sum_{i=1}^{N}}w_{i}\left(x_{i},x_{0}\right)y_{i},\label{eq:winhorst_size_dist-1}
\end{equation}
\par\end{center}

\noindent where
\noindent \begin{center}
\begin{equation}
w_{i}\left(x_{i},x_{0}\right)=\dfrac{K\left(x_{i},x_{0}\right)}{{\textstyle \sum_{i=1}^{N}}K\left(x_{i},x_{0}\right)}\label{eq:winhorst_size_dist}
\end{equation}
\par\end{center}

\noindent is the normalized kernel built using the local information
from the training sample, and $N$ is the number of objects in the
training sample. The kernel weighting function $K\left(x_{i},x_{0}\right)$
is chosen to have a Gaussian form:
\noindent \begin{center}
\begin{equation}
K\left(x_{i},x_{0}\right)=\exp\left(-\dfrac{1}{2h^{2}}\left\Vert x_{i}-x_{0}\right\Vert ^{2}\right),\label{eq:kernel_NW}
\end{equation}
\par\end{center}

\noindent where $h$ is the bandwidth scale that defines the region
of parameter space in which the data is averaged, and $\left\Vert x_{i}-x_{0}\right\Vert $
is the euclidean distance between the data points from the training
and test samples. A more detailed discussion about the NW estimator
can be found in \citet{hardle_1990} and \citet{RePEc:bes:jnlasa:v:102:y:2007:m:june:p:761-761}.

\noindent In this work, the training sample is composed of spectroscopic
quasars, and the distance is calculated between the colors of the
spectroscopic quasars and photometric quasars. Finally, for each photometric
quasar its photometric redshift is calculated considering all the
spectroscopic redshifts of the training sample by using the equation
\citep{2007MNRAS.382.1601W,2018ApJ...859...20T}:
\noindent \begin{center}
\begin{equation}
z_{photo}={\textstyle \sum_{i=1}^{N}}w_{i}\:z_{spec,i}.\label{eq:photo_z_NW}
\end{equation}
\par\end{center}

\subsection{Quasar training sample \label{sec:quasar_training_sample}}

To provide a training sample on which to use the NW estimator, a spectroscopic
quasar catalog is necessary. For this purpose, we used the quasar
catalog introduced in Section \ref{sec:qso_sdss_wise}. In total,
our training set contains 328956 quasars with $1.2\leq z\leq5.5$
to determine photometric redshifts using the NW estimator (see Table
\ref{tab:properties_samples}). 

\begin{table*}
\noindent \begin{centering}
\caption{The performance of the photometric redshifts for the quasar training
sample. The statistics are given for each subsample.\label{tab:training_sample}}
\begin{tabular}{ccccccccc}
\hline 
{\tiny{}Sample} & {\tiny{} Number of quasars} & {\tiny{} $\left\langle \triangle z\right\rangle $} & {\tiny{} $\sigma(\triangle z)$} & {\tiny{} $R_{0.3}$} & {\tiny{} $\left\langle \triangle z_{\textrm{norm}}\right\rangle $} & {\tiny{} $\sigma(\triangle z_{\textrm{\textrm{norm}}})$} & {\tiny{}$R_{0.1}^{\textrm{norm}}$} & {\tiny{}$R_{0.2}^{\textrm{norm}}$}\tabularnewline
 &  &  &  &  &  &  &  & \tabularnewline
\hline 
{\tiny{}Bo\"otes (All)} & {\tiny{}1193} & {\tiny{}-0.060 } & {\tiny{}0.35 } & {\tiny{}0.76 } & {\tiny{}-0.013 } & {\tiny{}0.10 } & {\tiny{}0.78 } & {\tiny{}0.94 }\tabularnewline
{\tiny{}Bo\"otes ($\textrm{err}\leq0.2$)} & {\tiny{}535} & {\tiny{}-0.015 } & {\tiny{}0.22 } & {\tiny{}0.87 } & {\tiny{}-0.0008 } & {\tiny{}0.071 } & {\tiny{}0.87 } & {\tiny{}0.97 }\tabularnewline
{\tiny{}Bo\"otes ($\textrm{err}\leq0.3$)} & {\tiny{}689} & {\tiny{}-0.033 } & {\tiny{}0.25 } & {\tiny{}0.84 } & {\tiny{}-0.006 } & {\tiny{}0.080 } & {\tiny{}0.84 } & {\tiny{}0.97 }\tabularnewline
{\tiny{}Bo\"otes ($\textrm{err}\leq0.5$)} & {\tiny{}881} & {\tiny{}-0.04 } & {\tiny{}0.26 } & {\tiny{}0.82 } & {\tiny{}-0.007 } & {\tiny{}0.084 } & {\tiny{}0.82 } & {\tiny{}0.96 }\tabularnewline
{\tiny{}$20\%$ Training sample} & {\tiny{}65573} & {\tiny{}-0.006 } & {\tiny{}0.28 } & {\tiny{}0.81 } & {\tiny{}0.005 } & {\tiny{}0.09 } & {\tiny{}0.82 } & {\tiny{}0.95 }\tabularnewline
\hline 
\end{tabular}
\par\end{centering}
$\;$

\emph{Notes:} {\tiny{}The experiments all used the same training sample
that does not include any spectroscopic quasars from the NDWFS-Bo\"otes
field, with the exception of the experiment where the Bo\"otes quasars
are included in the training sample.}{\tiny \par}
\end{table*}

\subsection{Redshift estimation \label{sec:redshift_estimation}}

The distance between the color indices ($u-g_{\textrm{PS}}$, $u-r_{\textrm{PS}}$,
$g_{\textrm{PS}}-r_{\textrm{PS}}$, $r_{\textrm{PS}}-i_{\textrm{PS}}$,
$i_{\textrm{PS}}-z_{\textrm{PS}}$, $z_{\textrm{PS}}-y_{\textrm{PS}}$,
$y_{\textrm{PS}}-[3.6\,\mu\textrm{m}]$, $[3.6\,\mu\textrm{m}]-[4.5\,\mu\textrm{m}]$)
of the spectroscopic quasars from the training sample and the corresponding
photometric objects are used as inputs to build the kernel matrix
as given in eq. \ref{eq:kernel_NW}. An important decision in building
the kernel is the choice of the bandwidth scale $h$. With smaller
values of $h$ nearby data points have more weight, while larger values
of $h$ result in an increasing contribution of distant data points.
Finally, the kernel functions determined using eq. \ref{eq:photo_z_NW}
are used as weights in the computation of the photometric redshift.

\begin{figure}[tp]
\begin{centering}
\includegraphics[bb=0bp 0bp 387bp 390bp,clip,scale=0.65]{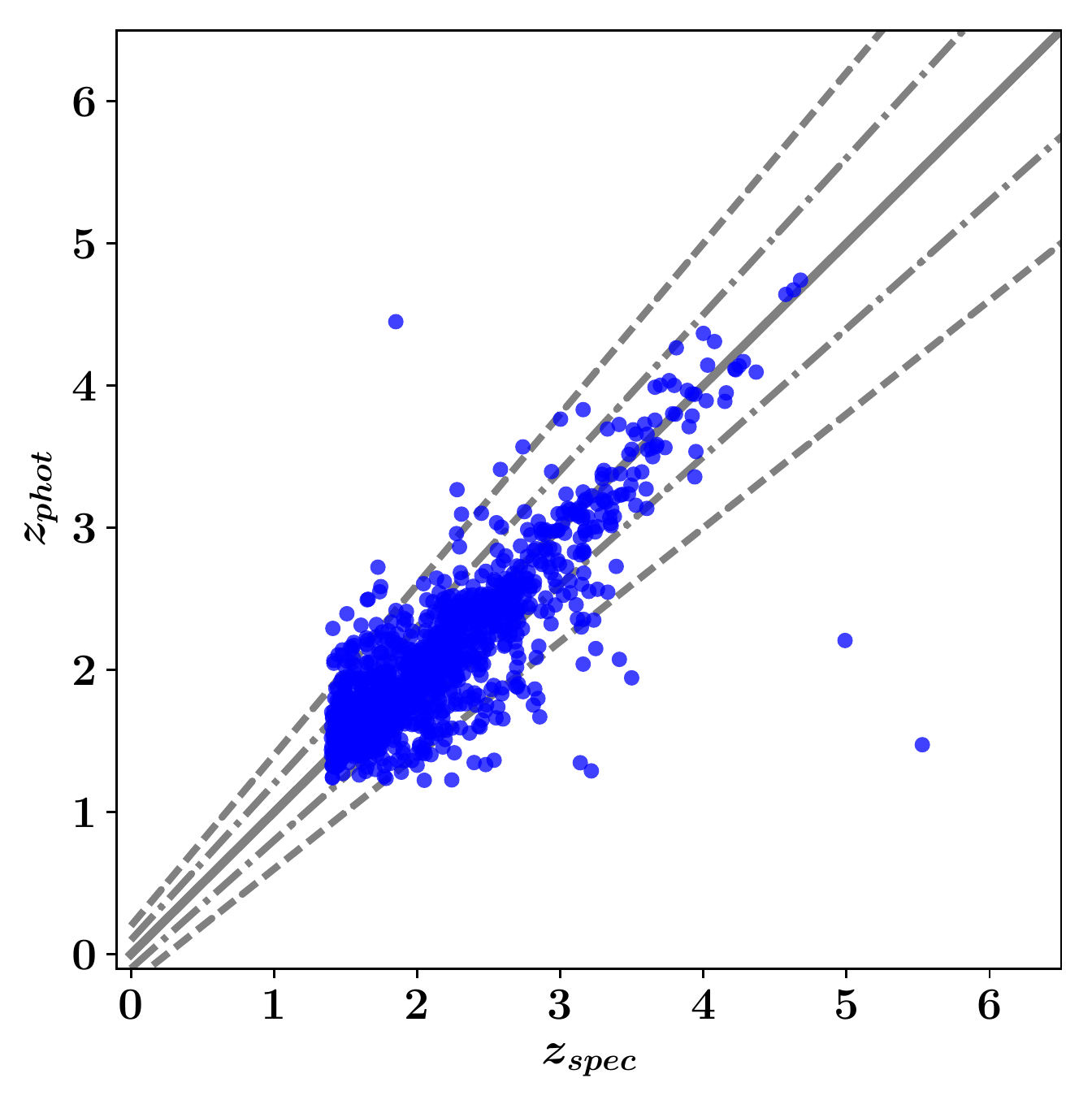}
\par\end{centering}
\centering{}\centering\caption{\label{fig:redshift_comparison_nw} Photometric $z_{\textrm{photo}}$
versus spectroscopic $z_{\textrm{spec}}$ redshift for spectroscopic
quasars in the NDWFS-Bo\"otes region. The photometric redshifts are
obtained using the NW kernel regression. The grey line indicates the
one-to-one $z_{\textrm{NW}}=z_{\textrm{spec}}$ relation, and the
dashed-dotted and dashed lines indicate the $z_{\textrm{NW}}-z_{\textrm{spec}}=\pm0.10\times(1+z_{\textrm{spec}})$
and $z_{\textrm{NW}}-z_{\textrm{spec}}=\pm0.20\times(1+z_{\textrm{spec}})$
relations, respectively. }
\end{figure}

\noindent The quality of the photometric redshifts determined with
the NW estimator is investigated utilizing two quasar samples. The
first sample is composed of 1193 quasars from  the quasar training
set described in Section \ref{sec:qso_sdss_wise} that are located
in the NDWFS-Bo\"otes field. The second sample is a subsample selected
randomly from the training set. We measure the performance of the
photometric redshifts in the samples using the following statistics:
\begin{itemize}
\item mean of the difference between photometric and spectroscopic redshifts,
$\left\langle \triangle z\right\rangle =\left\langle (z_{\textrm{phot}}-z_{\textrm{spec}})\right\rangle $,
unclipped;
\item standard deviation of the $\triangle z$, $\sigma(\triangle z)$;
\item fraction of quasars with $\left|\triangle z\right|<0.3$, $R_{0.3}$;
\item mean of normalized $\triangle z$, $\left\langle \triangle z_{\textrm{norm}}\right\rangle =\left\langle (z_{\textrm{phot}}-z_{\textrm{spec}})/(1+z_{\textrm{spec}})\right\rangle $,
unclipped;
\item standard deviation of the normalized bias, $\sigma(\triangle z_{\textrm{\textrm{norm}}})$;
\item fraction of quasars with $\left|\triangle z_{\textrm{norm}}\right|<0.1$
$\left(\left|\triangle z_{\textrm{norm}}\right|<0.2\right)$, $R_{0.1}^{\textrm{norm}}$
($R_{0.2}^{\textrm{norm}}$).
\end{itemize}
These results are summarized in Table \ref{tab:training_sample}.
We tested practically all the values in the range $0.01\leq h\leq0.5$,
and found that $h=0.09$ gives the best performance and least scatter
for the Bo\"otes spectroscopic quasars (see Table \ref{tab:training_sample}).
This $h$ value is slightly higher than the $h=0.05$ used in previous
estimations of photometric redshifts using the NW estimator (e.g.,
\citealt{2007MNRAS.382.1601W,2018ApJ...859...20T}). Figure \ref{fig:redshift_comparison_nw}
shows the photometric versus spectroscopic redshifts for the spectroscopic
sample of Bo\"otes quasars. At low-z, the dispersion of the photometric
is slightly higher in comparison with the high-z estimations. This
is expected as at low-z there are less spectral features for the NW
method to exploit and predict trends on the training sample. 

\begin{figure}[tp]
\begin{centering}
\includegraphics[clip,scale=0.28]{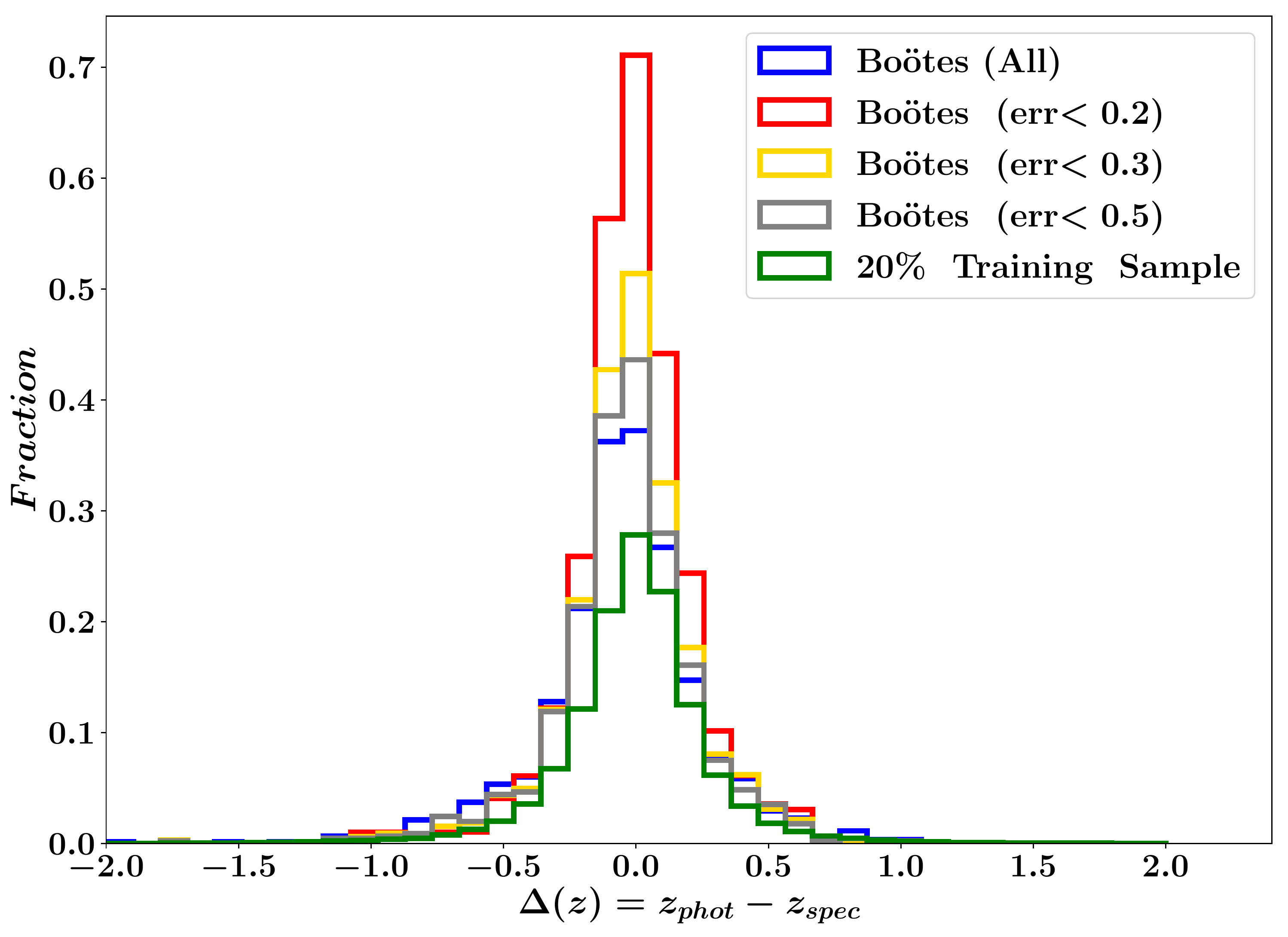}
\par\end{centering}
\centering{}\caption{\label{fig:redshift_comparison} Normalized histogram of the bias
$\bigtriangleup z=z_{\textrm{phot}}-z_{\textrm{spec}}$ between photometric
and spectroscopic redshifts for different samples. The phometric redshifts
are obtained using the NW kernel regression. Around $76\%$ of the
spectroscopic quasars in the Bo\"otes field have photometric redshifts
that are correct within $\left|\bigtriangleup z\right|=0.3$ (see
Table \ref{tab:training_sample}).}
\end{figure}

Figure \ref{fig:redshift_comparison} presents the normalized histogram
of the bias $\bigtriangleup z=z_{\textrm{phot}}-z_{\textrm{spec}}$
between photometric and spectroscopic redshifts for the different
experiments listed in Table \ref{tab:training_sample}. The first
row lists the result for the sample of Bo\"otes quasars. This shows
that the majority of Bo\"otes quasars have good redshift estimations.
The number of outliers is small with $76\%$ of the quasars having
photometric redshifts that do not differ more than $\left|\triangle z\right|\leq0.3$
from their spectroscopic redshifts. A good comparison for the redshift
accuracy of the Bo\"otes quasars can be obtained using a validation
sample that is randomly selected from a training sample. This validation
sample contains 65573 quasars, and its size is $20\%$ of the total
training sample. The result for this sample is listed in the last
row of Table \ref{tab:training_sample}. The Bo\"otes sample performs
worse than the $20\%$ of the training sample in terms of bias, scatter,
and fractions of quasars with correct redshifts. In order to better
understand the reasons for the difference in performance between the
two samples we examine the performance of the NW method as a function
of the photometric errors in the SDSS/Pan-STARRS1 bands. By restricting
the Bo\"otes sample only to quasars with photometric errors in the
SDSS/Pan-STARRS1 bands smaller than $err\leq\left[\,0.2,\,0.3,\,0.5\,\right]$,
respectively, its performance gets closer to that of the $20\%$ training
sample, and this further improves for the case when the photometric
errors are smaller or equal to 0.3 (Table \ref{tab:training_sample},
third-fourth rows).

Finally, we compared the performance of the NW regression kernel with
other photometric-redshift algorithms. For instance, \citet{2001AJ....122.1151R}
and \citet{2004ApJS..155..243W} reported that $70\%$ and $83\%$
of their predicted photometric redshifts are correct within $\left|\delta z\right|\leq0.3$.
These authors used empirical methods that used the color-redshift
relations to derive photometric redshifts using early SDSS data. \citet{2017AJ....154..269Y}
considered the asymmetries in the relative flux distributions of quasars
to estimate quasar photometric redshifts obtaining an accuracy of
$R_{0.2}^{\textrm{norm}}=87\%$ for SDSS/WISE quasars. \citet{2019MNRAS.485.4539J}
employed the SVM and XGBoost \citep{2016arXiv160302754C} algorithms
to achieve an average accuracy of $R_{0.2}^{\textrm{norm}}\simeq89\%$
for the photometric redshifts of their Pan-STARRS1/WISE quasars. \citet{2017ApJ...851...13S}
used SDSS/WISE adjacent flux ratios to train the RF and SVM methods
to obtain average results of $R_{0.2}^{\textrm{norm}}\simeq93\%$
for their photometric redshifts. In summary, despite the fact that
all these algorithms employed different samples and redshift intervals,
their accuracy is consistent with the results obtained in this work
using the NW regression kernel. Finally, it is also important to mention
that the reason for using the NW regression is its better performance
for our quasar sample in comparison with other ML methods. For example,
the RF regression underperforms in all the statistics indicated in
Table \ref{tab:training_sample}. Using the RF estimator, for the
Bo\"otes spectroscopic quasars (the \emph{All }sample in Table \ref{tab:training_sample})
we obtained less accurate photometric redshifts with $R_{0.3}=0.70$
and $R_{0.2}^{\textrm{norm}}=0.90$. Therefore, we calculated our
photometric redshifts using the NW regression kernel.

\noindent 

\noindent 

\section{Quasar sample \label{subsec:final_qso_sample}}

\subsection{\textbf{Final sample}}

We used the NW regression algorithm to assign a photometric redshift
to each photometric quasar detected by LOFAR. To eliminate potential
contamination by low-z galaxies, we restricted our quasar sample only
to point sources. The SDSS photometry pipeline\footnote{https://www.sdss.org/dr14/algorithms/classify/}
classifies an object as point-like (star) or extended (galaxy) source
based on the difference between its \texttt{PSF} and \texttt{cMODELMAG}
magnitudes\footnote{https://www.sdss.org/dr14/algorithms/magnitudes/}.
Various methods have been proposed to perform the morphological star-galaxy
separation with photometric data by adding Bayesian priors to the
aforementioned magnitude differences \citep{2002ApJ...579...48S},
and using decision tree classifiers \citep{2011AJ....141..189V}.
Here we employed a criterion derived directly from the Bo\"otes spectroscopic
quasars by considering the difference $\triangle_{mag}$ between the
\texttt{PSF} and \texttt{cMODELMAG }magnitudes in the SDSS-$r$ band
as a function of redshift. Spectroscopic quasars were binned according
to their redshifts to calculate the magnitude difference as the quantile
that contains $95\%$ of the quasars in the corresponding bin. The
redshift intervals match those utilized to derive the luminosity function
in Section \ref{sec:qlf}. Objects with magnitude differences less
than the $\triangle_{mag}$ value in their corresponding redshift
bin were considered to be point sources, and are included in our RSQs
sample. Figure \ref{fig:psf_cmodel_diff} shows the magnitude differences
as a function of redshift. Finally, we restricted our RSQs sample
to magnitudes of $16.0\leq i_{\textrm{PS}}\leq23.0$ to avoid extrapolation
beyond the range of the quasar training sample (see Table \ref{tab:properties_samples}).
The resulting catalog consists of 17924 objects, with 104 sources
having a LOFAR counterpart within a radius of $2^{\prime\prime}$.

\noindent 
\begin{figure}[tp]
\begin{centering}
\includegraphics[clip,scale=0.33]{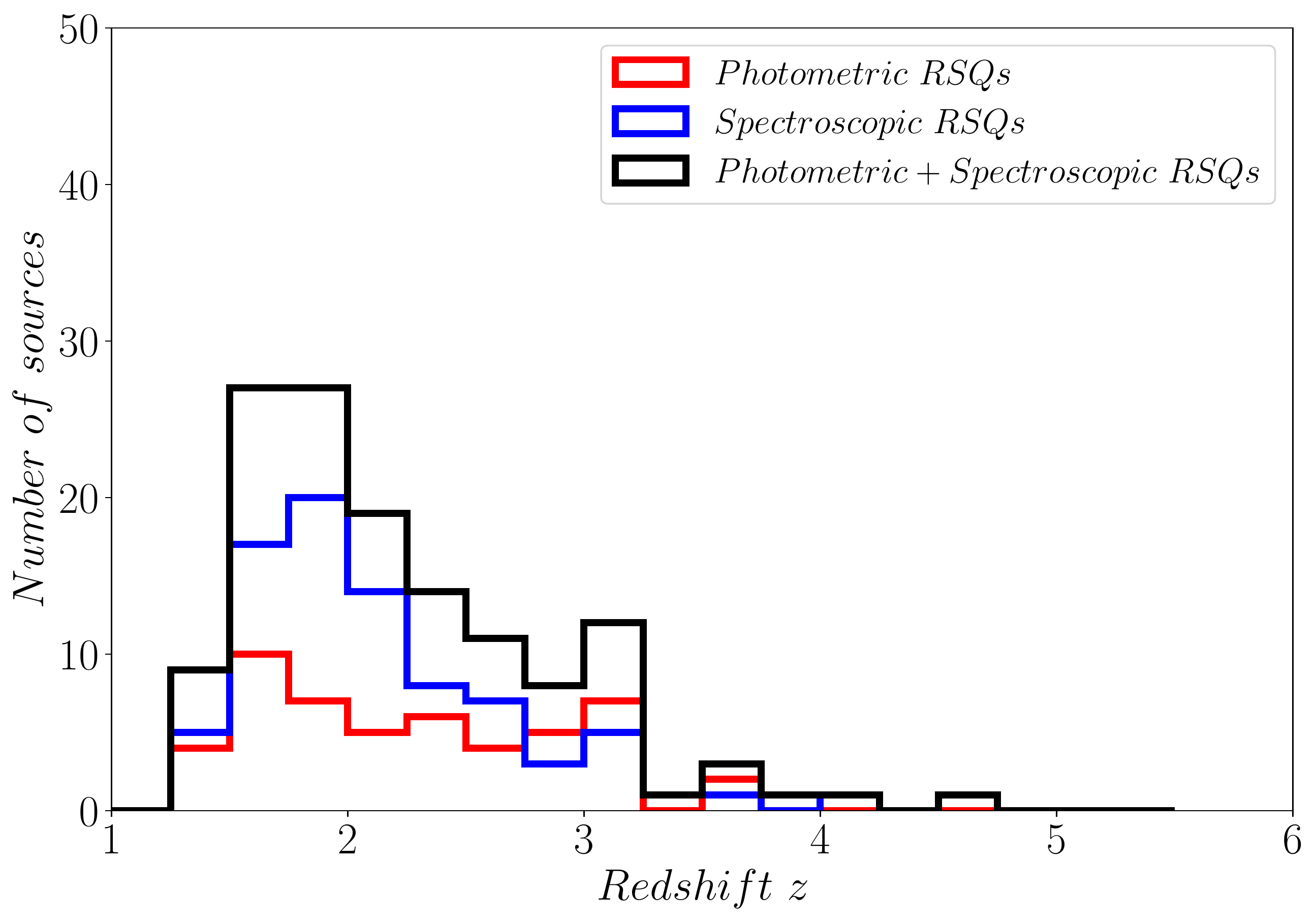}
\par\end{centering}
\centering{}\centering\caption{\label{fig:redshift_distribution} Redshift distribution of photometric
(red) and spectroscopic (blue) RSQs. Also, the combined redshift distribution
(black) of photometric and spectroscopic RSQs is plotted.}
\end{figure}

\noindent 
\begin{figure}[tp]
\begin{centering}
\includegraphics[clip,scale=0.39]{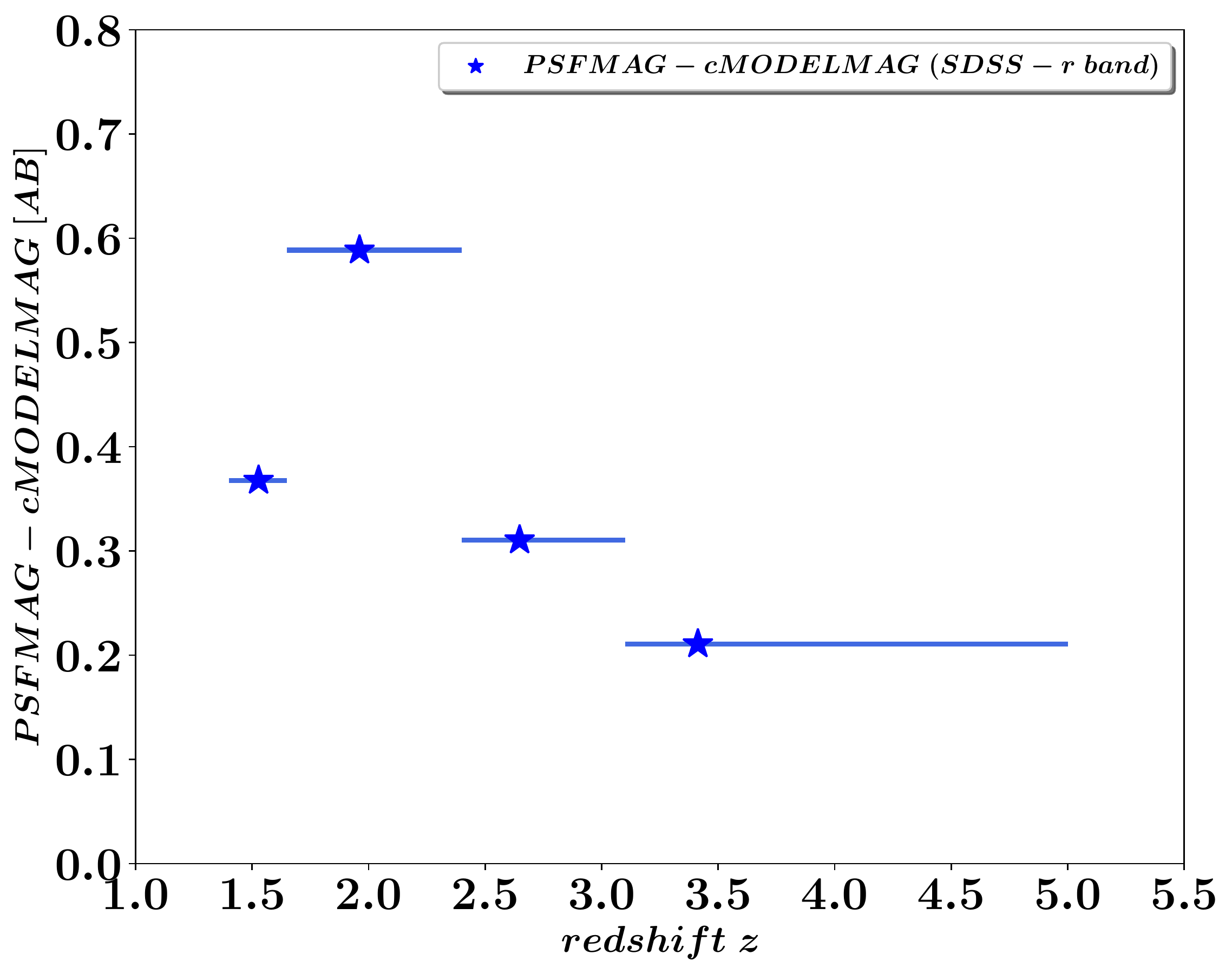}
\par\end{centering}
\centering{}\centering\caption{\label{fig:psf_cmodel_diff}\emph{ }Difference between the \texttt{PSF}
and \texttt{cMODELMAG }magnitudes in the $r$ band to separate point-like
and extended sources as a function of redshift. The difference values
are calculated considering the quantile that contains $95\%$ of the
quasars in the corresponding redshift bin. The redshift intervals
match those used to derive the luminosity function in Section \ref{sec:qlf}.}
\end{figure}

\noindent 
\begin{figure*}[tp]
\begin{centering}
\includegraphics[clip,scale=0.6]{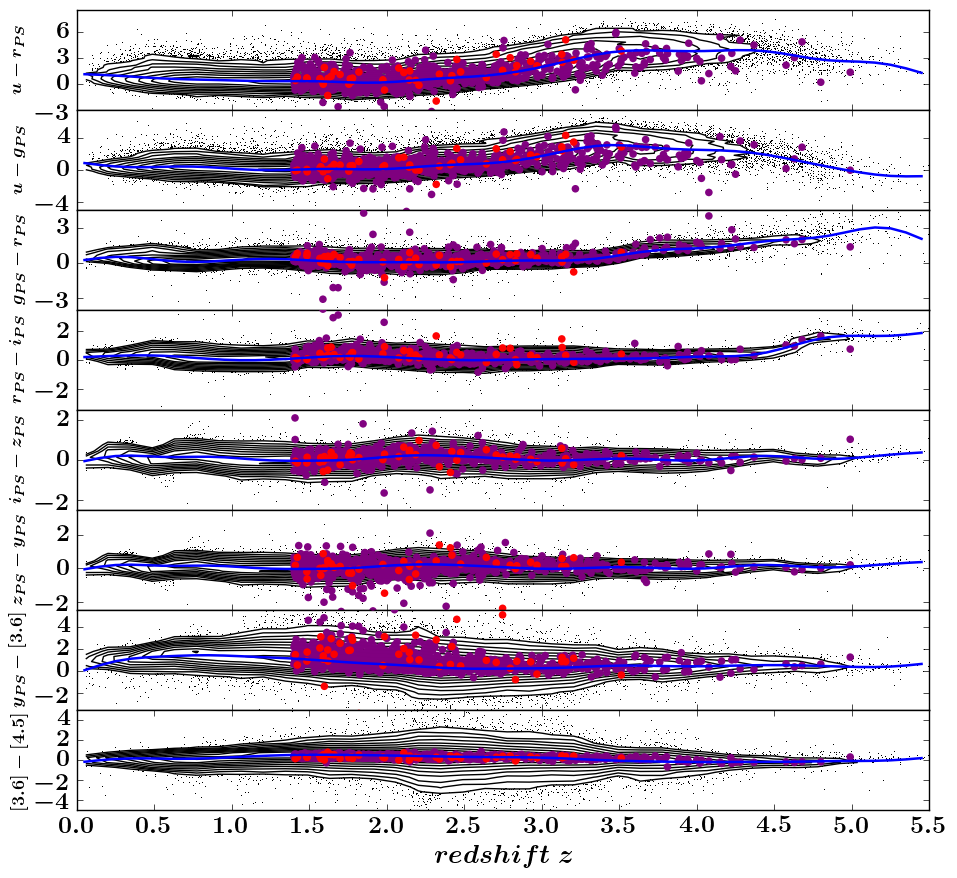}
\par\end{centering}
\centering{}\centering\caption{\label{fig:qso_colors_z}\emph{ }Quasar colors versus redshift for
different quasar samples between $z=1.4$ and $z=5.5$. \emph{Red
points}: RSQs with photometric redshifts; \emph{purple points}: Bo\"otes
spectroscopic quasars; \emph{black contours} and points: the color
distributions of the quasar training sample from Section \ref{sec:qso_sdss_wise};
\emph{blue lines:} mean color\textendash redshift relations derived
from the quasar training sample. }
\end{figure*}

After the morphological cut, we carefully inspected the false-color
RGB images (R=$B_{W}$, G=$R$, B=$I$) of the photometric quasars
detected with $5\sigma$ significance in the LOFAR mosaic overlaid
with radio contours to reject blended sources, artifacts, and spurious
matches to the radio data. The majority of discarded sources corresponded
to objects that are spurious matches to the LOFAR data. The final
RSQs catalog resulting from our selection contains 47 photometric
quasars and 83 spectroscopic quasars. In Figure \ref{fig:colors_sdss},
it can be seen that the colors of the photometric quasars agree well
with those of the $z>1.4$ quasars in the training and NDWFS-Bo\"otes
quasar samples. The importance of the use of the LOFAR data is clear
as it allows for a significant reduction in the number of stellar
contaminants in the sample of objects taken as quasars by the ML algorithms.
The initial sample is reduced from $\sim18000$ objects to only 47
RSQs detected at $5\sigma$ significance in the LOFAR mosaic. This
represents a reduction of two orders of magnitude in the number of
contaminants. 

\begin{figure}[tp]
\begin{centering}
\includegraphics[clip,scale=0.33]{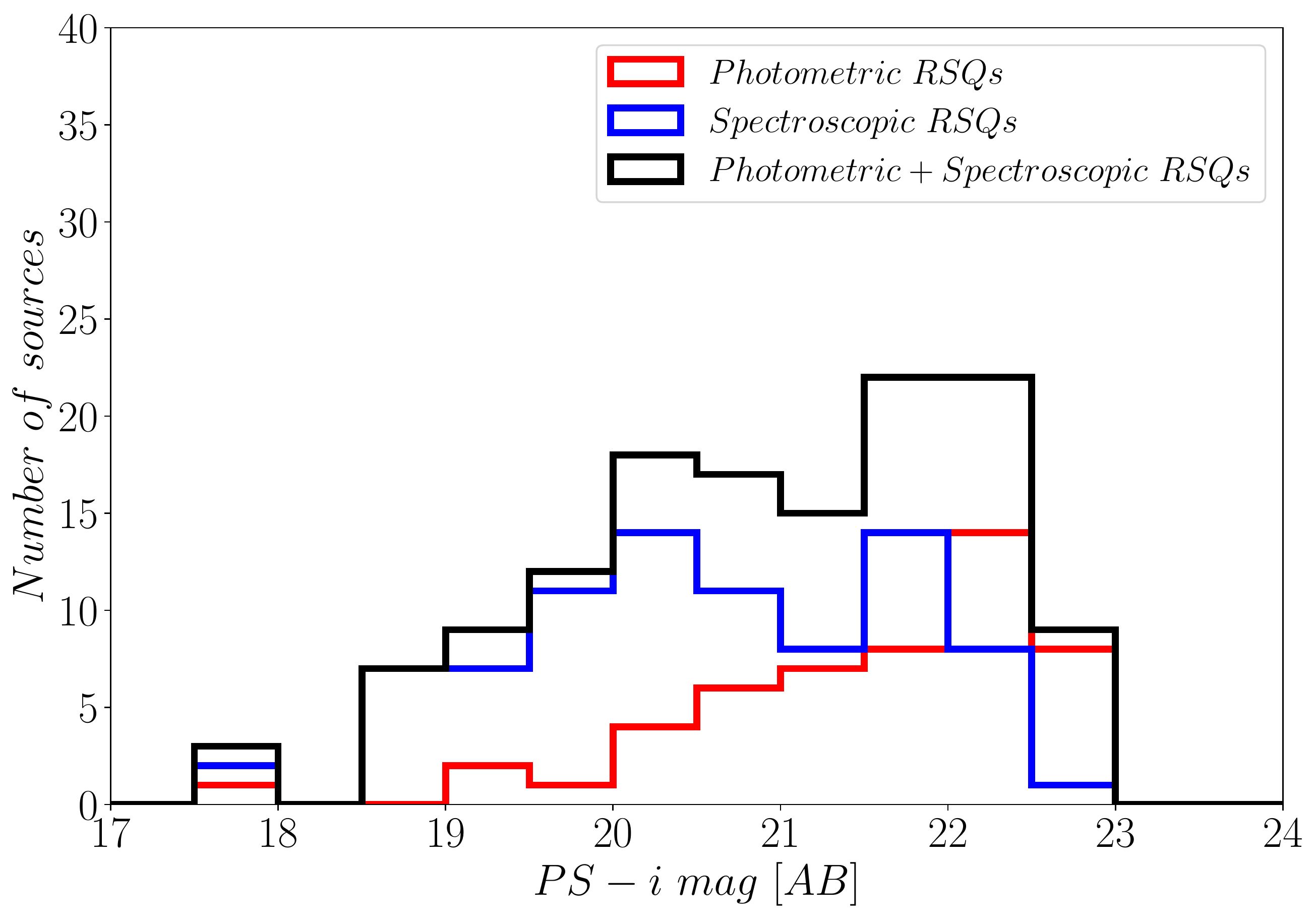}
\par\end{centering}
\begin{centering}
\includegraphics[clip,scale=0.33]{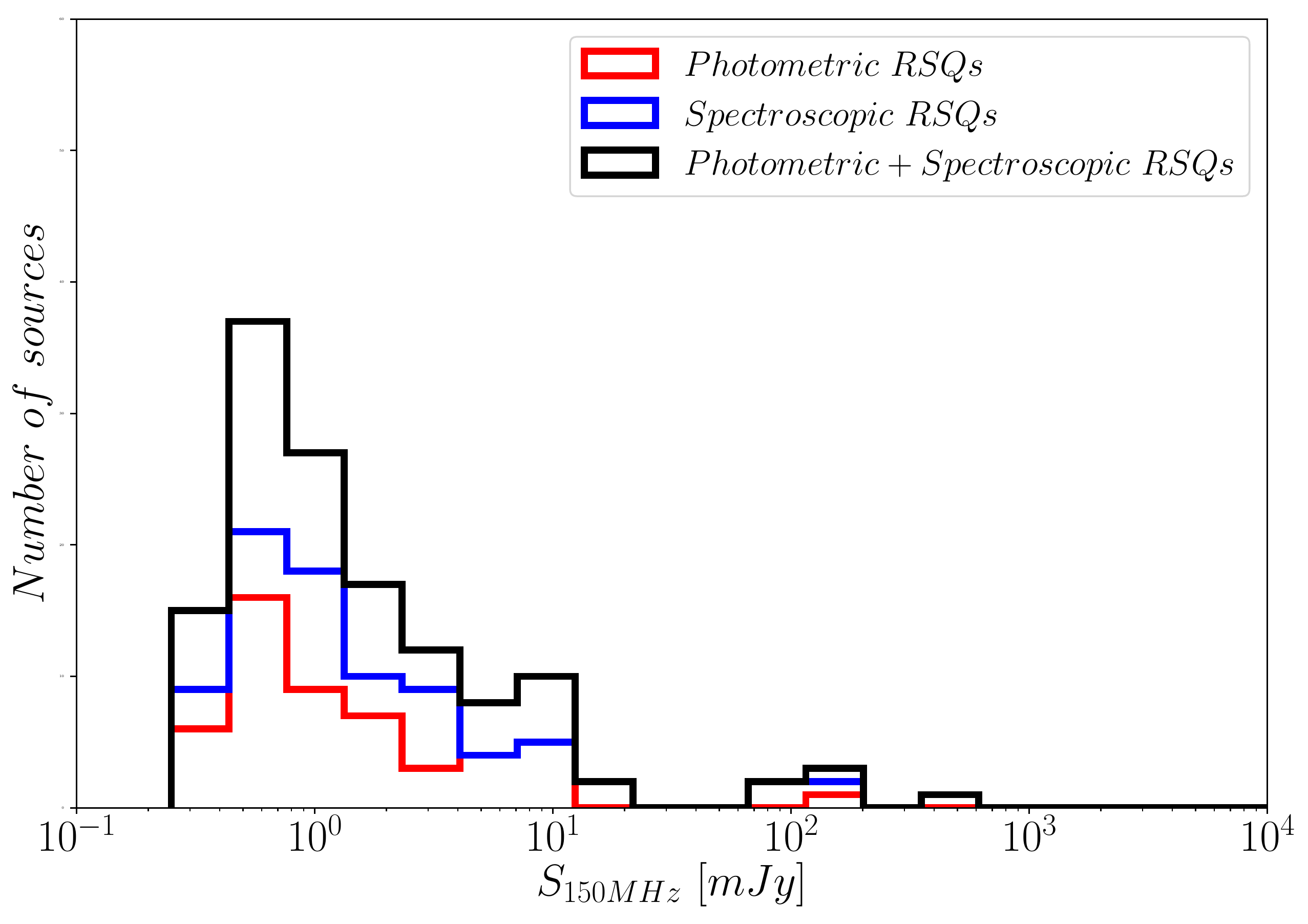}
\par\end{centering}
\centering{}\centering\caption{\label{fig:mag_flux_distribution} $i_{\textrm{PS}}$ (top) and total
flux $S_{\textrm{150MHz}}$ (bottom) distributions of photometric
(red) and spectroscopic (blue) RSQs. Also, the combined redshift distribution
(black) of photometric and spectroscopic RSQs is plotted. The method
employed to select the quasars is described in Section \ref{subsec:classification_algorithms}. }
\end{figure}

\begin{figure}[tp]
\centering{}\includegraphics[bb=0bp 0bp 772bp 764bp,clip,scale=0.31]{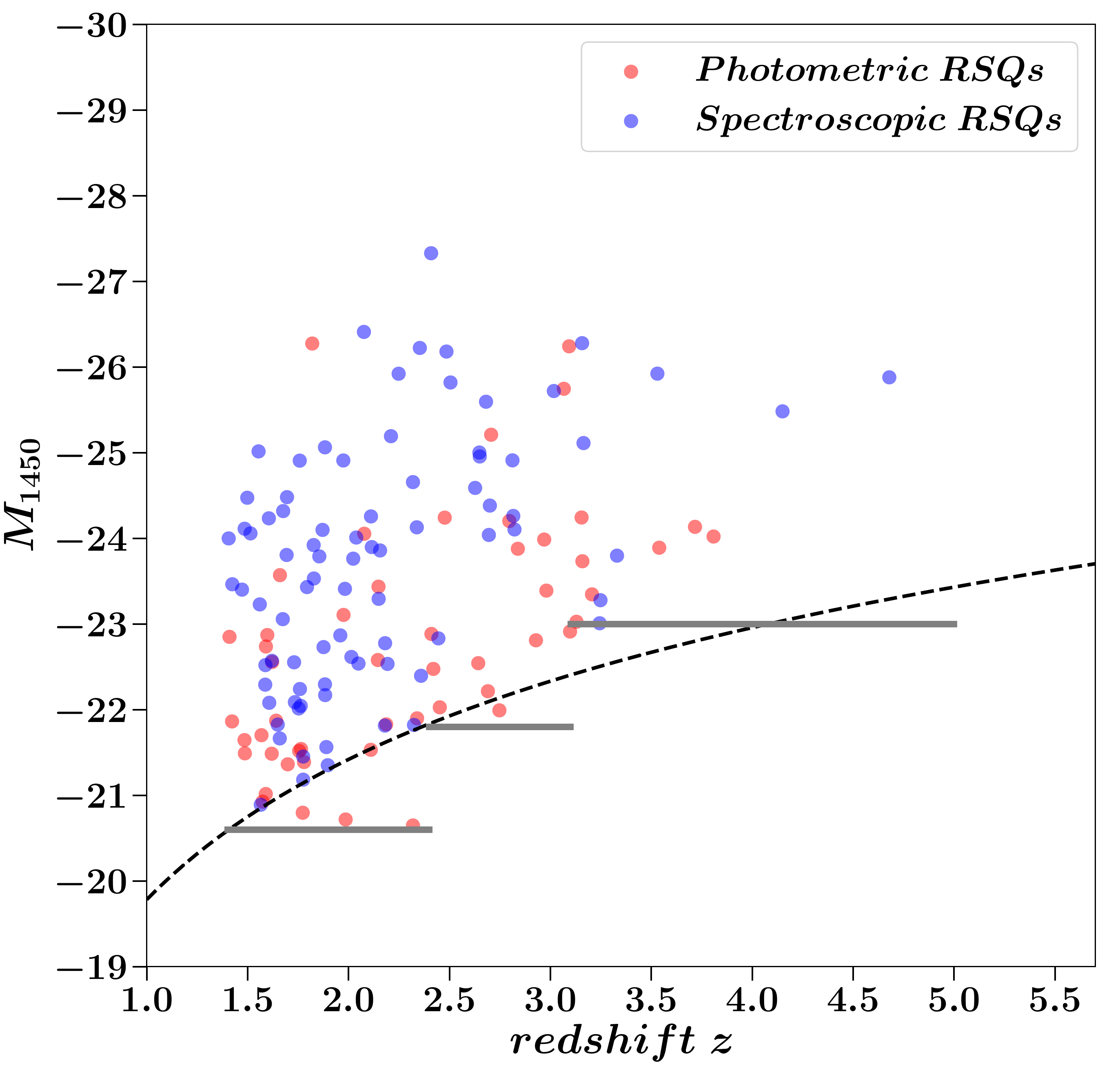}\centering\caption{\label{fig:M1450_redshift} Absolute magnitude $M_{\textrm{1450}}$
versus redshift for photometric (red) and spectroscopic (blue) RSQs
in our sample. To minimize incompleteness due to incomplete $M_{\textrm{1450}}$
bins while retaining the maximum numbers of quasars for estimating
the luminosity function, we consider only RSQs with $M_{\textrm{1450}}\leq\left[-20.6,-21.8,-23.0\right]$
at $1.4\leq z<2.4$, $2.4\leq z<3.1,$ and $3.1\leq z<5.0$; respectively
. The dashed line denotes the magnitude limit $i_{\textrm{PS}}=23.0$.
This limit is calculated assuming a quasar continuum described by
a power-law with slope $\alpha=-0.5$ with no emission line contribution
or $\textrm{Ly}_{\alpha}$ forest blanketing.}
\end{figure}

\begin{figure}[tp]
\centering{}\includegraphics[bb=0bp 0bp 788bp 771bp,clip,scale=0.33]{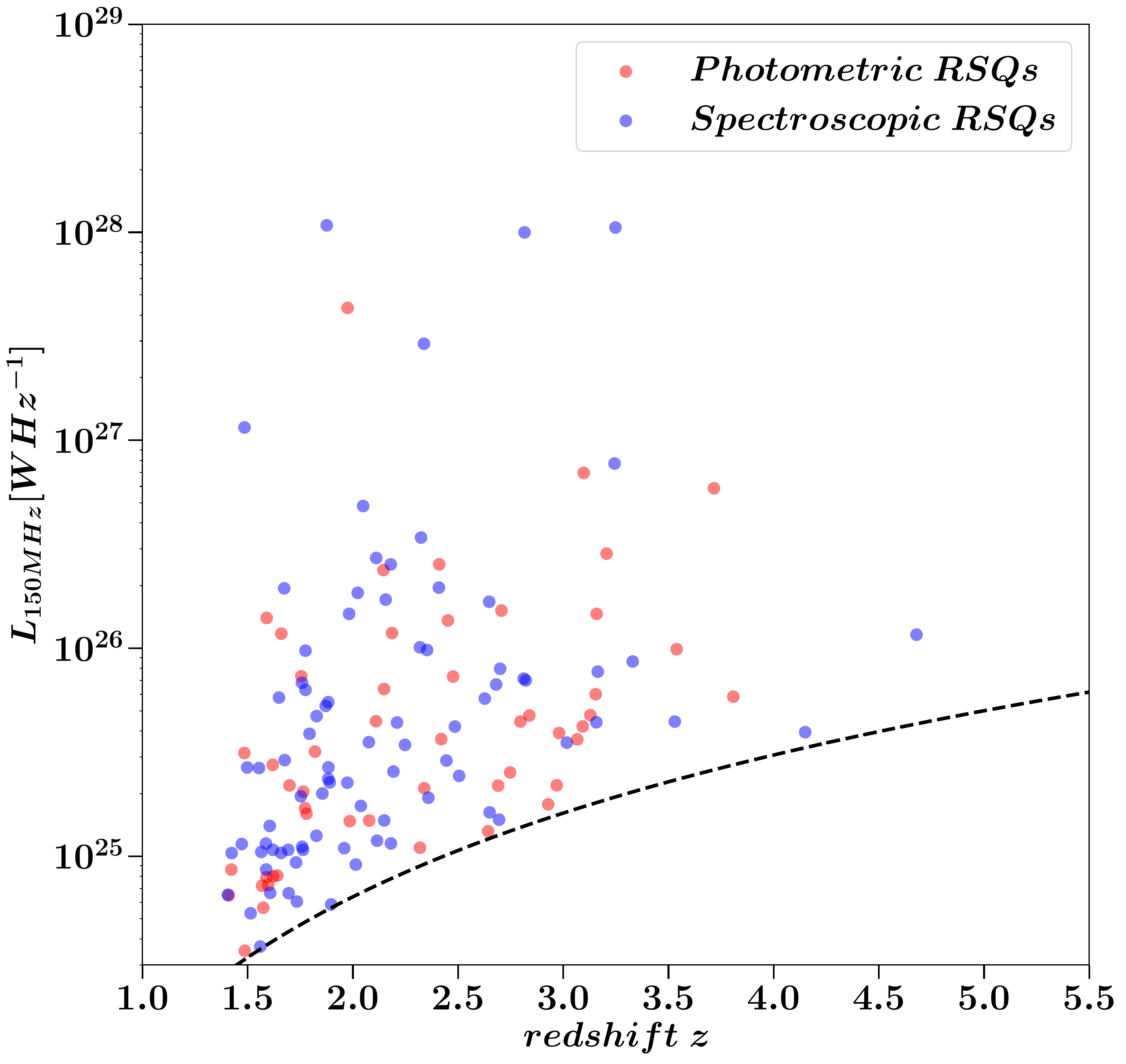}\centering\caption{\label{fig:L150_redshift} Rest frame absolute luminosity density
at 150 MHz versus redshift for photometric (red) and spectroscopic
(blue) RSQs in our sample. The solid line denotes the $5\sigma$ flux
limit ($275\:\mu\textrm{Jy}$) of the Bo\"{o}tes observations presented
by \citet{2018arXiv180704878R}.}
\end{figure}

In Figure \ref{fig:qso_colors_z}, we compare the colors of the photometric
quasars with those of the training sample and Bo\"otes spectroscopic
quasars as functions of redshift. The color-redshift spaces occupied
by photometric RSQs are in good agreement with those of Bo\"otes
spectroscopic quasars and the quasar training sample. Figure \ref{fig:redshift_distribution}
displays the redshift distribution of photometric and spectroscopic
quasars. At $z\lesssim2.8$, the majority of the quasars in our sample
are spectroscopic. Considering that $77\%$ of the photometric quasars
have photometric errors smaller than $err\leq0.5$, we conclude that
the accuracy of their photometric redshifts is similar or slightly
worse to that of the Bo\"otes sample with photometric errors that
are $err\leq0.5$ (see Table \ref{tab:training_sample}). The $i_{\textrm{PS}}$-band
magnitude and 150 MHz flux distributions of the RSQs sample are presented
in Figure \ref{fig:mag_flux_distribution}, while the absolute magnitude
and radio luminosity are displayed in Figures \ref{fig:M1450_redshift}
and \ref{fig:L150_redshift}, respectively. The absolute magnitudes
are calculated using the K-correction discussed in Section \ref{sec:simuqso}.
The majority of RSQs (130 in total) in our sample are unresolved or
resolved into single-component radio sources in the LOFAR-Bo\"otes
mosaic with a resolution of $\sim5^{\prime\prime}$. In our sample,
only 11 quasars present radio-morphologies consistent with a core-jet
structure. Appendix \ref{sec:appendix_B} presents a selection of
false-color RGB images of spectroscopic and photometric RSQs from
our final sample. The catalog of spectrocospic and photometric RSQs
are only available in electronic form\footnote{The catalogs can be accessed at the CDS via anonymous ftp to cdsarc.u-strasbg.fr
(130.79.128.5) or via http://cdsweb.u-strasbg.fr/cgi-bin/qcat?J/A+A/ }.

\noindent 

\noindent 

\noindent 

\subsection{Spectral energy distribution of photometric quasars \label{subsec:sed_photo_quasars}}

\noindent In this section, we used the deep photometry available in
the NDWFS-Bo\"otes field to calculate the spectral energy distributions
(SEDs) of the photometric quasars. This additional deep photometry
provides an additional piece of information to confirm the validity
of the photometrically selected sources selected using the ML algorithms
as robust candidate quasars. For this purpose, we used the I-band
matched photometry catalog presented by \citet{2007ApJ...654..858B}.
The $5\sigma$ limiting AB magnitudes for the relevant filters are
provided in Table \ref{tab:filters_bootes}. Each filter image was
convolved to a common pixel scale, so that all the point spread functions
(PSFs) images matched a Moffat profile. In these images, photometry
was computed with SExtractor \citep{1996AAS..117..393B} using the
$I$-band as the detection band. This catalog contains more than two
millions of $I$-band selected sources.

\noindent 
\begin{table}
\noindent \centering{}\caption{Characteristics of the filters used in the NDWFS-Bootes field and
their $3\sigma$ AB limiting magnitudes. The aperture limits were
computed using a within $2^{''}$ diameter aperture for the optical
and near-infrared filters, $12^{''}$ for the NUV band, $4^{''}$
and $6^{''}$ for the IRAC and MIPS bands, respectively. \label{tab:filters_bootes}}
\begin{tabular}{cccc}
\hline 
{\tiny{}Filter} & {\tiny{}Central wavelength} & {\tiny{}FWHM} & {\tiny{}Depth}\tabularnewline
 & {\tiny{}{[}\AA{]}} & {\tiny{}{[}\AA{]}} & {\tiny{}{[}AB,$\:5\sigma${]}}\tabularnewline
\hline 
{\tiny{}GALEX/NUV } & {\tiny{}2329} & {\tiny{}796} & {\tiny{}25.5}\tabularnewline
{\tiny{}$U_{\textrm{spec}}$ } & {\tiny{}3590} & {\tiny{}540} & {\tiny{}25.2}\tabularnewline
{\tiny{}$B_{w}$ } & {\tiny{}4111} & {\tiny{}1275} & {\tiny{}25.4 }\tabularnewline
{\tiny{}$R$ } & {\tiny{}6407 } & {\tiny{}1700} & {\tiny{}25.0 }\tabularnewline
{\tiny{}$I$ } & {\tiny{}7540} & {\tiny{}1915} & {\tiny{}24.9 }\tabularnewline
{\tiny{}$Z_{\textrm{Subaru}}$ } & {\tiny{}8204} & {\tiny{}1130} & {\tiny{}24.1 }\tabularnewline
{\tiny{}$Y$ } & {\tiny{}9840} & {\tiny{}420} & {\tiny{}23.1}\tabularnewline
{\tiny{}$J$ } & {\tiny{}12493} & {\tiny{} 1787} & {\tiny{}22.9 }\tabularnewline
{\tiny{}$H$} & {\tiny{}16326} & {\tiny{}3071 } & {\tiny{}21.5}\tabularnewline
{\tiny{}$K$ } & {\tiny{}22147} & {\tiny{}4213} & {\tiny{}20.5}\tabularnewline
{\tiny{}$K_{s}$ } & {\tiny{}21453 } & {\tiny{}3220} & {\tiny{}20.7}\tabularnewline
{\tiny{}IRAC/CH1 } & {\tiny{}35465} & {\tiny{}7432} & {\tiny{}21.9}\tabularnewline
{\tiny{}IRAC/CH2 } & {\tiny{}45024} & {\tiny{}10097} & {\tiny{}21.5}\tabularnewline
{\tiny{}IRAC/CH3 } & {\tiny{}57156} & {\tiny{}13912} & {\tiny{}19.6}\tabularnewline
{\tiny{}IRAC/CH4 } & {\tiny{}78556} & {\tiny{}28312} & {\tiny{}19.6}\tabularnewline
{\tiny{}MIPS24 } & {\tiny{}234715} & {\tiny{}53245} & {\tiny{}18.3}\tabularnewline
\hline 
\end{tabular}
\end{table}

\begin{figure*}[tp]
\centering{}\includegraphics[bb=0bp 20bp 840bp 600bp,scale=0.4]{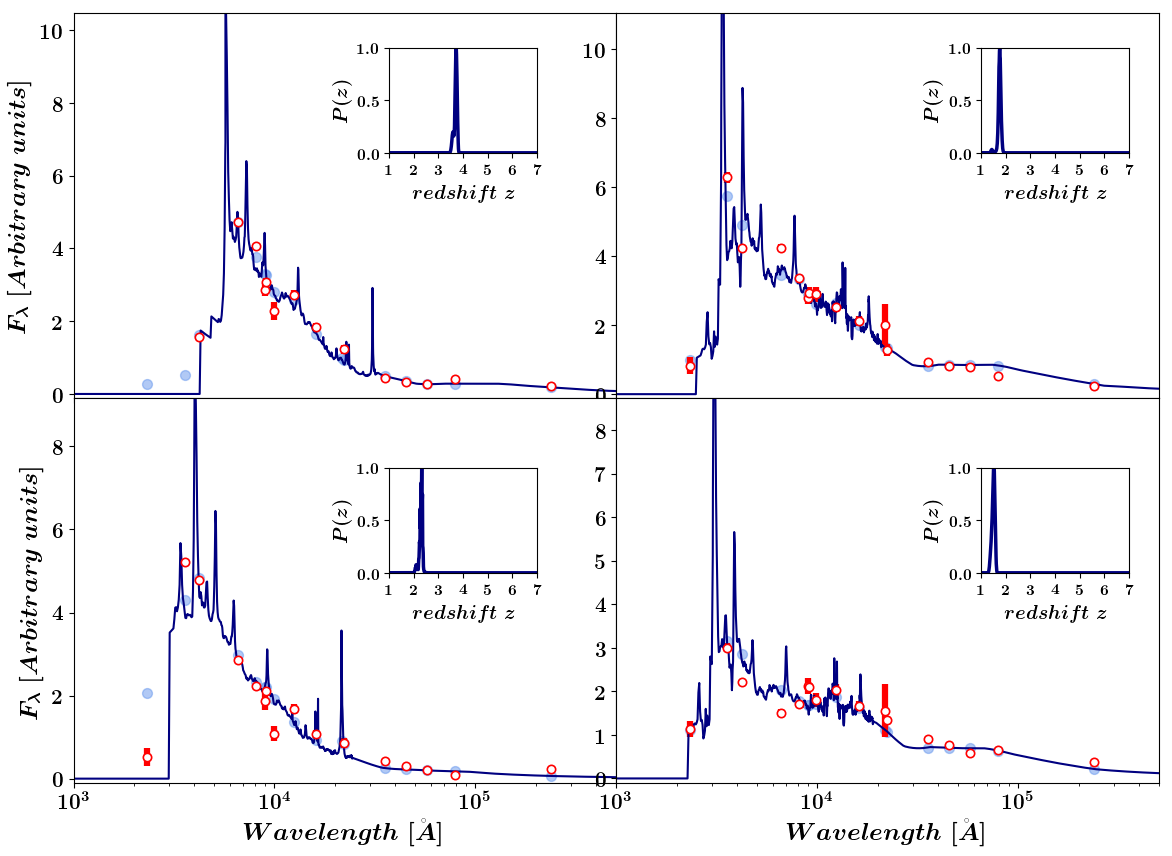}\centering\caption{\label{fig:SEDs_qso} The spectral energy distributions (SEDs) of
four photometric quasars identified using our ML algorithms (see Section
\ref{subsec:classification_algorithms}). The NDWFS-Bo\"{o}tes photometry
is used to calculate the SEDs. In each case the best-fit quasar template
(as derived from the EAZY calculation) is also plotted. Red circles
are the photometric points and the blue circles indicate the predicted
photometry by the best-fit template. The probability density distributions
(PDFs) for each object are shown in the small inset. These PDFs strongly
suggest that these objects are quasars located at $z>1.4$.\textbf{ }}
\end{figure*}

\begin{figure}[tp]
\begin{centering}
\includegraphics[clip,scale=0.33]{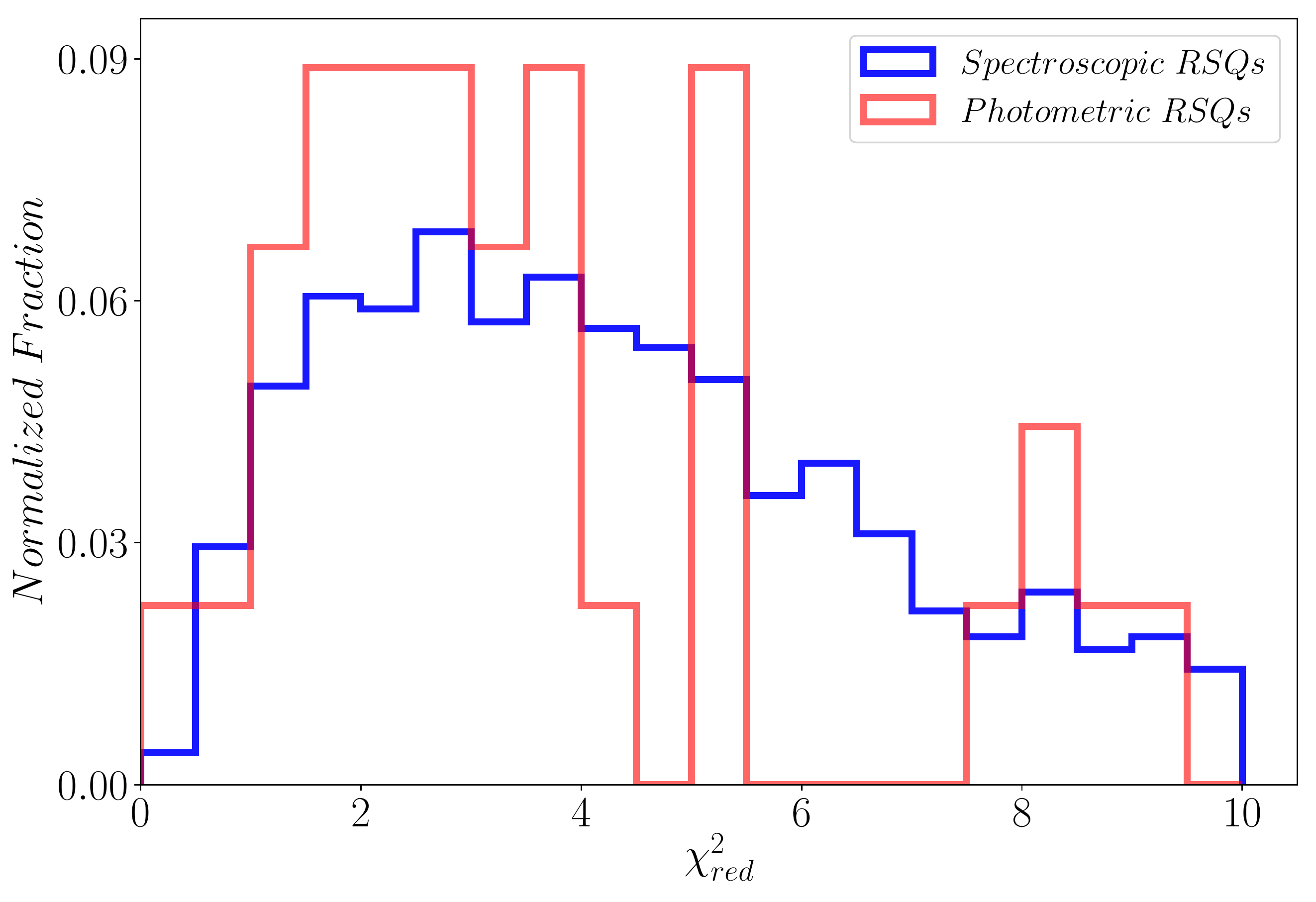}
\par\end{centering}
\centering{}\centering\caption{\label{fig:quis_zphoto_dist} Normalized $\chi_{red}^{2}$ distributions
of \emph{all} NDWFS-Bo\"otes spectroscopic quasars with $z\geq1.4$
(blue) and photometric (red) RSQs. See Section \ref{subsec:sed_photo_quasars}
for more details.}
\end{figure}

\noindent For each object, the goodness-of-fit, $\chi^{2}$, is estimated
through the fitting of the object's photometric data to a SED quasar
template. We used a compilation of quasar templates from the literature.
This includes the composite quasar templates by \citet{1990A&A...227..385C},
\citet{2001AJ....122..549V} and \citet{2006A&A...457...79G}, respectively;
and the type-1 quasar templates by \citet{2009ApJ...690.1250S} (pl\_I22491\_10\_TQSO1\_90,
pl\_QSOH, pl\_QSO\_DR2, pl\_TQSO1). To account for dust extinction
in the quasar hosts, we modify the quasar templates using the \citet{2000ApJ...533..682C}
starburst, \citet{1992ApJ...395..130P} Small Magellanic Cloud (SMC),
and the \citet{2004MNRAS.348L..54C} extinction laws. The extinction
was applied to each template according to a grid $E(B-V)=\left[0.0,0.05,0.1,0.15,0.2,0.25,0.3,0.4,0.5\right]$.
We performed these fittings using the photometric redshift code EAZY
\citep{2008ApJ...686.1503B}. For all the SED libraries, the zero
points were calculated using the standard procedure of fitting the
observed SEDs of a sample of spectroscopic objects, fixing their redshift
to the known spectroscopic redshift (see \citealt{2006A&A...457..841I}
for more details). The spectroscopic sample used to determine the
zero points are the NDWFS-Bo\"otes quasars with $z\geq1.4$. Due
to the large number of templates in the quasar template library, only
single template fits were considered. Figure \ref{fig:SEDs_qso} shows
the SED fitting to four photometric quasars in our sample. There is
a good agreeement between the NDWFS-Bo\"otes photometry and the quasar
templates. Figure \ref{fig:quis_zphoto_dist} displays the distribution
of the reduced goodness\textendash of-fit values, $\chi_{red}^{2}=\chi^{2}/N$,
where $N$ is the number bands in which the quasar is detected, for
NDWFS-Bo\"otes spectroscopic and photometric quasars. We determined
the probability that both samples are from the same parent population
by doing a Kolmogorov-Smirnov (K-S) test. The K-S test indicates that
there is a probability of $27\%$ that both samples are being drawn
from the same distribution. Thus, we cannot reject the hypothesis
that the distributions of the two samples are the same. This hypothesis
also cannot be rejected if the K-S test is applied to the redshift
distributions. In this case, we found that the hypothesis cannot be
rejected at the $95\%$ level.

\noindent 

\noindent 

\subsection{Cross-matching to X-ray data \label{subsec:xray_match}}

Our quasar sample was correlated with the 5-ks deep Chandra catalog
of the NDWFS-Bo\"otes field XBo\"otes \citep{2005ApJS..161....9K}
using a $2^{\prime\prime}$ matching radius. A total of 44 spectroscopic
quasars and 6 photometric quasars are detected by the Chandra satellite.
The high detection fraction of spectroscopic quasars is not unexpected,
as a X-ray detection in the XBo\"otes catalog was one of the criteria
used for spectroscopic follow-up of AGN candidates in the AGES survey
( see \citet{2012ApJS..200....8K}, Section 2). In conclusion, deeper
X-ray exposures are required to detect the rest of the quasars in
our sample.

\subsection{Density of photometric quasars}

In this section, we compare the density of photometric quasars in
our sample with previous works. After the morphological cut, but before
cross-matching with the LOFAR catalog the density of our sample is
of 1927 photometric quasars per square degree. This number could be
considered high in comparison with previous results. For example,
\citet{2015ApJS..219...39R} using a Bayesian kernel density algorithm
found a surface density of $\sim51$ optical/mid-infrared photometric
quasars per square degree. \citet{2019MNRAS.485.4539J} employed the
XGBoost and SVM classification algorithms to obtain a density of $\sim38$
Pan-STARRS/WISE photometric quasars per square degree. Both works
considered similar ranges in optical magnitude and redshift. Two reasons
can explain our higher density of photometric quasars. Firstly, we
do not restrict our mid-infrared photometry using any quality flag
(only limiting the sample to objects with $m_{\textrm{AB}}\geq15$),
as is done in the aforementioned works. Secondly, our relatively deep
mid-infrared observations were obtained with Spitzer-IRAC, while the
majority of mid-infrared imaging used by \citet{2015ApJS..219...39R}
comes from shallower WISE observations, and \citet{2019MNRAS.485.4539J}
employed only WISE data. Hence, it is only for the purposes of comparison
in this section that we restricted our sample only to photometric
quasars detected in WISE, and keep only those that fulfill the WISE
photometry flags used in \citet{2015ApJS..219...39R}. This reduced
sample contains 538 objects detected by WISE, which corresponds to
a density of $\sim58$ photometric quasars per square degree. This
density is in agreement with the aforementioned works, and the reason
for the apparent difference in the number of photometric quasars is
the depth of the IRAC imaging, and our treatment of the quality flags
of the IRAC photometry. 

\subsection{LOFAR and wedge-based mid-infrared selection of quasars \label{subsec:lofar_irac}}

\begin{figure}[tp]
\centering{}\includegraphics[bb=0bp 0bp 772bp 764bp,clip,scale=0.37]{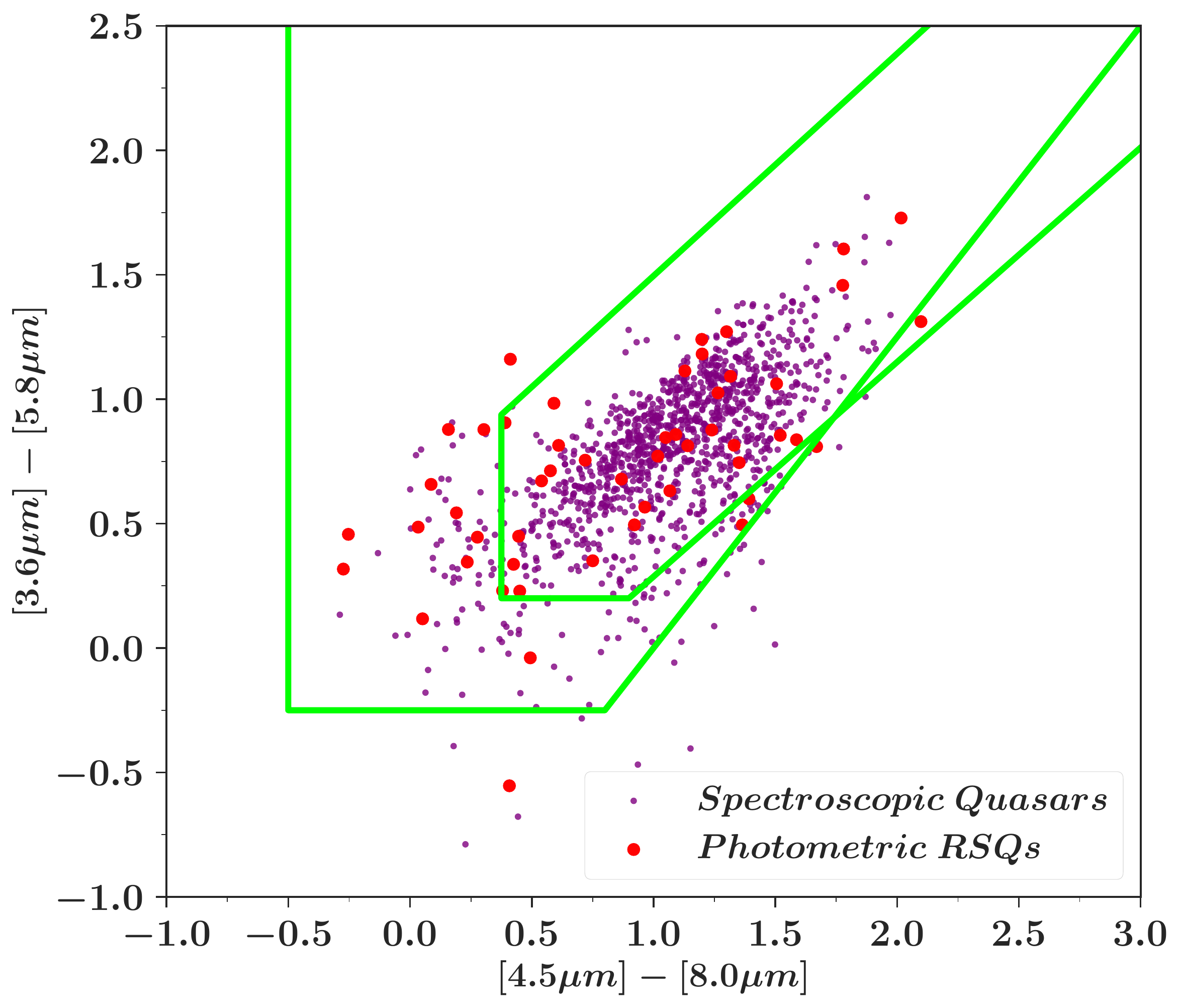}\centering\caption{\label{fig:irac_color_space} Mid-infrared colors for photometric
and spectroscopic quasars in the Bo\"otes field. The photometric
RSQs are plotted as red circles, while the spectroscopic quasars are
indicated by purple circles. The light green lines are the mid-infrared
color cuts proposed by \citet{2007AJ....133..186L} and \citet{2012ApJ...748..142D}.}
\end{figure}

In this section, we compare the LOFAR and wedge-based mid-infrared
selection \citep{2005ApJ...631..163S}  for objects classified as
quasars by the ML classification algorithms. Firstly, we used the
mid-infrared color cuts proposed by \citet{2007AJ....133..186L} and
\citet{2012ApJ...748..142D} to identify the presence of AGN-heated
dust in our photometric RSQs. Figure \ref{fig:irac_color_space} shows
the mid-infrared colors for different quasar samples in the NDWFS-Bo\"otes
field. The mid-infrared colors of the photometric RSQs are in good
agreement with those of spectroscopic quasars, and the majority of
both spectroscopic and RSQs are located within the region delimited
by the \citet{2012ApJ...748..142D} color cuts. To be exact, $89.93\%$
($70.59\%$) of the 1192 (47) spectroscopic quasars (photometric RSQs)
with redshifts $z>1.4$ are located within the region delimited by
the \citet{2012ApJ...748..142D} color cuts, $98.15\%$ ($94.12\%$)
reside in the region common to the \citet{2007AJ....133..186L} and
\citet{2012ApJ...748..142D} color cuts, and only $1.85\%$ ($5.88\%$)
are located outside the boundaries delimited by the aforementioned
wedge-based mid-infrared color cuts. Considering the following points
regarding our photometric RSQs: i) these objects are identified as
quasar by the ML algorithms; and ii) their optical and mid-infrared
colors are similar to those of spectroscopic quasars. We are confident,
therefore, that our sample of photometric RSQs is composed mainly
of real quasars and the number of contaminants is minimal. Moreover,
these points show that utilizing ML algorithms trained using optical
and infrared photometry and combined with LOFAR data is a very efficient
and robust way to identify quasars.

\begin{figure*}[tp]
\centering{}\includegraphics[clip,scale=0.38]{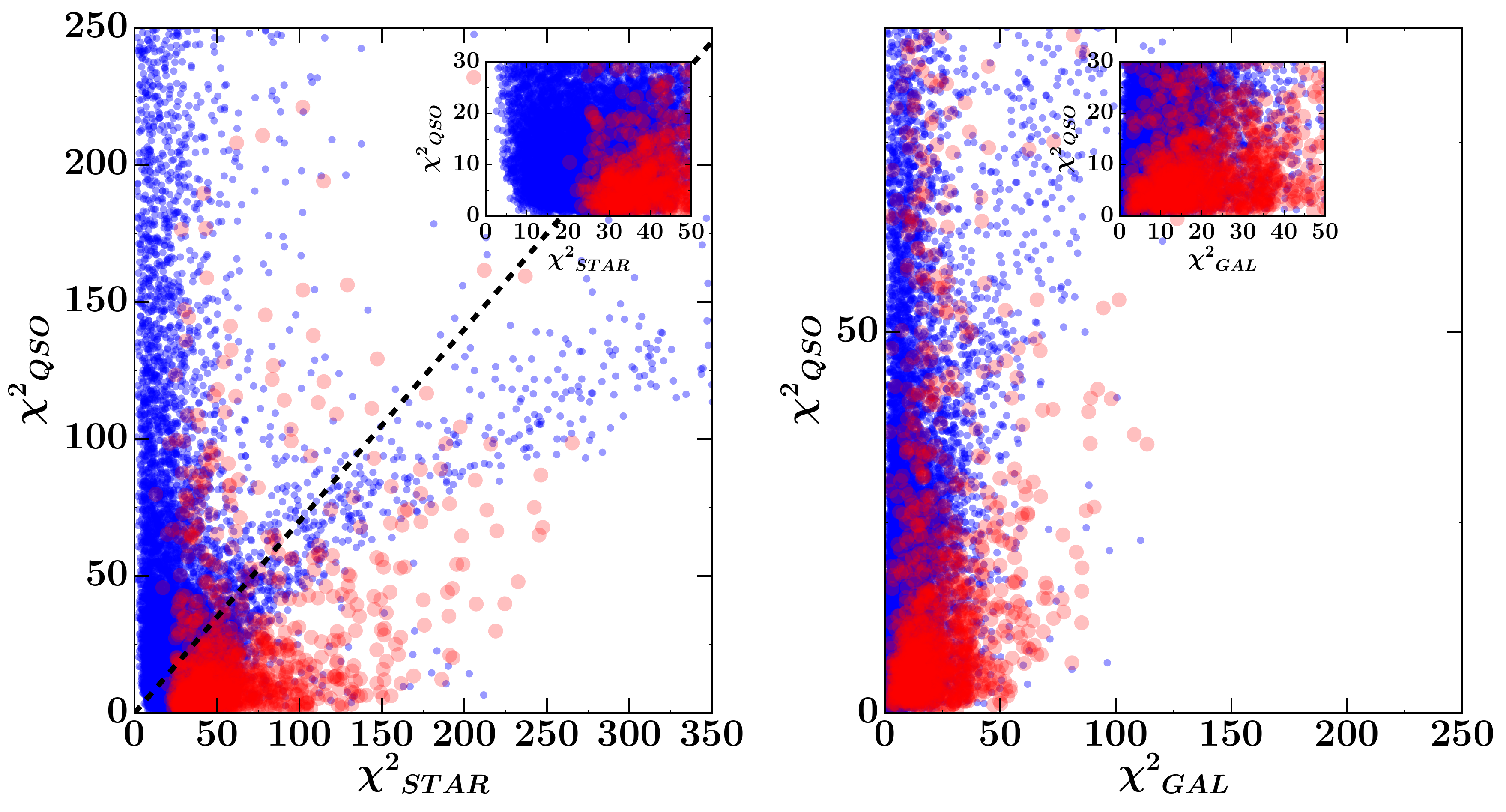}\centering\caption{\label{fig:qui_distribution} Comparison of the $\chi^{2}$ values
for quasar, galaxy, and star template fitting to the photometry of
spectroscopic quasars and candidate quasars in the NDWFS-Bo\"otes
field.}
\end{figure*}

\noindent To investigate the wedge-based mid-infrared selection of
quasars without radio detections, we first needed to establish the
nature of these objects as robust quasars candidates or contaminants
(i.e., stars or galaxies). For this purpose, we followed a classification
method based on the goodness\textendash of-fit, $\chi^{2}$, estimated
through the fitting of the object's photometric data to a given type
of SED template (quasar, galaxy, and stellar) (e.g., \citealt{2006A&A...457..841I,2015ApJ...813...53M}).
This fitting assigns to each object three $\chi^{2}$ values: $\chi_{\textrm{QSO}}^{2}$,
$\chi_{\textrm{GAL}}^{2}$, and $\chi_{\textrm{STAR}}^{2}$, which
corresponds to the fitting of the object photometry against the quasar,
galaxy, and stellar templates, respectively. In the SED fittings,
we utilize only the SDSS, Pan-STARRS1, and IRAC $3.6\,\mu\textrm{m}/4.5\,\mu\textrm{m}$
bands as this was the photometry used to classify them originally
as quasars by the ML algorithms. This also represents a scenario where
there is SDSS, Pan-STARRS1, and LOFAR coverage, but shallow or incomplete
mid-infrared data to perform a wedge-based mid-infrared selection
of faint quasars (e.g., \citealt{2005ApJ...631..163S,2012ApJ...754..120M}).
The three template libraries used in this analysis are as follows.
For the galaxy library, we use the latest version of the \citet{2008ApJ...686.1503B}
SED templates that include nebular emission lines; while for the star
library we select the \citet{2000ApJ...542..464C} SED templates.
For the quasar template set, we use a compilation of quasar templates
from the literature. This includes the composite quasar templates
by \citet{1990A&A...227..385C}, \citet{2001AJ....122..549V} and
\citet{2006A&A...457...79G}, respectively; and the type-1 quasar
templates are the same as those used in Section \ref{subsec:sed_photo_quasars}.
To account for dust extinction in the galaxies, we modified the galaxy
templates using the \citet{2000ApJ...533..682C} starburst and \citet{1992ApJ...395..130P}
Small Magellanic Cloud (SMC) extinction laws using a $E(B-V)$ grid
(see Section \ref{subsec:sed_photo_quasars}). The spectroscopic samples
used to determine the zero points are the quasars, galaxies, and stars
of the training sample presented in Section \ref{subsec:training_sample}.
We compute the $\chi^{2}$ values using the photometric redshift code
EAZY \citep{2008ApJ...686.1503B}, as described in Section \ref{subsec:sed_photo_quasars}.
From the $\chi^{2}$ distribution of spectroscopic quasars in Bo\"otes,
we defined empirical $\chi^{2}$ cuts to separate quasars from stars
and galaxies. Figure \ref{fig:qui_distribution} displays the comparison
of the $\chi^{2}$ values resulting from the quasar, galaxy, and star
template fitting to the photometry of spectroscopic and photometric
quasars in the NDWFS-Bo\"otes field. We found that the empirical
cuts $\chi_{\textrm{STAR}}^{2}\geq22$, $\chi_{\textrm{QSO}}^{2}\leq\chi_{\textrm{STAR}}^{2}\times0.33+8.33$
and $\chi_{\textrm{GAL}}^{2}\geq2.5$ select the majority of quasars
and reject an important fraction of likely stars and galaxies. Based
on the cuts described before, we selected 957 out of a  total of 1193
spectroscopic quasars; and 4935 of 17829 photometric quasars. From
these, 1423 photometric quasars are found to be located in the region
delimited by the \citet{2012ApJ...748..142D} color cuts. The analysis
of the mid-infrared selection was limited to the \citet{2012ApJ...748..142D}
wedge as it is expected that the majority of quasars will be located
within this region. Next we compare the number of photometric quasars
selected by the \citet{2012ApJ...748..142D} wedge to the expected
number of quasars. The expected number of quasars can be determined
using the following expression:

\begin{equation}
N_{{\scriptscriptstyle \textrm{QSO}}}=A\times\iint\,\Phi^{*}\left(M_{1450}^{*},z\right)V_{c}\left(z\right)\,dz\,dM_{1450},\label{eq:number_qso}
\end{equation}

\noindent where $A$ is the survey area, $\Phi^{*}\left(M_{1450}^{*},z\right)$
is the quasar luminosity function, and $V_{c}\left(z\right)$ is the
comoving volume. We estimated the number of total (radio- detected
and undetected) faint quasars using the results by \citet{2018AJ....155..110Y}
and eq. \ref{eq:number_qso}. These authors studied the luminosity
function of faint quasars between $0.5<z<4.5$ in a $1.0\;\textrm{deg}^{2}$
field within the VIMOS VLT Deep Survey \citep{2013A&A...559A..14L}.
For redshifts $z>3.5$, \citet{2018AJ....155..110Y} did not fit the
luminosity function due to the low number of quasars in their sample;
therefore, our analysis is limited to the redshift range $1.4\lesssim z\lesssim3.5$.
According to \citet{2018AJ....155..110Y}, the expected number of
faint quasars at $1.4\lesssim z\lesssim3.5$ in a $9.3\;\textrm{deg}^{2}$
region like the NDWFS-Bo\"otes field is approximately $N_{{\scriptscriptstyle \textrm{QSO}}}\sim1060$.
However, in the redshift range $1.4\leq z\leq3.5$ there are more
than 1100 spectroscopic quasars, and 1380 photometric quasars fulfill
the \citet{2012ApJ...748..142D} color cuts. Counting only the photometric
quasars and not known spectroscopic quasars, the number is higher
than the expected number of quasars. This indicates some degree of
contamination in the sample of photometric quasars selected using
the wedge-based mid-infrared selection. Most likely these contaminants
are compact galaxies or AGNs that mimic the colors of quasars, as
the majority of stars are expected to be located outside the wedge
define by the \citet{2012ApJ...748..142D} color cut. Additionally,
comparing these numbers against those of the LOFAR selection, it is
likely that there are more contaminants in the wedge-based mid-infrared
selected sample than in the LOFAR selected sample. For a case like
this, the addition of Euclid \citep{2011arXiv1110.3193L} near-infrared
imaging ($JH$ bands) in the ML classification process will be useful
in eliminating some of these likely contaminants.

The results of this section demonstrated that in the cases where a
of lack of deep and complete mid-infrared coverage needed to perform
a wedge-based mid-infrared selection of quasars, ML algorithms trained
with optical and infrared photometry combined with LOFAR data provide
an excellent approach for obtaining samples of quasars. Moreover,
considering this and the results obtained in Section \ref{subsec:final_qso_sample},
where with LOFAR data we are not only able to eliminate the stellar
contamination in our quasar sample, but also to reduce the number
of contaminants by two orders of magnitude. It is clear that the use
of LOFAR data to select quasars has a great potential for compiling
samples of quasars. 

\noindent 

\noindent 

\section{Luminosity Function of radio-selected quasars\label{sec:Section5}}

In the following subsections, we describe the steps required to compute
the luminosity function of RSQs.

\subsection{Selection completeness and accuracy of photometric redshifts \label{sec:selection_function-2}}

\begin{figure}[tp]
\begin{centering}
\includegraphics[clip,scale=0.37]{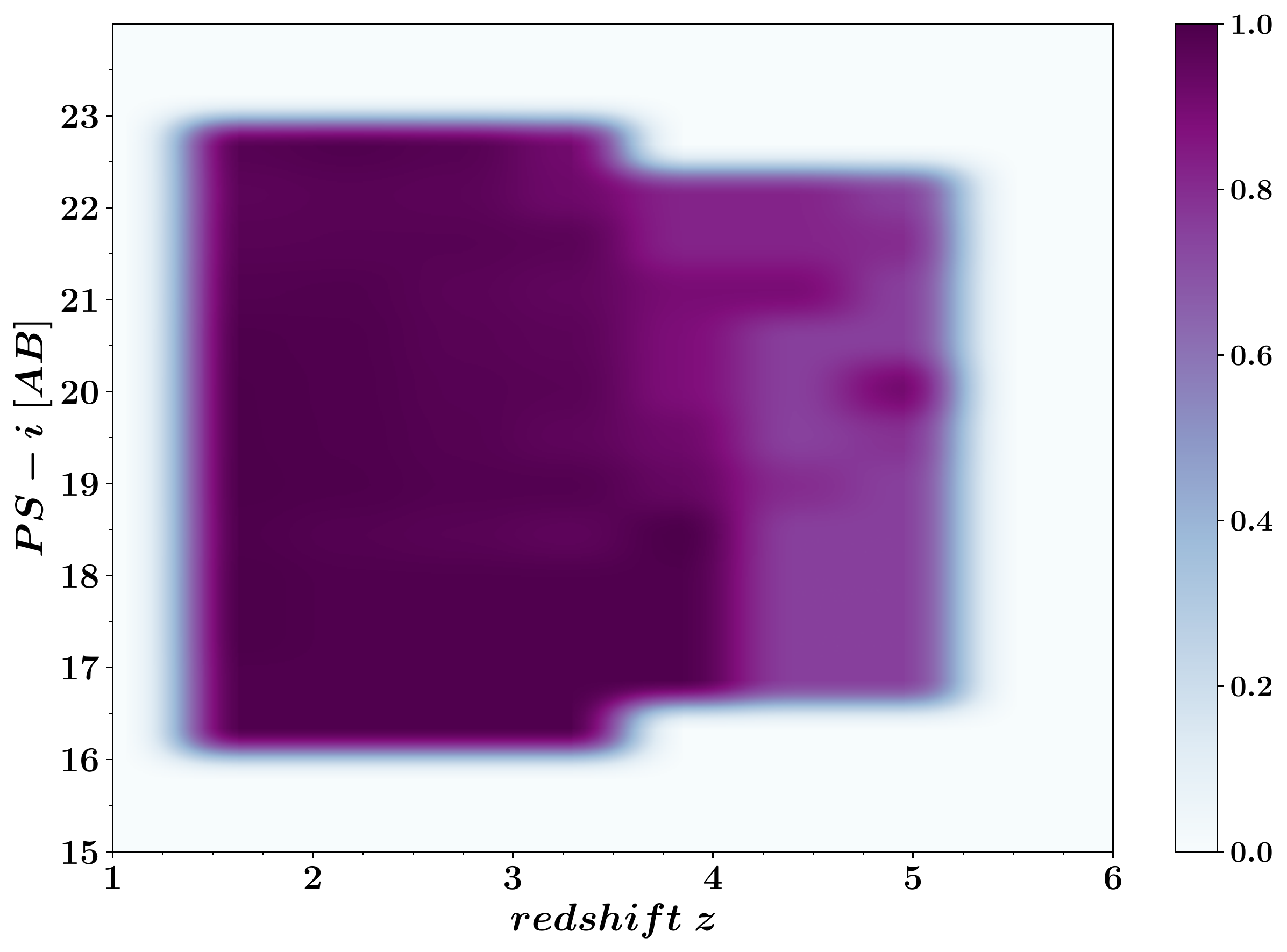}
\par\end{centering}
\centering{}\centering\caption{\label{fig:selection_completeness} The selection completeness, $P_{\textrm{comp}}\left(\,i_{\textrm{PS}},z\,\right)$,
binned according to magnitude $i_{\textrm{PS}}$ and redshift $z$,
in bins of size $\triangle i_{\textrm{PS}}=0.5$, and $\triangle z=0.3,$
respectively. }
\end{figure}

An important step in measuring the luminosity function of quasars
is to account (and correct) for the different sources of incompleteness
that could bias the quasar counts. In our analysis, we need to consider
the completeness of our sample selection and the accuracy of the photometric
redshifts. The selection completeness, $P_{\textrm{comp}}\left(\,i_{\textrm{PS}},z\,\right)$,
is the fraction of quasars that were successfully identified as quasars
by the classification algorithms, as function of magnitude $i_{\textrm{PS}}$
and redshift. $P_{\textrm{comp}}\left(\,i_{\textrm{PS}},z\,\right)$
is derived from the data themselves as follows. First, the spectroscopic
quasar sample introduced in Section \ref{subsec:training_sample}
is binned according to magnitude $i_{\textrm{PS}}$ and redshift $z$,
in bins of size $\triangle i_{\textrm{PS}}=0.5$, and $\triangle z=0.3,$
respectively. The quasars in each bin are separated into two subsamples.
The first subsample is created by randomly sampling without replacement
all quasars in the bin, while the second subsample includes all the
quasars that were not sampled. The sizes of the first and second subsamples
are 20\% and 80\% of quasars in the $i_{\textrm{PS}}-z$ bin, respectively.
Having done this for all bins, the corresponding samples are combined
to create final training (80\%) and target (20\%) samples. The final
result is the uniform and randomly sampled separation of the training
sample into two subsamples in the $i_{\textrm{PS}}-z$ plane. The
main advantage of this binning scheme is that it provides an unbiased
and efficient way to map locally the selection completeness obtained
using the classification algorithms as a function of magnitude and
redshift. The second subsample is used as the training sample for
the classification algorithms, while the first subsample has the role
of target sample and it is employed to derive the selection completeness.
Figure \ref{fig:selection_completeness} shows the selection completeness,
$P_{\textrm{comp}}\left(\,i_{\textrm{PS}},z\,\right)$.

\noindent 

\begin{figure}[tp]
\begin{centering}
\includegraphics[clip,scale=0.39]{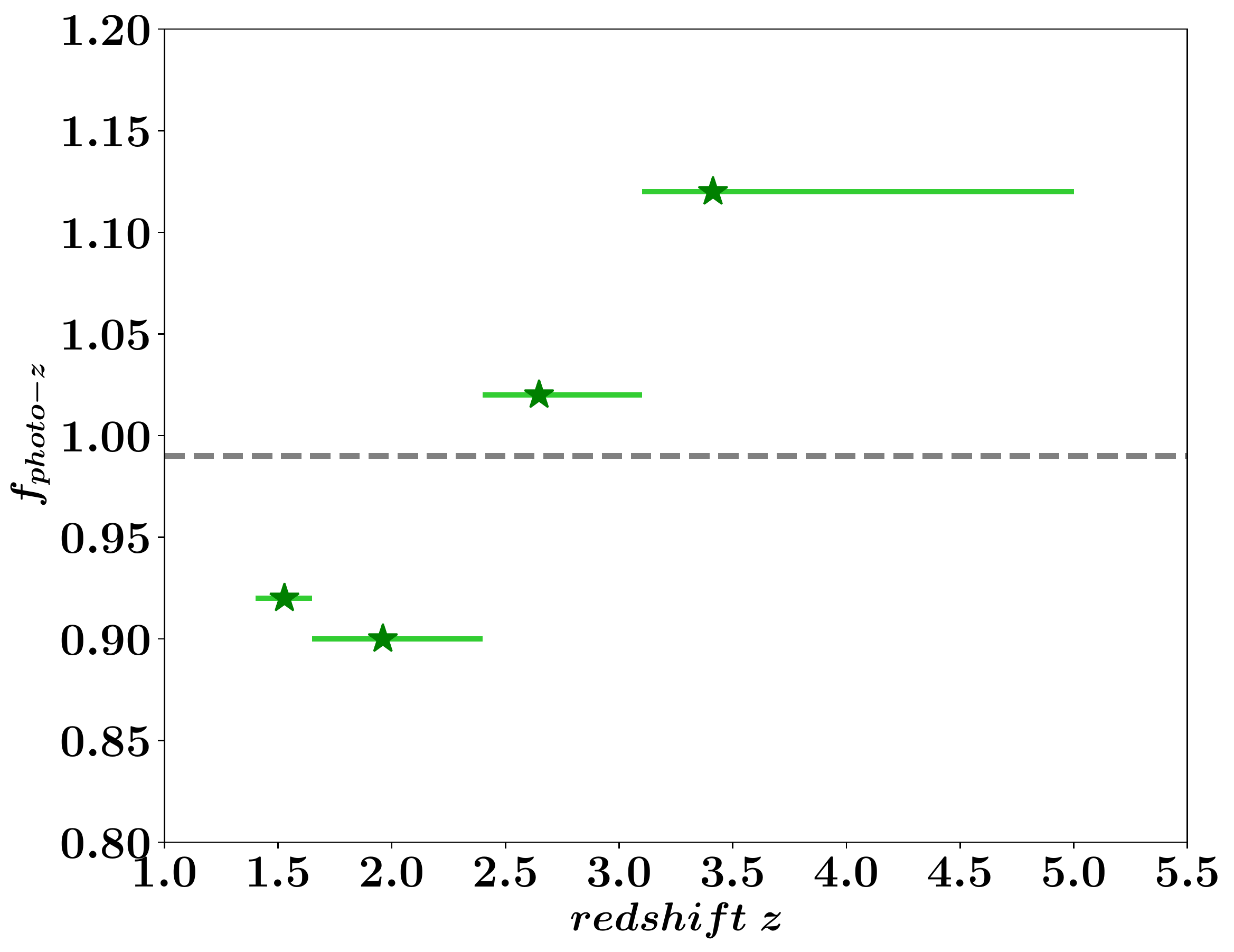}
\par\end{centering}
\centering{}\centering\caption{\label{fig:f_z_correction}\emph{ }Ratio between the number of spectroscopic
quasars with correctly and incorrectly assigned redshifts, $f_{\textrm{photo-z}}$,
as a function of redshift. The redshift intervals match those used
to derive the luminosity function in Section \ref{sec:qlf}. The gray
dashed line denotes the median value of $f_{\textrm{photo-z}}$.}
\end{figure}

In addition to correcting for selection incompleteness in our sample,
we also need to correct for the accuracy of the photometric redshifts
determined using the NW regression method. For this purpose, we determine
the expected number of spectroscopic quasars to have photometric redshifts
correctly and incorrectly assigned within a redshift bin using the
NW method. This is done following a similar approach to determining
the selection completeness. First, the spectroscopic quasar sample
introduced in Section \ref{subsec:training_sample} is divided into
the same redshift bins used to derive the luminosity function in Section
\ref{sec:qlf}. In each bin, the quasars contained in that bin are
randomly separated to create samples with sizes of 20\% and 80\% of
all quasars in the bin, respectively. The corresponding samples from
all the bins are combined to create final training (80\%) and target
(20\%) samples. The training sample is used to train the NW regression
method, while the target sample is utilized to determine the expected
number of spectroscopic quasars with correctly and incorrectly assigned
redshifts within the boundaries of the redshift bins. The ratio between
the number of spectroscopic quasars with correctly and incorrectly
assigned redshifts, $f_{\textrm{photo-z}}$, provides an estimate
of the excess of photometric quasars with incorrectly assigned photometric
redshifts within a redshift bin (see Figure \ref{fig:f_z_correction}).
This ratio is used as a correction factor for each photometric quasar
within the corresponding redshift bin. The derived correction factors
have a median factor of $f_{\textrm{photo-z}}\backsimeq1.0$, with
the $1.65<z<2.4$ redshift bin having the smallest value with $f_{\textrm{photo-z}}=0.90$.

\noindent 

\subsection{Simulated Quasar Spectra \label{sec:simuqso}}

In order to calculate the the K-correction (see Section \ref{sec:k_corr}),
we construct a synthetic quasar library that is an accurate representation
of the quasar demography. The variety in the quasar spectral features
(UV continuum slope, emission-line EW, and intervening HI absorbers
along the line of sight) determine the range in quasar colors. It
is important that these spectral features are taken into account to
obtain reliable simulated quasar spectra. These spectra are later
incorporated into a synthetic quasar library that allow us to compute
the K-correction for a given redshift. We explain the procedure followed
to build our synthetic quasar library below.

\noindent Following several authors \citep{1990AA...228..299M,1999AJ....117.2528F,2006AJ....131.2766R,2013ApJ...768..105M},
the synthetic quasar spectra are built using a Monte-Carlo (MC) approach.
We perform MC simulations to generate quasar spectra adopting a broken
power-law $\left(f_{\lambda}\propto\lambda^{-\alpha_{\lambda}}\right)$
for the UV continuum at $1100$ \AA. The slope values are
drawn from a Gaussian distribution, with mean values of $\left\langle \alpha_{\lambda}\right\rangle =-1.7$
for $\lambda<1100$ \AA \ \citep{2002ApJ...565..773T}, and
$\left\langle \alpha_{\lambda}\right\rangle =-0.5$ for $\lambda>1100$ \AA \
 \citep{2001AJ....122..549V}, both with standard deviations of $\sigma=0.30$.
We bin BOSS quasars by their luminosity to obtain the distribution
for the parameters (wavelength, EW, FWHM) of emission lines. This
allows us to recover the intrinsic emission line mean and dispersion
as function of luminosity, as well as reproducing empirical trends
such as the Baldwin effect \citep{1977ApJ...214..679B}. We again
assume gaussianity when the emission line features are added to the
quasar continuum. For each template spectrum, the intergalactic absorption
that gives rise to the $\textrm{Ly}_{\alpha}$ forest is included
by creating sightlines in a MC fashion adopting the prescription of
neutral absorbers by \citet{1999ApJ...518..103B}. The spectrum is
then convolved with our filter passbands to obtain the colors for
each synthetic quasar. A Gaussian error is added to the photometry
of the mock quasars with a $\sigma$ derived from the photometric
errors of the real magnitudes that match the simulated ones. This
error is combined in quadrature with the photometric calibration errors
of Pan-STARRS1 \citep{2012ApJ...750...99T}. In Figure \ref{fig:ndwfs_filters},
we show a synthetic spectrum from our quasar library.

\subsection{K-correction \label{sec:k_corr}}

\begin{figure}[tp]
\begin{centering}
\includegraphics[clip,scale=0.32]{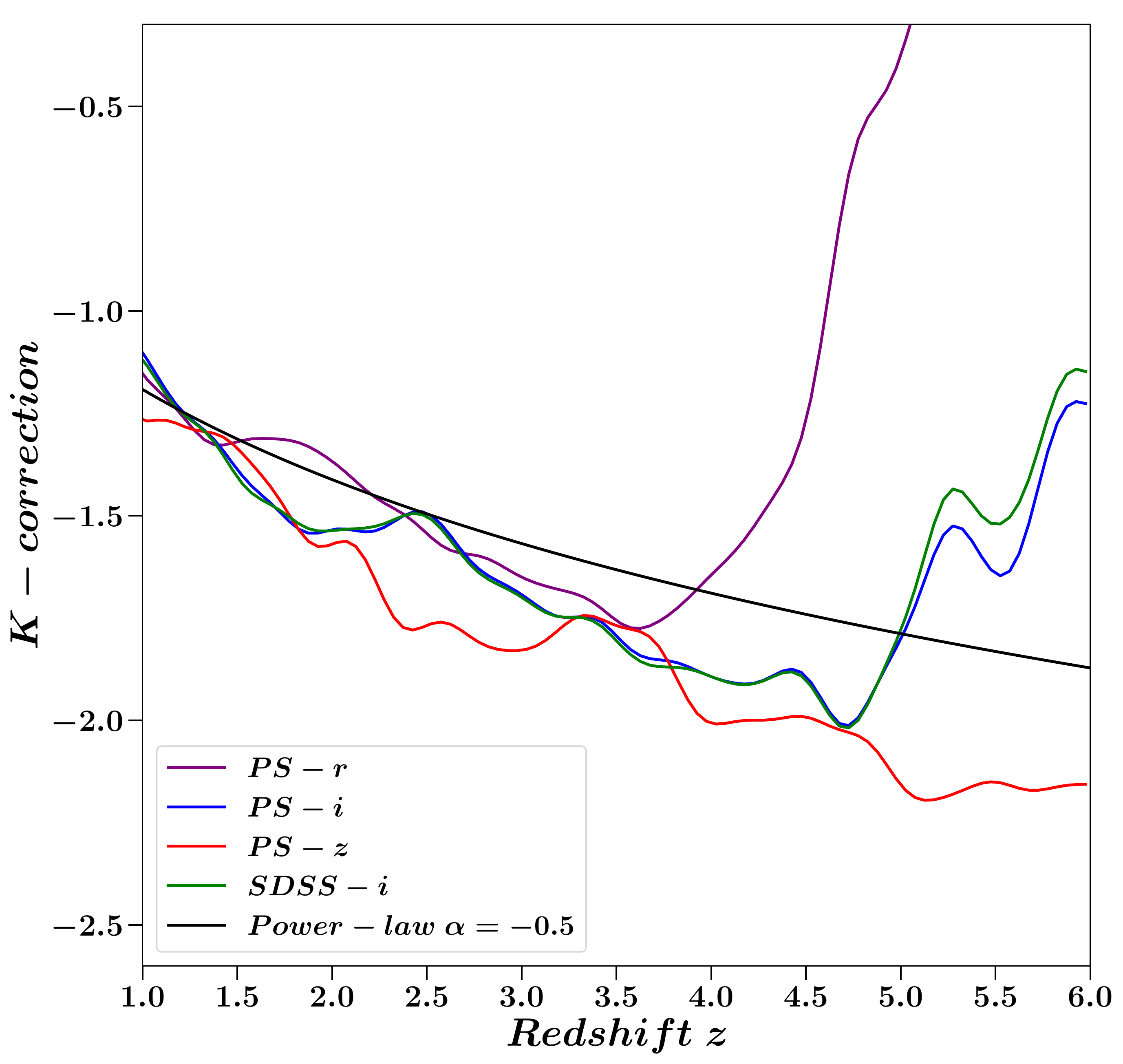}
\par\end{centering}
\centering{}\centering\caption{\label{fig:Kcorr_z} K-correction for different filters determined
using our simulated quasar spectra. The Pan-STARRS1 $r_{\textrm{PS}}$
and $i_{\textrm{PS}}$ filters are indicated by blue and green, while
the red and cyan are the expected K-corrections for the Pan-STARRS1
$z_{\textrm{PS}}$ and SDSS-$i$ bands. The solid black line is the
K-correction assuming a power-law with slope $\alpha=-0.5$ with no
emission line contribution or $\textrm{Ly}_{\alpha}$ forest blanketing.
At $z\gtrsim3.7$, the difference between the $r_{\textrm{PS}}$ and
$i_{\textrm{PS}}$ bands becomes more significant as the $\textrm{Ly}_{\alpha}$
line moves in or out of the filters. The same situation occurs at
$z\gtrsim3.7$, but for the $i_{\textrm{PS}}$ and $z_{\textrm{PS}}$
bands.}
 
\end{figure}

Usually, the luminosity functions of quasars are expressed in the
absolute magnitude at rest-frame $1450$ \AA, $M_{\textrm{1450}}$,
which provides a good measurement of the quasar continuum in a region
without strong emission lines (e.g., \citealt{2006AJ....131.2766R,2009MNRAS.399.1755C,2012ApJ...755..169M}).
To derive $M_{\textrm{1450}}$ for the Bo\"otes quasars, we use the
apparent magnitude $m_{\textrm{X}}$ in a fiducial filter as a proxy:
\begin{equation}
M_{\textrm{1450}}=m_{\textrm{X}}-5\,\log\left(d_{\textrm{L}}/10\right)-K_{\textrm{X}},\label{eq:}
\end{equation}

\noindent where $d_{\textrm{L}}\left(z\right)$ is the luminosity
distance in parsecs, and $K_{\textrm{X}}$ is the K-correction which
allows us to convert the magnitudes of distant objects in a given
bandpass filter into an equivalent measurement into their rest-frame.
Using our synthetic quasar library described in Section \ref{sec:simuqso},
the K-correction can be determined from the difference between apparent
magnitudes $m_{\textrm{X}}$ and $m_{\textrm{1450}}$, 
\noindent \begin{center}
\begin{equation}
K_{\textrm{X}}=m_{\textrm{X}}-m_{\textrm{1450}}-2.5\,\log\left(1+z\right),\label{.}
\end{equation}
\par\end{center}

\noindent with $m_{\textrm{1450}}$ calculated using a top-hat filter
of width $50$ \AA. Figure \ref{fig:Kcorr_z} displays the
K-correction obtained for five different filters, and the expected
result from a quasar that has only a power-law continuum and no emission
line contribution or $\textrm{Ly}_{\alpha}$ forest blanketing. The
K-correction curves between $1.0\leq z\leq6.0$ are obtained by calculating
their average value in redshift bins of size $\triangle z=0.1$. At
$z\gtrsim3.7$, the difference between the $r_{\textrm{PS}}$ and
$i_{\textrm{PS}}$ bands becomes more significant as the $\textrm{Ly}_{\alpha}$
line moves or exits the filters. The same situation occurs at $z\gtrsim4.7$,
but for the $i_{\textrm{PS}}$ and $z_{\textrm{PS}}$ bands. Therefore,
we estimate $M_{\textrm{1450}}$ using K-corrections selected to minimize
any bias caused by the redshifting of the $\textrm{Ly}_{\alpha}$
emission line. For $z<3.7$ quasars, we use a K-correction based on
the $r_{\textrm{PS}}$ band, while for the intervals $3.7\leq z\leq4.7$
and $z>4.7$ K-corrections based on the $i_{\textrm{PS}}$ and $z_{\textrm{PS}}$
bands, respectively, are employed.

\noindent 

\subsection{Quasar Luminosity function \label{sec:qlf}}

We construct the luminosity functions for all quasars in our radio-matched
sample using the classical $1/V_{\textrm{max}}$ method \citep{1968ApJ...151..393S}
for flux limited samples. The main advantage of this method is that
an assumption about the underlying model of the luminosity function
is not required. The estimator adopted to compute the comoving quasar
density in a certain luminosity bin is:
\noindent \begin{center}
\begin{equation}
{\displaystyle \begin{array}{c}
{\displaystyle \Phi\left(L\right)=\frac{1}{\triangle L}\,\overset{n}{\sum_{i=1}}\left(F\left(S_{\textrm{150MHz}}\right)\times P_{\textrm{comp}}\left(\,i_{\textrm{PS}},z\,\right)\right.}\\
{\displaystyle \left.\times P_{\textrm{comp}}\left(\,i_{\textrm{PS}},z\,\right)\times f_{\textrm{photo-z}}\left(z\right)\times V_{\textrm{max},i}\right)^{-1},}
\end{array}}\label{eq:-5-1}
\end{equation}
\par\end{center}

\noindent where $n$ is the number of quasars in the luminosity bin,
$V_{\textrm{max},i}$ is the is the maximum comoving volume in which
a quasar would be observable and included in our sample, $\triangle L$
is the luminosity bin width, $F\left(S_{\textrm{150MHz}}\right)$
is the radio-catalog completeness of the LOFAR-Bo\"otes mosaic \citep{2018arXiv180704878R}.
$P_{\textrm{comp}}\left(\,i_{\textrm{PS}},z\,\right)$ is the selection
completeness, and $f_{\textrm{photo-z}}\left(z\right)$ is the accuracy
of the NW photometric redshifts derived in Section \ref{sec:selection_function-2}.
Since our quasar sample is built using a radio-optical survey, we
calculate $V_{\textrm{max}}$ using the maximum redshift at which
the flux of a quasar with a certain luminosity lies above the corresponding
flux limit \citep{2005MNRAS.357.1267C,2015MNRAS.449.2818T}, $z_{\textrm{max}}=\min\left(\,z_{\textrm{max}}^{\textrm{R}},\,z_{\textrm{max}}^{\textrm{O}}\,\right)$
, where $z_{\textrm{max}}^{\textrm{R}}$ and $z_{\textrm{max}}^{\textrm{O}}$
are the maximum redshifts according to the radio and optical flux
limits. 

\noindent We model the quasar luminosity function as a double power-law
in absolute magnitude $M_{1450}$ \citep{1995ApJ...438..623P}, 
\noindent \begin{center}
\begin{equation}
{\displaystyle \begin{array}{c}
{\displaystyle \Phi\left(M_{1450}\right)=\Phi^{*}\left(M_{1450}^{*}\right)}\\
{\displaystyle \times\left(10^{0.4(\alpha+1)(M_{1450}-M^{*})}+10^{0.4(\beta+1)(M_{1450}-M^{*})}\right)^{-1},}
\end{array}}\label{eq:-5}
\end{equation}
\par\end{center}

\noindent where $M_{1450}^{*}$ is the break magnitude, $\Phi^{*}$
the normalization constant, $\alpha$ is the faint-end slope, and
$\beta$ is the bright-end slope. We split the quasar sample into
four different redshift intervals between $1.4<z<5.0$ with a $M_{1450}$
bin size equal to $\triangle M_{1450}=1.2\,\textrm{mag}$. The redshift
intervals and $M_{1450}$ bin size are selected to avoid incompleteness
effects in the luminosity function calculations (see Figure \ref{fig:M1450_redshift}).
Due to the small number of quasars in each luminosity bin ($N_{\textrm{QSO}}<50$),
the error bars are calculated assuming the low-statistics limit, using
the $84.13\%$ confidence Poisson upper limits and lower limits from
\citet{1986ApJ...303..336G}. Finally, we use the $M_{\textrm{1450}}$
lower limits indicated in Figure \ref{fig:M1450_redshift} to avoid
incompleteness effects in the calculation of the luminosity function.

\begin{figure*}[tp]
\begin{centering}
\includegraphics[clip,scale=0.8]{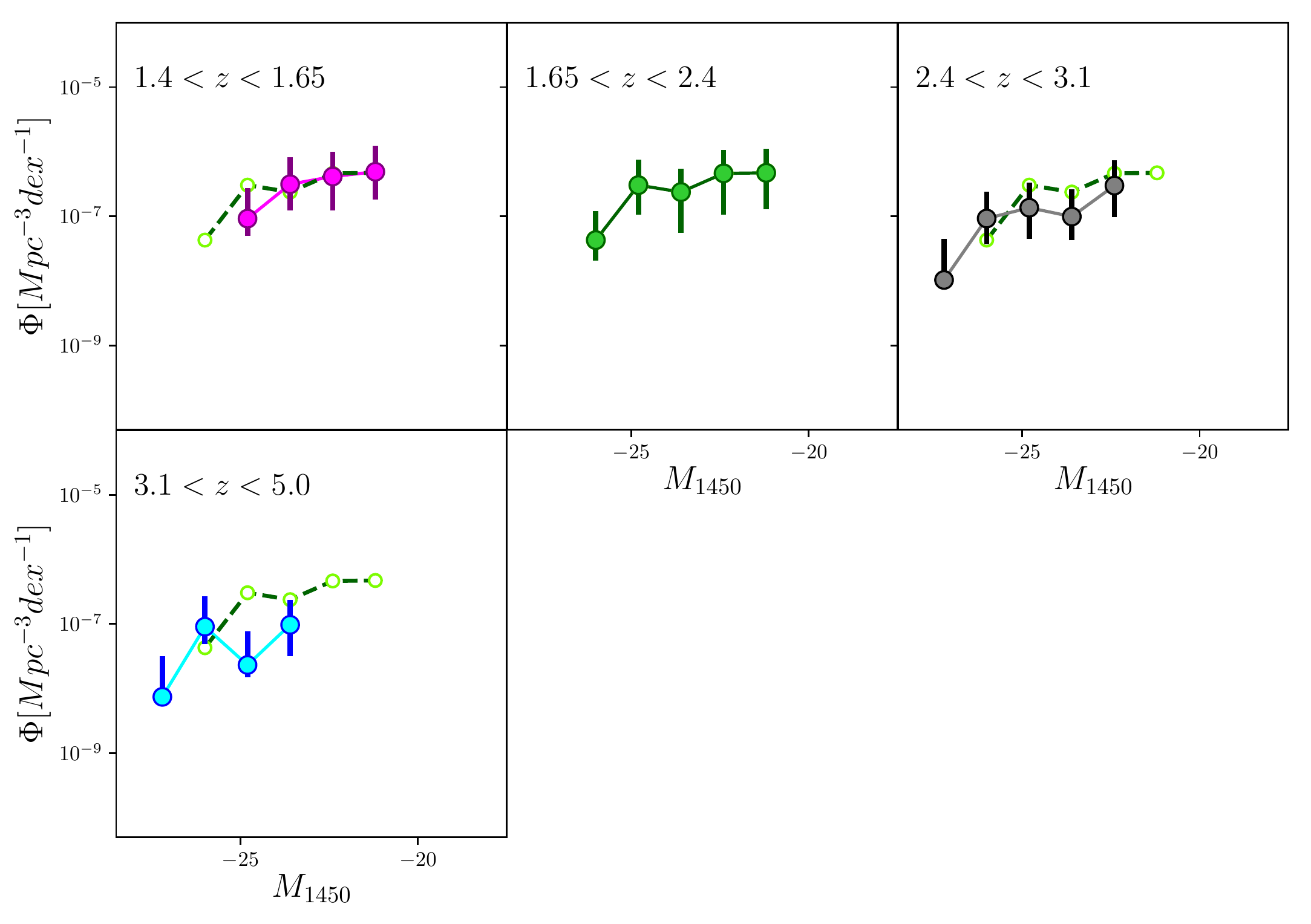}
\par\end{centering}
\centering{}\centering\caption{\label{fig:lf_nofitting} Rest-frame $M_{1450}$ binned luminosity
functions of our Bo\"{o}tes RSQ samples (colored circles) for five
non-overlapping redshift bins between $1.4<z<5.0$. In each panel,
we show as a reference the luminosity function at $1.65<z<2.4$.}
 
\end{figure*}

Figure \ref{fig:lf_nofitting} shows the luminosity function measurements
in four subpanels with one for each redshift interval. We use a total
of 83 spectroscopic and 47 photometric quasars to estimate the quasar
luminosity function. The resulting binned luminosity functions are
tabulated in Table \ref{tab:lf_values}. The luminosity function in
the range $1.65<z<2.4$ is plotted as a reference in all the subpanels.
This reference indicates that the space density of  RSQs is higher
at $1.65<z<2.4$ in comparison to the other redshift intervals, that
is the comoving space density of RSQs reaches a maximum between $1.65<z<2.4$.
The comoving space density of RSQs is discussed with further detail
in Section \ref{sec:density_evolution}. Additionally, a good continuity
between the points of the faint and bright ends is obtained in the
five redshift intervals.

\begin{table*}
\caption{Binned luminosity functions for RSQs between $1.4<z<5.0$. \label{tab:lf_values}}

\begin{centering}
\begin{tabular}{ccccccc}
 &  &  &  &  &  & \tabularnewline
\hline 
{\tiny{}Redshift range } & {\tiny{}$z_{\textrm{median}}$} & {\tiny{}$M_{1450}^{*}$ bin center} & {\tiny{}$\log\left(\Phi\right)^{a}$ } & {\tiny{}$\left.\sigma_{low}\right.^{b}$} & {\tiny{}$\left.\sigma_{upp}\right.^{b}$} & {\tiny{}$N^{c}$}\tabularnewline
\hline 
{\tiny{}$1.4<z<1.65$} & {\tiny{}1.57 } & {\tiny{}$-24.8$ } & {\tiny{}$-7.03$ } & {\tiny{}$0.42$} & {\tiny{}$1.81$} & {\tiny{}$3$}\tabularnewline
 &  & {\tiny{}$-23.6$} & {\tiny{}$-6.50$ } & {\tiny{}$1.89$ } & {\tiny{}$5.01$} & {\tiny{}$6$}\tabularnewline
 &  & {\tiny{}$-22.4$} & {\tiny{}$-6.38$} & {\tiny{}$2.90$} & {\tiny{}$5.79$} & {\tiny{}$11$}\tabularnewline
 &  & {\tiny{}$-21.2$} & {\tiny{}$-6.30$} & {\tiny{}$3.09$} & {\tiny{}$7.53$} & {\tiny{}$7$}\tabularnewline
\hline 
{\tiny{}$1.65<z<2.4$} & {\tiny{}1.92} & {\tiny{}$-26.0$ } & {\tiny{}$-7.36$ } & {\tiny{}$0.22$} & {\tiny{}$0.77$} & {\tiny{}$4$}\tabularnewline
 &  & {\tiny{}$-24.8$} & {\tiny{}$-6.52$ } & {\tiny{}$1.99$ } & {\tiny{}$4.54$} & {\tiny{}$8$}\tabularnewline
 &  & {\tiny{}$-23.6$} & {\tiny{}$-6.62$} & {\tiny{}$1.82$} & {\tiny{}$3.08$} & {\tiny{}$18$}\tabularnewline
 &  & {\tiny{}$-22.4$} & {\tiny{}$-6.33$} & {\tiny{}$3.58$} & {\tiny{}$5.95$} & {\tiny{}$19$}\tabularnewline
 &  & {\tiny{}$-21.2$} & {\tiny{}$-6.23$} & {\tiny{}$3.42$} & {\tiny{}$6.41$} & {\tiny{}$13$}\tabularnewline
\hline 
{\tiny{}$2.4<z<3.1$} & {\tiny{}2.70} & {\tiny{}$-27.2$ } & {\tiny{}$-7.70$ } & {\tiny{}$0.02$} & {\tiny{}$0.34$} & {\tiny{}$1$}\tabularnewline
 &  & {\tiny{}$-26.0$ } & {\tiny{}$-7.14$ } & {\tiny{}$0.55$} & {\tiny{}$1.47$} & {\tiny{}$3$}\tabularnewline
 &  & {\tiny{}$-24.8$} & {\tiny{}$-6.83$} & {\tiny{}$0.91$} & {\tiny{}$1.97$} & {\tiny{}$9$}\tabularnewline
 &  & {\tiny{}$-23.6$} & {\tiny{}$-7.39$} & {\tiny{}$0.56$} & {\tiny{}$1.66$} & {\tiny{}$5$}\tabularnewline
 &  & {\tiny{}$-22.4$} & {\tiny{}$-6.64$} & {\tiny{}$2.01$} & {\tiny{}$4.35$} & {\tiny{}$9$}\tabularnewline
\hline 
{\tiny{}$3.1<z<5.0$} & {\tiny{}3.25} & {\tiny{}$-27.2$ } & {\tiny{}$-8.13$ } & {\tiny{}$0.01$} & {\tiny{}$0.24$} & {\tiny{}$1$}\tabularnewline
 &  & {\tiny{}$-26.0$ } & {\tiny{}$-7.04$ } & {\tiny{}$0.41$} & {\tiny{}$1.78$} & {\tiny{}$2$}\tabularnewline
 &  & {\tiny{}$-24.8$} & {\tiny{}$-7.63$} & {\tiny{}$0.08$} & {\tiny{}$0.54$} & {\tiny{}$2$}\tabularnewline
 &  & {\tiny{}$-23.6$} & {\tiny{}$-6.99$} & {\tiny{}$0.70$} & {\tiny{}$1.44$} & {\tiny{}$10$}\tabularnewline
\hline 
\end{tabular}
\par\end{centering}
$\;$

\emph{Notes:} $\left.\right.^{a}$ {\tiny{}$\Phi$ is in units of
$1\times10^{-7}\:\textrm{Mpc}^{-3}\,\textrm{mag}^{-1}$.}$\left.\right.^{b}$
{\tiny{}$\sigma_{low}$ and $\sigma_{upp}$ are in units of $1\times10^{-7}\:\textrm{Mpc}^{-3}\,\textrm{mag}^{-1}$.
}$\left.\right.^{c}$ {\tiny{}$N$ is the number of quasars in the
corresponding $M_{1450}$ bin. }{\tiny \par}
\end{table*}

\noindent 

\section{Results \label{sec:Section6}}

\subsection{Model-fitting \label{sec:qlf-models}}

The observed evolution of the AGN luminosity function has traditionally
been studied using luminosity \citep{2000MNRAS.317.1014B,2009MNRAS.399.1755C,2006AJ....131.2766R},
density \citep{2013ApJ...768..105M,2010AJ....139..906W,2015ApJ...798...28K},
and even hybrid luminosity-density models \citep{2003ApJ...598..886U,2005A&A...441..417H,2013ApJ...773...14R,2016A&A...587A..41P}.
These studies have found that the evolution of the luminosity function
of quasars can be described by a pure luminosity evolution (PLE) model
at $z<2.2$ \citep{2009MNRAS.399.1755C}, while a combined luminosity
evolution and density evolution (LEDE) model can describe its evolution
at $z\gtrsim2.2$ \citep{2009MNRAS.399.1755C,2013ApJ...773...14R,2016A&A...587A..41P}.
The PLE introduces the redshift-dependence of the break magnitude
using the following second-order polynomial \citep{2009MNRAS.399.1755C}

\begin{equation}
M_{1450}^{*}\left(z\right)=M_{1450}^{*}\left(z=0\right)-2.5\,\left(k_{1}z+\,k_{2}z^{2}\right),
\end{equation}
while the LEDE model introduces the redshift-dependence in the normalization
and break magnitude using the following log-linear ansatz:

\begin{equation}
\log\left(\Phi^{*}\right)=\log\left[\Phi^{*}\left(z=z_{p}\right)\right]+c_{1}\,\left(z-z_{p}\right),
\end{equation}

\begin{equation}
M_{1450}^{*}\left(z\right)=M_{1450}^{*}\left(z=z_{p}\right)\,+c_{2}\,\left(z-z_{p}\right),
\end{equation}

\noindent where $z_{p}$ is the pivot redshift. Following previous
works (e.g., \citealt{2013ApJ...773...14R}), we employ the PLE model
to fit our binned luminosity function for redshift intervals $z<2.4$,
while at $z>2.4$ we use the LEDE model with $z_{p}=2.4$.

\begin{figure*}[tp]
\begin{centering}
\includegraphics[clip,scale=0.8]{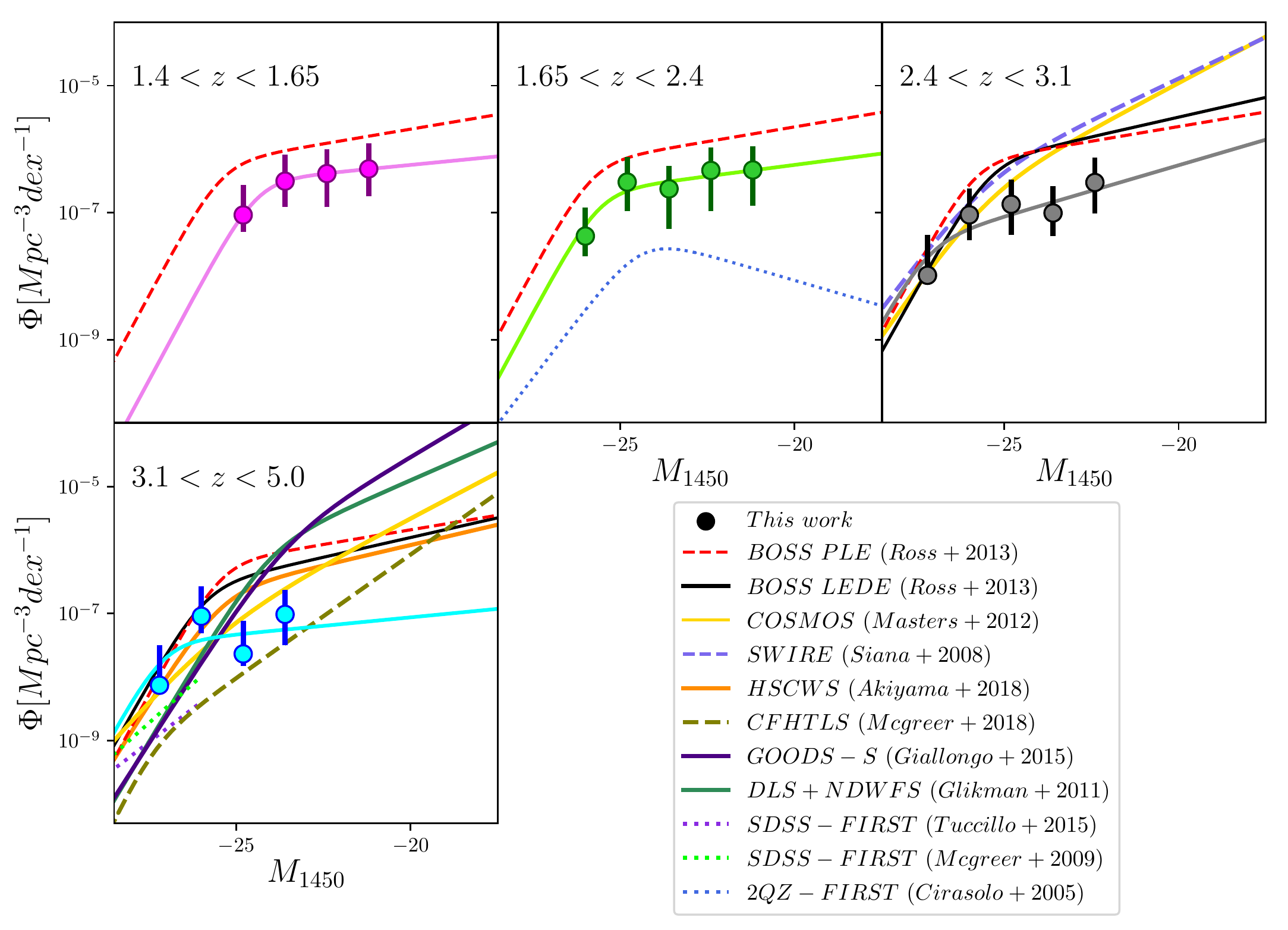}
\par\end{centering}
\centering{}\centering\caption{\label{fig:lf_fittings} Rest-frame $M_{1450}$ binned luminosity
functions of our Bo\"{o}tes RSQ samples (colored circles) for four
non-overlapping redshift intervals between $1.4<z<5.0$. The lines
show the corresponding best-fit models in each redshift bin. The best-fitting
parameters for each fit are presented in Table \ref{tab:fitting_results}.
For comparison, we show the QLFs from previous works \citep{2005MNRAS.357.1267C,2008ApJ...675...49S,2009AJ....138.1925M,2011ApJ...728L..26G,2012ApJ...755..169M,2013ApJ...773...14R,2015AAA...578A..83G,2015MNRAS.449.2818T,2018PASJ...70S..34A,2018AJ....155..131M}
measured over similar redshift intervals. The single-power law fits
by \citet{2009AJ....138.1925M} and \citet{2015MNRAS.449.2818T} are
plotted in the range $-29\leq M_{\textrm{1450}}\leq-26$, which is
the original range where they were measured.}
 
\end{figure*}

We use the $\chi^{2}$ minimization to fit the luminosity function
data points in each redshift bin to the corresponding models described
above. Because of the relatively small area ($\sim9.3\:\textrm{deg}^{2}$)
of the Bo\"otes field, there are only a few bright quasars in our
sample. This implies that the bright-end slope $\beta$ will be determined
with high-uncertainty due to small number statistics. Therefore, we
fix the bright-end slope $\beta$ to the values reported by \citet{2013ApJ...773...14R}
in their study of the quasar luminosity function using SDSS-DR9/BOSS
data \citep{2012ApJS..203...21A}. Additionally, we fix the parameters
$\left(k_{1}\:,k_{2}\right)$ and $\left(c_{1}\:,c_{2}\right)$ in
the PLE and LEDE models, respectively, to the values obtained by \citet{2013ApJ...773...14R}.
The parameter values from \citet{2013ApJ...773...14R} are chosen
to match our redshift intervals. For the LEDE model, we use the parameter
values corresponding to their S82 sample. Using the parameters from
their DR9 sample produces similar results. Finally, the best-fit parameters
and their associated uncertainties are summarized in Table \ref{tab:fitting_results}.
The corresponding best-fit model is shown with a colored line in each
subpanel of Figure \ref{fig:lf_fittings}. The models have a good
agreement with the binned luminosity function. 

\begin{table*}
\caption{Parametric model best-fit parameters and uncertainties. See Section
\ref{sec:qlf} for more details about the models used. \label{tab:fitting_results}}

\begin{centering}
\begin{tabular}{cccccccc}
 &  &  &  &  &  &  & \tabularnewline
\hline 
{\tiny{}Model } & {\tiny{}Redshift range } & {\tiny{}$\alpha$ } & {\tiny{}$\beta$} & {\tiny{}$M_{1450}^{*}(z=0)$} & {\tiny{}$k_{1}$} & {\tiny{}$k_{2}$} & {\tiny{}$\Phi^{*}$}\tabularnewline
\hline 
 &  & {\tiny{}{[}$\textrm{faint-end}${]} } & {\tiny{}{[}$\textrm{bright-end}${]} } & {\tiny{}{[}$\textrm{mag}${]} } &  &  & {\tiny{}{[}$\textrm{Mpc}^{-3}\,\textrm{mag}^{-1}${]} }\tabularnewline
{\tiny{}PLE} & {\tiny{}$1.4<z<1.65$} & {\tiny{}$-1.13\pm0.10$ } & {\tiny{}$-3.55$ } & {\tiny{}$-20.99\pm0.10$} & {\tiny{}$1.293$} & {\tiny{}$-0.268$} & {\tiny{}$3.28\times10^{-7}\pm4.03\times10^{-9}$}\tabularnewline
{\tiny{}PLE} & {\tiny{}$1.65<z<2.4$ } & {\tiny{}$-1.18\pm0.15$ } & {\tiny{}$-3.55$ } & {\tiny{}$-21.87\pm0.43$} & {\tiny{}$1.293$} & {\tiny{}$-0.268$} & {\tiny{}$2.25\times10^{-7}\pm8.80\times10^{-8}$}\tabularnewline
{\tiny{}PLE} & {\tiny{}$1.4<z<2.4$ } & {\tiny{}$-1.19\pm0.16$ } & {\tiny{}$-3.55$ } & {\tiny{}$-20.99\pm0.25$} & {\tiny{}$1.293$} & {\tiny{}$-0.268$} & {\tiny{}$3.61\times10^{-7}\pm1.15\times10^{-7}$}\tabularnewline
\hline 
 &  &  &  & {\tiny{}$M_{1450}^{*}(z=2.4)$} & {\tiny{}$c_{1}$} & {\tiny{}$c_{2}$} & \tabularnewline
\hline 
{\tiny{}LEDE} & {\tiny{}$2.4<z<3.1$} & {\tiny{}$-1.40\pm0.16$} & {\tiny{}$-3.51$ } & {\tiny{}$-26.98\pm1.38$} & {\tiny{}$-0.689$} & {\tiny{}$-0.809$} & {\tiny{}$6.17\times10^{-8}\pm6.20\times10^{-8}$}\tabularnewline
{\tiny{}LEDE} & {\tiny{}$3.1<z<5.0$} & {\tiny{}$-1.13\pm0.65$} & {\tiny{}$-3.51$ } & {\tiny{}$-26.41\pm2.06$} & {\tiny{}$-0.689$} & {\tiny{}$-0.809$} & {\tiny{}$1.40\times10^{-7}\pm2.41\times10^{-7}$}\tabularnewline
{\tiny{}LEDE} & {\tiny{}$2.4<z<5.0$} & {\tiny{}$-1.29\pm0.18$} & {\tiny{}$-3.51$ } & {\tiny{}$-26.12\pm0.58$} & {\tiny{}$-0.689$} & {\tiny{}$-0.809$} & {\tiny{}$1.63\times10^{-7}\pm1.07\times10^{-7}$}\tabularnewline
\hline 
\end{tabular}
\par\end{centering}
$\;$

\emph{Note:} Parameters without errors are kept fixed during the fitting.
\end{table*}

\noindent 

\noindent 
\begin{figure*}[tp]
\begin{centering}
\includegraphics[clip,scale=0.6]{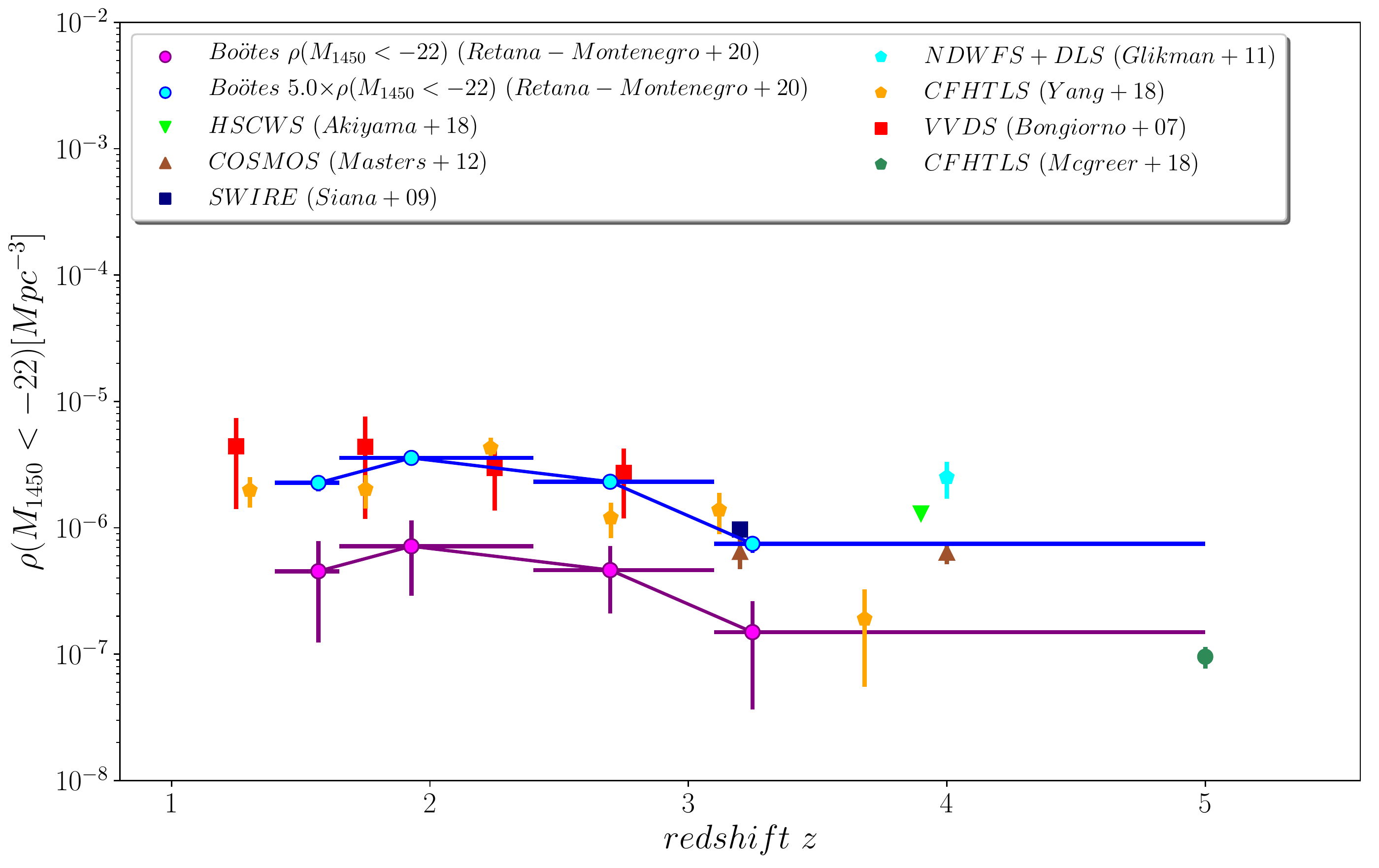}
\par\end{centering}
\centering{}\centering\caption{\label{fig:spatial_density} Spatial density of RSQs with $M_{1450}<-22$
as a function of redshift compared to the space density of faint quasar
samples $(M_{1450}<-22)$ from the literature. The spatial density
of RSQs is indicated by purple circles, while estimates from the literature
\citep{2007A&A...472..443B,2008ApJ...675...49S,2011ApJ...728L..26G,2012ApJ...755..169M,2018PASJ...70S..34A,2018AJ....155..110Y,2018AJ....155..131M}
are represented by the corresponding symbols in the legend box. We
also plot the spatial density of RSQs scaled by a factor of $5.0$
$(1/0.20)$ (blue circles).}
 
\end{figure*}

\subsection{Comparison to previous works \label{sec:comparison_previouswork}}

In Figure \ref{fig:lf_fittings}, we also compare our best-fit models
with previous works. The SDSS-III/BOSS luminosity function \citep{2013ApJ...773...14R}
was estimated at $2.2<z<3.5$ employing a uniform sample of 22301
quasars over an area of $2236\:\:\textrm{deg}^{2}$. Additionally,
\citet{2013ApJ...773...14R} investigated the evolution of the QLF
using a combination of SDSS \citep{2006AJ....131.2766R}, boss21+MMT
and BOSS Stripe 82 datasets to over a redshift range of $0.3<z<4.75$.
\citet{2013ApJ...773...14R} fitted their QLF data using a PLE model
at $z<2.2$, while their fittings at $z>2.2$ were carried out employing
a LEDE model. Furthermore, we compare our results with previous surveys
of faint quasars $\left(M_{\textrm{1450}}\leq-22.0\right)$ \citep{2008ApJ...675...49S,2011ApJ...728L..26G,2015AAA...578A..83G,2015ApJ...813...53M,2018PASJ...70S..34A,2018AJ....155..131M}.
The survey area of these studies ranges from $170\:\textrm{arcmin}^{2}$
\citep{2015AAA...578A..83G} to $339.8\:deg^{2}$ \citep{2018PASJ...70S..34A}.
Finally, we also consider the results of SDSS and 2QZ \citep{2001MNRAS.322L..29C,2004MNRAS.349.1397C}
radio-selected samples \citep{2005MNRAS.357.1267C,2009AJ....138.1925M,2015MNRAS.449.2818T}
using FIRST \citep{1995ApJ...450..559B}. It is clear that the number
density of our RSQs at all redshifts considered is lower in comparison
with that of samples of optically bright and faint quasars, which
are composed of both radio-detected and radio-undetected objects.
Naturally, the number density of RSQs is higher than the density of
SDSS/2QZ FIRST-selected samples \citep{2005MNRAS.357.1267C,2009AJ....138.1925M,2015MNRAS.449.2818T},
which are composed mainly of radio-loud quasars with fluxes $S_{1.4\textrm{GH}z}>1.0\:\textrm{mJy}$.
This flux limit corresponds to a LOFAR flux of $S_{150\textrm{MH}z}>4.80\:\textrm{mJy}$,
assuming a spectral index of $\alpha=-0.7$. As can be seen in the
bottom panel of Figure \ref{fig:mag_flux_distribution}, the fraction
of quasars fluxes with $S_{150\textrm{MH}z}>4.80\:\textrm{mJy}$ is
just $18\%$ of the total number of RSQs in our sample. Therefore,
the lower number densities presented by SDSS/2QZ FIRST-selected samples
are expected.

\begin{figure*}[tp]
\begin{centering}
\includegraphics[clip,scale=0.4]{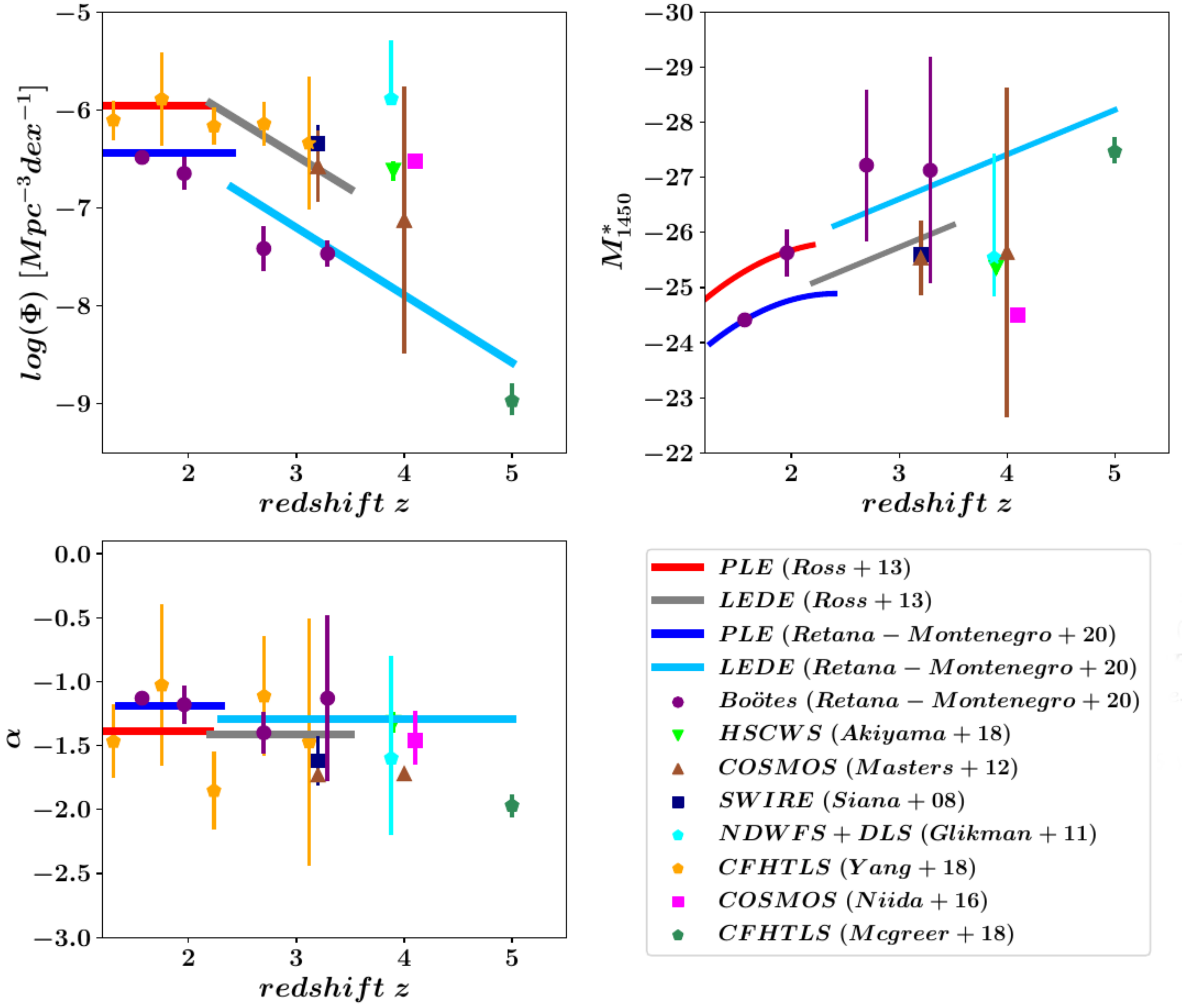}
\par\end{centering}
\centering{}\centering\caption{\label{fig:parameter_evolution} Best-fit quasar luminosity function
parameters as a function of redshift. Our results are indicated by
purple circles, while estimates from the literature \citep{2008ApJ...675...49S,2011ApJ...728L..26G,2012ApJ...755..169M,2016ApJ...832..208N,2018PASJ...70S..34A,2018AJ....155..110Y,2018AJ....155..131M}
are represented by the corresponding symbols in the legend box. The
red and gray lines represent the PLE and LEDE models from \citet{2013ApJ...773...14R},
and the blue and dark cyan lines our PLE ($1.4<z<2.4$) and LEDE ($2.4<z<5.0$)
models listed on Table \ref{tab:fitting_results}. For clarity, we
shift vertically the PLE and LEDE models from \citet{2013ApJ...773...14R}
by a factor of $+0.1$ in the third panel.}
 
\end{figure*}

We compare our best-fit parameters with previous studies at different
redshifts to constrain the evolution of the luminosity function of
RSQs. In Figure \ref{fig:parameter_evolution}, we compare the normalization
constant $\log\left(\Phi^{*}\right)$, the break magnitude $M_{1450}^{*}$,
and the faint-end slope $\alpha$ with previous values reported for
faint quasars as a function of redshift. We also plot the PLE and
LEDE models by \citet{2013ApJ...773...14R}, as well as our PLE and
LEDE models. From Figure \ref{fig:parameter_evolution}, we find the
following trends:
\begin{enumerate}
\item Our $\log\left(\Phi^{*}\right)$ values are lower in comparison with
those of other samples of faint quasars \citep{2008ApJ...675...49S,2011ApJ...728L..26G,2012ApJ...755..169M,2016ApJ...832..208N,2018AJ....155..110Y,2018PASJ...70S..34A}.
For redshifts $z<2.4$, the normalization constant within uncertainties
is consistent with a PLE evolutionary trend. At $z>2.4$, $\log\left(\Phi^{*}\right)$
seems to decrease following a linear-log trend reminiscence of a LEDE
evolution, and similar to that of the faint quasars.%For redshifts , it is hard to draw conclusions as the two values are not consistent between their respective. %At ,  seems to decrease following a linear-log trend reminiscence of a LEDE evolution, and similar to that of %the faint quasars.%This is in disagreement with PLE, in which the space density of QSOs peaks at the same redshift at every %luminosity. 
\item The break magnitude $M_{1450}^{*}$ seems to get brighter with increasing
redshift, a trend that is consistent with previously estimates \citep{2008ApJ...675...49S,2011ApJ...728L..26G,2015ApJ...813...53M,2016ApJ...832..208N}. 
\item The faint-end slope $\alpha$ does become steeper with increasing
redshift with a mean value of $\alpha=-1.15$ at $z<2.4$, while at
$z>2.4$ the mean value is $\alpha=-1.26$. Our $\alpha$ values are
consistent within error bars with those reported previously by several
authors \citep{2008ApJ...675...49S,2011ApJ...728L..26G,2016ApJ...832..208N,2018PASJ...70S..34A,2018AJ....155..110Y}.
\end{enumerate}
\noindent 

\subsection{Density evolution of RSQs \label{sec:density_evolution}}

In the last 20 years, the evolution of quasar activity has been studied
in the UV/optical \citep{2001AJ....122.2833F,2003A&A...408..499W,2006AJ....131.2766R,2007AA...461...39F,2007A&A...472..443B,2009MNRAS.399.1755C,2011ApJ...728L..26G,2013ApJ...773...14R,2015ApJ...813...53M,2016ApJ...833..222J,2018AJ....155..131M,2018AJ....155..110Y},
X-ray \citep{2005A&A...441..417H,2005ApJ...624..630S,2015MNRAS.451.1892A,2015MNRAS.453.1946G,2015ApJ...804..104M},
infrared \citep{2006ApJ...638...88B,2008ApJ...675...49S,2011ApJ...728...56A},
and radio \citep{2003ApJ...591...43V,2006MNRAS.370.1034C,2006MNRAS.371..695C,2009AJ....138.1925M,2015MNRAS.449.2818T}.
Here, we study this evolution using the spatial density of quasars
\citep{2001AJ....122.2833F,2013ApJ...768..105M,2015MNRAS.449.2818T}

\begin{equation}
\rho\left(\,<M_{1450}\,,z\,\right)=\int_{-\infty}^{M_{1450}}\,\Phi\left(\,M_{1450}\,,\,z\,\right)\,dM,
\end{equation}

\noindent where $\Phi\left(\,M_{1450}\,,\,z\,\right)$ is the luminosity
function of quasars, and it is integrated over all quasars more luminous
than $M_{1450}$. The integration is performed using the binned luminosity
functions instead of using the best-fit luminosity functions, as it
avoids uncertainties related to the model fitting and the extrapolation
of the models. An upper limit of $M_{1450}=-22.0$ is selected for
the integration as it is the lowest luminosity limit that is common
between our work and other samples of faint quasars, when comparing
their spatial density.

Figure \ref{fig:spatial_density} displays the space density of RSQs
($M_{1450}\leq-22.0$) from our sample as a function of redshift,
along with other faint quasar samples from the literature. Our spatial
density estimates have lower values in comparison with those of other
samples. However, our spatial density values show a very similar redshift
evolution to that of faint quasars, and this trend continues towards
$z=5.0$. It is clear that a peak in quasar activity occurs around
$z\approx2.0$ according to our results and the surveys carried out
by \citet{2007A&A...472..443B} and \citet{2018AJ....155..110Y}.
This agrees with the picture provided by measurements of the luminosity
function using samples of bright quasars ($M_{1450}\leq-24.0$) (e.g.,
\citealt{2006AJ....131.2766R,2009MNRAS.399.1755C,2013ApJ...773...14R}).
However, the quasar samples by \citet{2018PASJ...70S..34A} and \citet{2011ApJ...728L..26G}
display space densities much higher than expected in comparison with
the trend suggested by our results and other samples of faint quasars
\citep{2007A&A...472..443B,2015ApJ...813...53M,2018AJ....155..110Y,2008ApJ...675...49S,2018AJ....155..131M}.
These higher spatial densities could be attributed to sample contamination
\citep{2018PASJ...70S..34A} or cosmic variance \citep{2011ApJ...728L..26G}.
Finally, and as expected, our values of the space density of RSQs
are significantly higher in comparison to those previously estimated
for RLQs. Samples of RLQs present space density values of just a few
$\textrm{Gpc}^{-3}$ (e.g., \citealt{2003ApJ...591...43V,2015MNRAS.449.2818T}).
This highlights the fact that deep LOFAR observations allow us to
detect the radio-emission of quasars that otherwise would be classified
as radio-quiet \citep{10.3389/fspas.2018.00005}.

\begin{figure*}[tp]
\begin{centering}
\includegraphics[clip,scale=0.6]{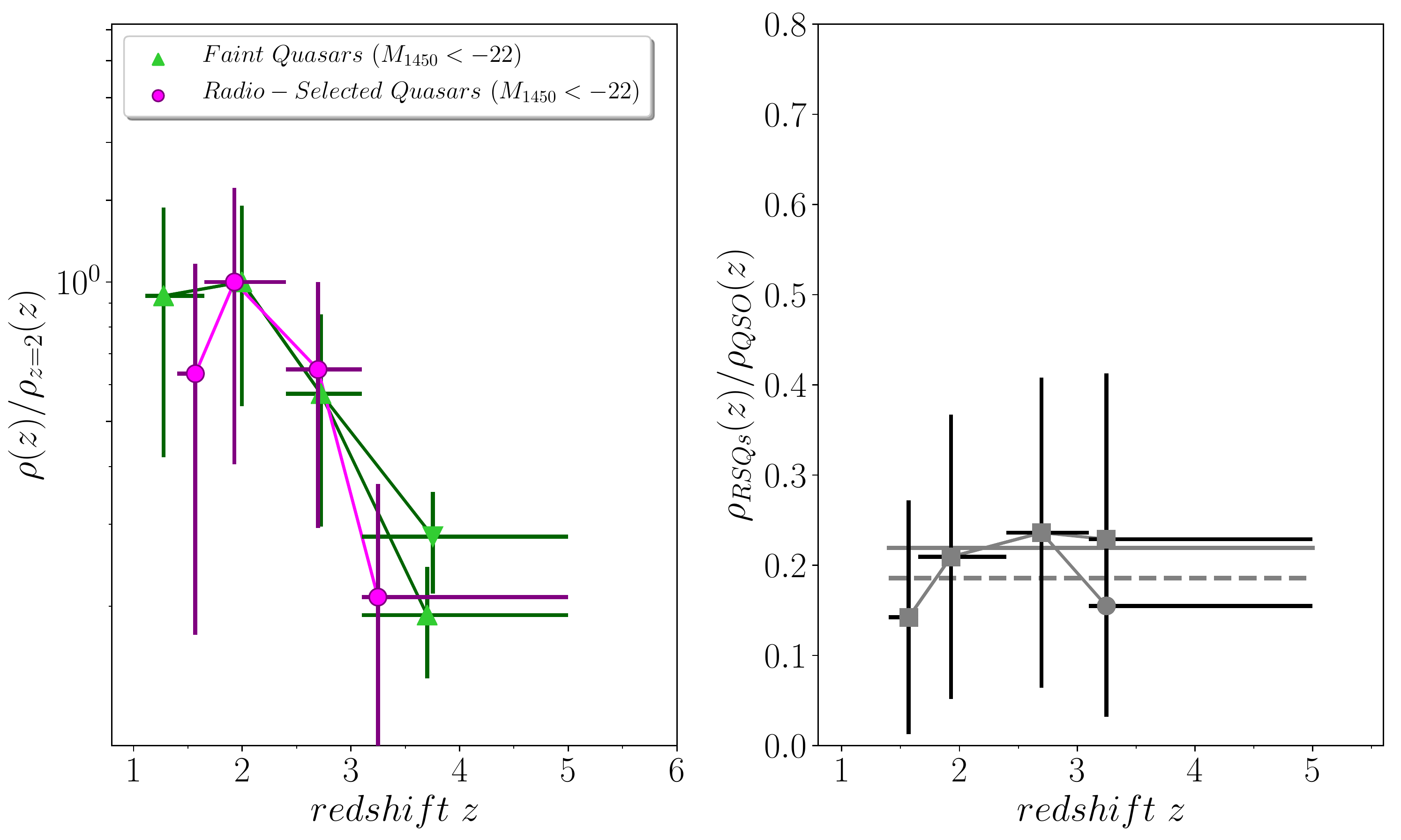}
\par\end{centering}
\centering{}\centering\caption{\label{fig:relative_spatial_density} \emph{Left panel:} Spatial densities,
normalized to $z\sim2$, as a function of redshift for optically faint
quasars and RSQs. Our results are indicated by fuchsia circles, while
the spatial density for faint quasars is determined from results reported
in the literature \citep{2007A&A...472..443B,2008ApJ...675...49S,2011ApJ...728L..26G,2012ApJ...755..169M,2018PASJ...70S..34A,2018AJ....155..110Y}
and is denoted by green triangles. The downward triangle indicates
the spatial density excluding the results by \citet{2011ApJ...728L..26G}
and \citet{2018PASJ...70S..34A}, while the value denoted the downward
triangle includes them. \emph{Right panel}: Relative fraction of RSQs
with respect to the spatial density of faint quasars as a function
of redshift. The ratio is calculated between overlapping redshift
bins. The gray solid line indicates the mean ratio of $0.22$ excluding
the works by \citet{2011ApJ...728L..26G} and \citet{2018PASJ...70S..34A},
while the dashed line denotes a mean ratio of $0.18$ including these
works. }
 
\end{figure*}

Having computed the spatial density for the relevant samples of faint
quasars and our RSQ sample, we calculate their normalized to $z\sim2$
spatial densities as a function of redshift. In the left panel of
Figure \ref{fig:relative_spatial_density}, we show the normalized
spatial density of RSQs and faint quasars with $M_{\textrm{1450}}<-22$.
It is clear that the space density of faint quasars and RSQs decreases
rapidly with redshift. From its maximum at $z\sim2$, the space density
of faint quasars (RSQs) declines between $z\simeq2$ and $z\simeq3$
by a factor of $1.74\pm0.84$ $(1.54\pm0.84)$, while it reduces further
from $z\simeq2$ to $z\simeq5.0$ by a factor of $5.23\pm1.41$ $(4.78\pm3.61)$.
At $z\sim1.5$, the normalized space-density ratio is $1.07\pm0.59$
$(1.57\pm1.15)$. Note the agreement within error bars between the
evolution of the space density of RSQs and that of other faint quasar
samples \citep{2007A&A...472..443B,2008ApJ...675...49S,2015ApJ...813...53M,2018AJ....155..110Y}.
Considering the works by \citet{2018PASJ...70S..34A} and \citet{2011ApJ...728L..26G},
the space density of faint quasars decreases from $z\simeq2$ to $z\simeq5.0$
by a factor of $3.54\pm0.87$.

Only a small fraction, less than $10\%$, of the quasars are classified
as radio-loud \citep{1989AJ.....98.1195K,2002AJ....124.2364I,2007ApJ...656..680J}.
However, RLQs are often associated with massive host galaxies (\citealt{2009ApJ...697.1656S,2017A&A...600A..97R},
and references therein), whose radio emission is produced by large
and powerful radio jets \citep{1994AJ....108..766B,2008MNRAS.390..595M}.
However, the dependency of the radio-loud fraction (RLF) of quasars
on redshift and luminosity is still a matter of debate. Some authors
have found that the RLF is a strong function of luminosity (e.g.,
\citealt{1993MNRAS.263..461P,1994AJ....108.1548L}) and redshift (e.g.,
\citealt{1986MNRAS.218..265P,1990MNRAS.244..207M,1992ApJ...391..560V}),
while others have found that it does not depend significantly on either
redshift or luminosity (e.g., \citealt{2000AJ....119.1526S,2003MNRAS.341..993C}).
\citet{2015AJ....149...61K} found that selection effects could be
biasing the conclusions about the evolution of the RLF. Using the
LoTSS survey \citep{2019A&A...622A...1S}, \citet{2019Anda} showed
that quasars exhibit a wide continuum of radio properties, with no
clear bimodality in the radio-loudness parameter.

In the context of this work, we compute the relative fraction of RSQs
with respect to the spatial density of faint quasars as a function
of redshift by dividing the spatial density of RSQs, $\rho_{\textrm{RSQs}}\left(z\right)$,
by the spatial density of faint quasars (radio-detected plus radio-undetected),
$\rho_{\textrm{QSO}}\left(z\right)$. Figure \ref{fig:relative_spatial_density}
displays the relative fraction of RSQs, $\rho_{\textrm{RSQs}}\left(z\right)/\rho_{\textrm{QSO}}\left(z\right)$,
as a function of redshift. The relative fraction of RSQs considering
the error bars and excluding the results of \citet{2011ApJ...728L..26G}
and \citet{2018PASJ...70S..34A} is roughly independent of redshift,
with a median value of $0.22\pm0.16$. This fraction is of $0.18\pm0.14$
considering the results of \citet{2011ApJ...728L..26G} and \citet{2018PASJ...70S..34A}.
In Figure \ref{fig:relative_spatial_density}, the spatial density
of RSQs is multiplied by a factor of $5.0\,(1/0.20)$ to compare it
with the spatial density of faint quasars. With the multiplicative
factor applied, the agreement between the two spatial densities is
good. Moreover, it highlights the similarity in the redshift evolution
of RSQs and faint quasars up to $z\sim5$. A fraction of $\sim0.20$
of RSQs with respect to faint quasars is relatively higher than the
fractions of $\sim0.10-0.15$ of RLQs with respect to the whole quasar
population previously estimated (e.g., \citet{1999ApJ...511..612G,2000AJ....119.1526S,2007ApJ...656..680J}).
However, this is not unexpected; as previously mentioned, our deep
LOFAR observations allow us to detect the radio-emission of a considerable
number of quasars that otherwise would be identified as radio-quiet.

Finally, our results for the spatial density of RSQs demonstrate that
the selection of quasars utilizing ML algorithms that combines optical
and infrared with LOFAR observations (see Section \ref{subsec:3.3 classification_results})
is very efficient and robust.

\noindent 

\section{Discussion \label{sec:Section7}}

In this section, we discuss several aspects related to RSQs such as:
the origins of their radio-emission, the environments where these
objects reside, and their location in spectroscopic parameter spaces.

\subsection{The origins of radio-emission in RSQs \label{sec:RSQs_radio_emission_origin}}

An important piece of information that is needed to understand the
origins of the radio-emission in RSQs could come from their observed
radio-morphologies. Although some of the brightest RLQs have double-lobed
radio morphologies, the majority of intermediate-luminosity RLQs show
core dominated radio-morphologies \citep{2006AJ....131..666D,2007AJ....133.1615L,2017MNRAS.466..921C}.
Approximately, $92\%$ of RSQs in our sample present compact radio-morphology
at the resolution of the LOFAR-Bo\"otes mosaic. A possible explanation
for the origin of radio-emission in these objects lies in the interaction
between outflows and the IGM, as first suggested by \citet{1992ApJ...396..487S}
to explain the low radio-emission in broad absorption-line quasars
(BALQSOs). In this mechanism, radio-emission originates from particles
accelerated on the shock fronts caused by the collision of uncollimated
central outflows with the IGM of the host galaxy \citep{2014MNRAS.442..784Z,2016MNRAS.455.4191Z}.
This scenario is supported by the observations of BALQSOs, as these
objects have intermediate radio-luminosities and the majority present
core dominated radio-morphologies \citep{2000ApJ...538...72B,2008MNRAS.391..246L,2011ApJ...743...71D,2018arXiv181107931M},
along with core-jet structures \citep{2015A&A...579A.109K} and lobes
\citep{2014MNRAS.440.2474W} in some instances. Assuming that the
absorbing troughs observed in the spectra of BALQSOs are caused by
uncollimated central outflows loaded into the broad emission-line
region (BELR), the low numbers of BAL systems in RLQs with double-lobed
radio morphology \citep{2011ApJ...743...71D,2013Ap&SS.345..355P,2014MNRAS.440.2474W}
can be attributed to the fact that in these quasars the central outflows
form collimated jets which are physically separated from the BELR.
In this scenario, the lack of radio-emission in quasars, traditionally
classified as radio-quiet, can be explained considering that in a
majority of cases the outflowing material is slowed down by a dense
interstellar clump and the formation of shock fronts is hindered.
In our LOFAR-Bo\"otes mosaic, we detect the radio-emission of many
quasars that in previous radio surveys would have remained undetected.
The radio-emission in these quasars could have originated from the
interaction between quasar outflows and the IGM. However, deeper optical
and low-frequency radio surveys, in addition to LOFAR sub-arcsecond
resolution observations \citep{2015A&A...574A.114V,2016MNRAS.461.2676M},
are needed to explore this mechanism in detail.

\subsection{The environment of RSQs \label{sec:RSQs_morphology_environment-1} }

A fraction of $92\%$ of RSQs in our sample could be classified as
compact steep-spectrum sources (CSS) according to their radio properties.
CSS sources are usually a fraction of $\sim10-30$ per cent in previous
radio surveys \citep{1982MNRAS.198..843P,1990A&A...231..333F,1998PASP..110..493O},
and they are characterized by their small projected linear sizes and
median steep radio spectrum ($\alpha<-0.77$, \citealt{10.1086/316162}).
The brighter RSQs in our sample have a steep spectral index distribution
with a median value of $\alpha\backsimeq-0.70$ \citep{10.3389/fspas.2018.00005},
and only $8\%$ of RSQs in our sample present morphologies consistent
with core-jet structures.

It has been suggested that CSS may be small either because they are
young and still in an early stage of their evolutionary path, eventually
developing into Fanaroff-Riley type-I/II \citep{1974MNRAS.167P..31F}
radio sources \citep{1995A&A...302..317F,2000MNRAS.319....8A,2000MNRAS.319..445S,2016AN....337...36C},
or because they are embedded in a very dense environment that frustrates
the propagation of the radio jets \citep{1984AJ.....89....5V,1986A&A...170...10F,1989A&A...217...44F,2007A&A...461..923O}.
The compactness of their radio-morphologies suggests that RSQs may
reside in host galaxies with a large supply of gas to fuel the early
stages of quasar activity. Ultimately, these scenarios will have to
be tested against high-resolution observations with submillimeter
and radio interferometers that can spatially resolve the host-galaxies
of RSQs and their synchrotron-dominated core-jets, respectively. These
observations will, in turn, help us to shed light on the complex interplay
between RSQs and their host-galaxies, and how quasar activity is triggered
in these systems.

\subsection{RSQs and their location in spectroscopic parameter spaces \label{sec:RSQs_quasar_evolution}}

The most striking features of Figure \ref{fig:relative_spatial_density},
excluding the results by \citet{2011ApJ...728L..26G} and \citet{2018PASJ...70S..34A}
from the analysis are: i) RSQs show evolutionary trends and declining
factors that are similar to those presented by faint quasars ($M_{\textrm{1450}}\leq-22.0$)
(see Figs. \ref{fig:spatial_density}, \ref{fig:parameter_evolution},
\ref{fig:relative_spatial_density}), and ii) the fact that RSQs may
compose to up $22\pm16\%$ of the total faint quasar population, a
fraction that within uncertainties is independent of redshift (see
Fig. \ref{fig:relative_spatial_density}). Interestingly, similar
decline factors in the space density of low- \citep{1994ApJ...421..412W,2009MNRAS.399.1755C,2016A&A...587A..41P}
and high- \citep{1995AJ....110...68S,1995AJ....110.2553K,2016A&A...587A..41P}
optical luminosity quasars, respectively, had been reported before.
In these works, high-luminosity quasars have declining factors of
$\simeq2-3$ between $z\approx2$ and $z\approx4$, while low-luminosity
quasars present steeper declining factors of $\simeq6-8$ between
the same redshift intervals. These factors are consistent with a downsizing
evolutionary scenario. In this scenario, high-luminosity quasars evolve
first at earlier epochs and reach their maximum space density at high-z,
while low-luminosity quasars predominantly evolve at later epochs
reaching their maximum space density at low-z (e.g., \citealt{2005A&A...441..417H,2005ApJ...624..630S,2009MNRAS.399.1755C}).
Declining factors similar to those of high-luminosity quasars had
been reported for RLQs samples \citep{1998ASPC..146...17H,2003ApJ...591...43V}. 

Since the early 2000's there has been substantial advances in our
understanding of quasars using the broad emission line properties
and their correlations. Probably, the most widely used broad emission
line correlations are the eigenvector 1 (E1) and $\textrm{C}_{\textrm{IV}}$
parameter spaces. The E1 parameter space started as the primary eigenvector
in the Principal Component Analysis performed by \citet{1992ApJS...80..109B},
where $\textrm{Fe}_{\textrm{II}}$ and $\textrm{H}_{\beta}$ emission
are related to line width. The generalization of this concept led
to the 4D Eigenvector 1 (4DE1) parameter space, with the addition
of the properties of the $\textrm{C}_{\textrm{IV}}$ and the soft
X-ray photon index \citep{2000ARA&A..38..521S,2000ApJ...536L...5S}.
The 4DE1 parameter space serves as a 4D equivalent of the 2D Hertzsprung-Russell
diagrams \citep{1909AN....179..373H,1914PA.....22..275R}. This parameter
space has revealed a principal sequence of quasars characterized by
the Eddington ratio, and since its introduction has become an important
tool for depicting the diversity of quasars and their evolutionary
states (see \citealt{2015FrASS...2....6S}, and references therein).
The $\textrm{C}_{\textrm{IV}}$ parameter space ($\textrm{C}_{\textrm{IV}}$
EW versus $\textrm{C}_{\textrm{IV}}$ blueshift) has been used to
study different quasar properties at high-z (e.g., \citealt{1999ASPC..162..395B,2007ApJ...666..757S,2011AJ....141..167R,2015AJ....149...61K,2016MNRAS.461..647C}).
In the context of the 4DE1, E1, and $\textrm{C}_{\textrm{IV}}$ parameter
spaces, several authors (\citealt{2003ApJ...597L..17S}, \citealt{2007ApJ...666..757S},
\citealt{2008MNRAS.387..856Z}, and \citealt{2011AJ....141..167R})
have determined that RLQs and RQQs are clustered at different locations
in their corresponding parameter spaces (see Fig. 14 in \citealt{2011AJ....141..167R}
and Fig. 3 in \citealt{2003ApJ...597L..17S}). In particular, \citet{2011AJ....141..167R}
and \citet{2015AJ....149...61K} demonstrated using the E1 and $\textrm{C}_{\textrm{IV}}$
spaces, which may trace the relative power of radiation line-driven
accretion disk winds \citep{2011AJ....141..167R}, that on average
RLQs present weaker radiation line-driven winds in comparison with
RQQs. These authors suggest that RLQs and RQQs are two parallel evolutionary
sequences, and possibly a series of spin and merge events \citep{2007ApJ...658..815S,2009AN....330..291S,2017ApJ...849....4S}
are responsible for the triggering of radio jets, and turning RQQs
into radio-loud.

Considering that RSQs present evolutionary trends similar to those
of both faint quasars and bright quasars, it is possible that faint
(radio and optically) RSQs could share properties of both RLQs and
RQQs. Thus, RSQs would occupy intermediate locations between RQQs
and RLQs in their corresponding E1, 4DE1 and $\textrm{C}_{\textrm{IV}}$
parameter spaces. Future spectroscopic studies of RSQs would be a
major step forward towards understanding radio-loudness. 

\section{Conclusions\label{sec:conclusions}}

In this work, we train three ML algorithms: RF, SVM, and Bootstrap
aggregation with optical and infrared imaging to compile a sample
of quasars in the $9.3\:\textrm{deg}^{2}$ Bo\"otes field. We eliminate
stellar and likely galaxy contaminants from our sample by requiring
a $5\sigma$ detection in deep LOFAR imaging by applying a morphological
criterium, respectively. The requirement of a $5\sigma$ LOFAR detection
does not only allow us to eliminate the stellar contamination in our
sample, but also to reduce the number of contaminants by two orders
of magnitude. The final sample consists of 130 quasars with either
spectroscopic or photometric redshifts in the range of $1.4\leq z\leq5.0$.
We estimate the photometric redshifts of the photometric quasars using
the NW kernel regression estimator \citep{doi:10.1137/1109020,10.2307/25049340}.
When comparing the predictions of this method to the spectroscopic
redshifts of 1193 Bo\"otes spectroscopic quasars, we find that $76\%$
of the quasars have photometric redshifts that are within $\left|\delta z\right|\leq0.3$
of their spectroscopic redshifts. The spectral energy distributions
calculated using deep photometry available for the NDWFS-Bo\"otes
field confirm the validity of the photometrically selected quasars
using the ML algorithms as robust candidate quasars. We demonstrate
that in cases of lack of deep and complete mid-infrared coverage needed
to perform a wedge-based mid-infrared selection of AGNs, the selection
of quasars using ML algorithms trained with optical and infrared photometry
in combination with LOFAR data is an effective approach for obtaining
samples of quasars. We compute the fraction of quasars missed due
to our selection (i.e. selection function) using a library of simulated
quasar spectra. The binned optical luminosity function of RSQs is
computed using the $1/V_{\textrm{max}}$ method \citep{1968ApJ...151..393S}
in five different redshift bins between $1.4\leq z\leq5.0$. These
luminosity functions are corrected for incompleteness due to the radio
observations and selection method employed. The parametric fits to
the binned optical luminosity function of RSQs are consistent with
a PLE evolution model at $z<2.4$, and a LEDE evolution at $z>2.4$.

We have studied the optical luminosity function of RSQs down to faint
luminosities of $M_{\textrm{1450}}=-22$. Previous studies were mostly
limited to bright RLQs with luminosities $M_{\textrm{1450}}=-26$
\citep{2005MNRAS.357.1267C,2006MNRAS.370.1034C,2009AJ....138.1925M,2015MNRAS.449.2818T}.
We find evidence that suggests that the faint-end slope $\alpha$
is becoming steeper with increasing redshift, as found by previous
measurements of the luminosity function of faint quasars \citep{2011ApJ...728L..26G,2015AAA...578A..83G}.
Our mean values of the faint-end slope are $\alpha=-1.15$ at $z<2.4$,
while at $z>2.4$ the mean value is $\alpha=-1.26$. We calculate
the space density of RSQs over $1.4\leq z\leq5.0$ and find an evolutionary
trend below and above the peak of their space density that is comparable
to that of faint quasars. By comparing the spatial density of RSQs
with that of faint quasars at similar redshifts, we find that RSQs
may compose to up $22\pm16\%$ of the total faint quasar population.
This fraction, within uncertainties, seems to remain constant with
redshift. We argue that considering the similarities in evolutionary
trends and declining factors between RSQs and faint quasars, the fainter
(optically and radio) RSQs may have properties of both RLQs and RQQs.
Finally, we discuss several aspects of RSQs such as: the origins of
their radio emission, the environments where these objects reside,
and their location in E1, 4DE1 and $\textrm{C}_{\textrm{IV}}$ parameter
spaces.

%Our luminosity function measurements are lower in comparison with those of the whole (radio-detected and radio-undetected) faint quasar population. Clearly, deeper radio imaging is %required to detect the radio emission of the fainter quasars.

Our work demonstrates the feasibility of studying the evolution of
RSQs using samples of quasars compiled with ML algorithms trained
with optical and infrared photometry combined with LOFAR data. Future
studies of the luminosity function of RSQs will benefit from the advent
of the new generation of wide-field radio (LOTSS: \citealt{2011JApA...32..557R,2017A&A...598A.104S,2019A&A...622A...1S};
EMU: \citealt{2011PASA...28..215N}), optical (LSST: \citealt{2002SPIE.4836...10T,2009arXiv0912.0201L};
DES: \citealt{2005IJMPA..20.3121F}), infrared (WFIRST: \citealt{2013arXiv1305.5422S};
Euclid: \citealt{2011arXiv1110.3193L}), and spectroscopic surveys
(EBOSS: \citealt{2016AJ....151...44D}; DESI: \citealt{2016arXiv161100036D}).

%The evolutionary trend in the luminosity function of quasars above the peak of their number density at z is a fundamental observable to understand the %early growth of SMBHs. 

\begin{acknowledgements}

ERM acknowledges financial support from NWO Top project, No. 614.001.006. HR acknowledges support from the ERC Advanced Investigator program NewClusters 321271. ERM  acknowledges the valuable suggestions and comments from H. Andernach. Moreover, we would like to thank the referee for valuable suggestions on the manuscript. \\

LOFAR, the Low Frequency Array designed and constructed by ASTRON, has facilities in several countries, that are owned by various parties (each with their own funding sources), and that are collectively operated by the International LOFAR Telescope (ILT) foundation under a joint scientific policy. The Open University is incorporated by Royal Charter (RC 000391), an exempt charity in England \& Wales and a charity registered in Scotland (SC 038302). The Open University is authorized and regulated by the Financial Conduct Authority.\\

Funding for the Sloan Digital Sky Survey IV has been provided by the Alfred P. Sloan Foundation, the U.S. Department of Energy Office of Science, and the Participating Institutions. SDSS- IV acknowledges support and resources from the Center for High-Performance Computing at the University of Utah. The SDSS web site is www.sdss.org. \\
SDSS-IV is managed by the Astrophysical Research Consortium for the Participating Institutions of the SDSS Collaboration including the Brazilian Participation Group, the Carnegie Institution for Science, Carnegie Mellon University, the Chilean Participation Group, the French Participation Group, Harvard-Smithsonian Center for Astrophysics, Instituto de Astrof\'{i}sica de Canarias, The Johns Hopkins University, Kavli Institute for the Physics and Mathematics of the Universe (IPMU) / University of Tokyo, Lawrence Berkeley National Laboratory, Leibniz Institut f\"{u}r Astrophysik Potsdam (AIP), Max-Planck-Institut f\"{u}r Astronomie (MPIA Heidelberg), Max-Planck-Institut f\"{u}r Astrophysik (MPA Garching), Max-Planck-Institut f\"{u}r Extraterrestrische Physik (MPE), National Astronomical Observatory of China, New Mexico State University, New York University, University of Notre Dame, Observat\'{o}rio Nacional / MCTI, The Ohio State University, Pennsylvania State University, Shanghai Astronomical Observatory, United Kingdom Participation Group, Universidad Nacional Aut\'{o}noma de M\'{e}xico, University of Arizona, University of Colorado Boulder, University of Oxford, University of Portsmouth, University of Utah, University of Virginia, University of Washington, University of Wisconsin, Vanderbilt University, and Yale University.
\\
The Pan-STARRS1 Surveys (PS1) have been made possible through contributions of the Institute for Astronomy, the University of Hawaii, the Pan-STARRS Project Office, the Max-Planck Society and its participating institutes, the Max Planck Institute for Astronomy, Heidelberg and the Max Planck Institute for Extraterrestrial Physics, Garching, The Johns Hopkins University, Durham University, the University of Edinburgh, Queen's University Belfast, the Harvard-Smithsonian Center for Astrophysics, the Las Cumbres Observatory Global Telescope Network Incorporated, the National Central University of Taiwan, the Space Telescope Science Institute, the National Aeronautics and Space Administration under Grant No. NNX08AR22G issued through the Planetary Science Division of the NASA Science Mission Directorate, the National Science Foundation under Grant No. AST-1238877, the University of Maryland, and Eotvos Lorand University (ELTE).
\\
This publication makes use of data products from the Wide-field Infrared Survey Explorer, which is a joint project of the University of California, Los Angeles, and the Jet Propulsion Laboratory/California Institute of Technology, funded by the National Aeronautics and Space Administration.

\end{acknowledgements}

\begin{appendix} \section{A sample of false color RGB (R=$B_W$, G=$R$, B=$I$) images} %\section{Title of the second appendix} 
\label{sec:appendix_B}

In this appendix, we present a sample of false color RGB (R=$B_{W}$,
G=$R$, B=$I$) images centered on spectroscopic and photometric quasars.
Each image covers $70^{\prime\prime}\times70^{\prime\prime}$, and
the inset size is $7^{\prime\prime}\times7{}^{\prime\prime}$. The
contours are $[3,5,7,9,11,13]\times\sigma$ times the local noise
level in the LOFAR (white) and FIRST (purple) images. 

\noindent 
\begin{figure*}[tp]
\begin{centering}
\includegraphics[clip,scale=0.92]{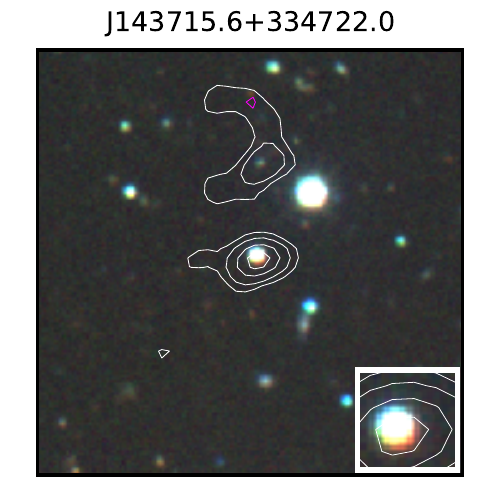}\includegraphics[clip,scale=0.92]{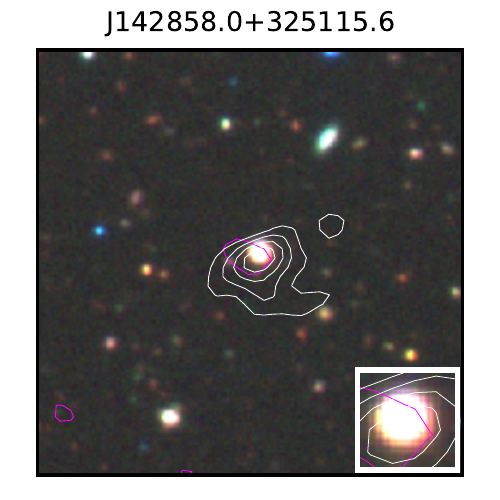}\includegraphics[clip,scale=0.92]{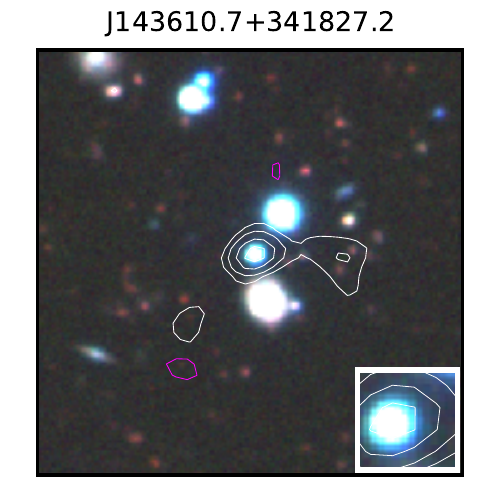}\includegraphics[clip,scale=0.92]{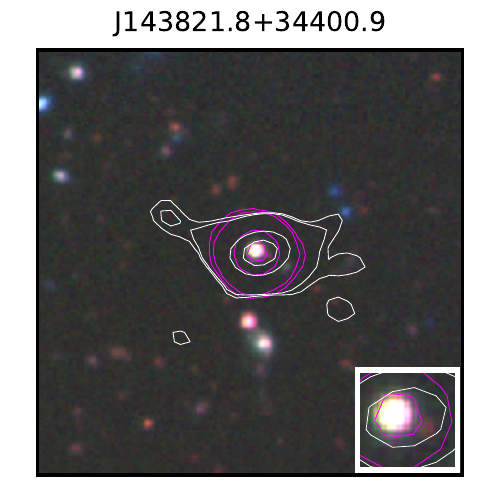}
\par\end{centering}
\begin{centering}
\includegraphics[clip,scale=0.92]{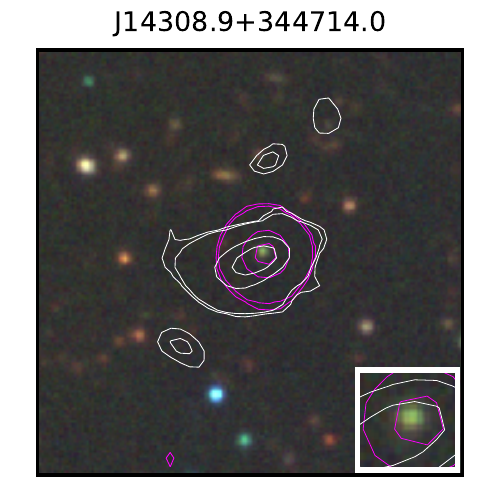}\includegraphics[clip,scale=0.92]{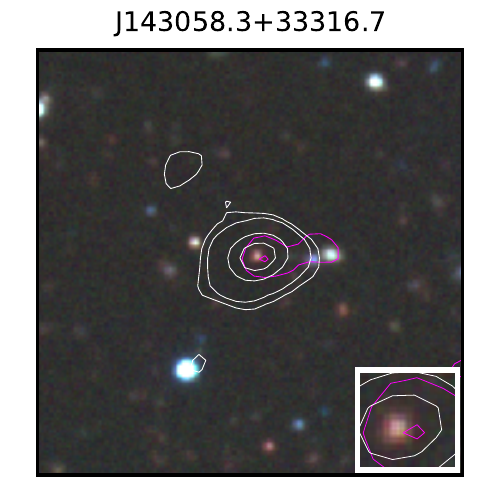}\includegraphics[clip,scale=0.92]{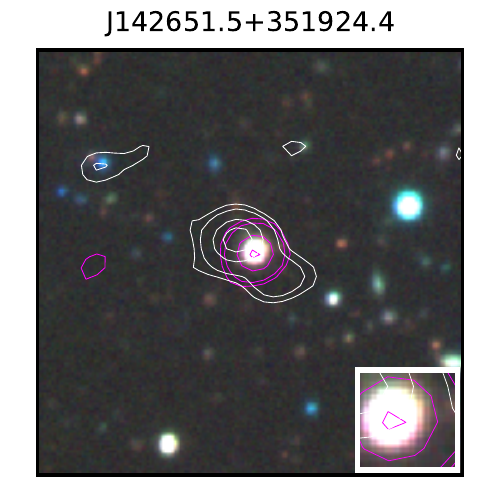}\includegraphics[clip,scale=0.92]{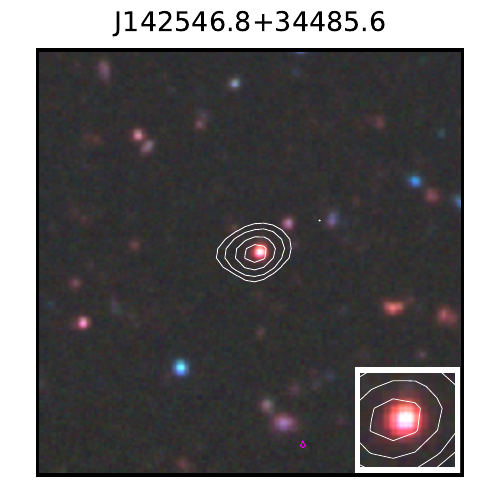}
\par\end{centering}
\centering{}\centering\caption{\label{fig:rgb_plots_spec_qso} A sample of false color RGB (R=$B_{W}$,
G=$R$, B=$I$) NDWFS images centered on spectroscopic quasars. Each
image covers $70^{\prime\prime}\times70^{\prime\prime}$, and the
inset size is $7^{\prime\prime}\times7{}^{\prime\prime}$. The contours
are $[3,5,7,9,11,13]\times\sigma$ times the local noise level in
the LOFAR (white) and FIRST (purple) images. }
 
\end{figure*}

\noindent 
\begin{figure*}[tp]
\begin{centering}
\includegraphics[clip,scale=0.92]{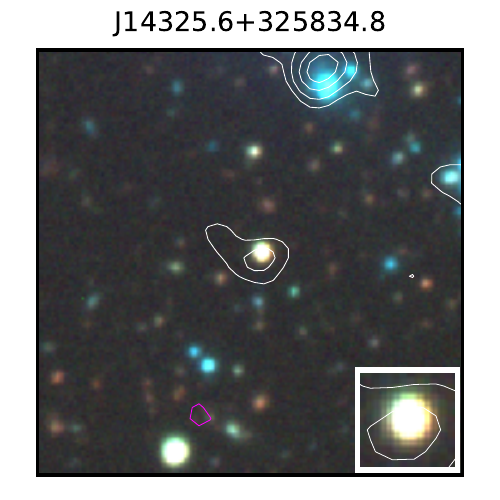}\includegraphics[clip,scale=0.92]{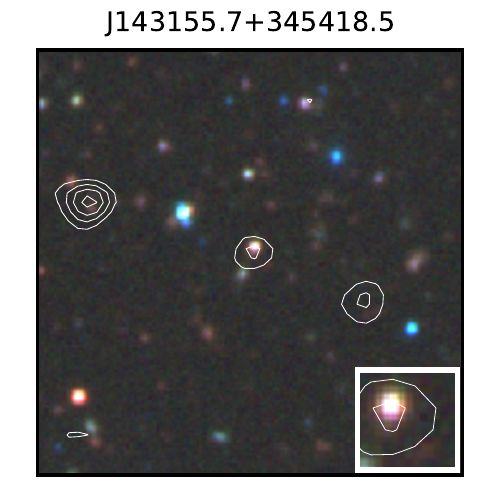}\includegraphics[clip,scale=0.92]{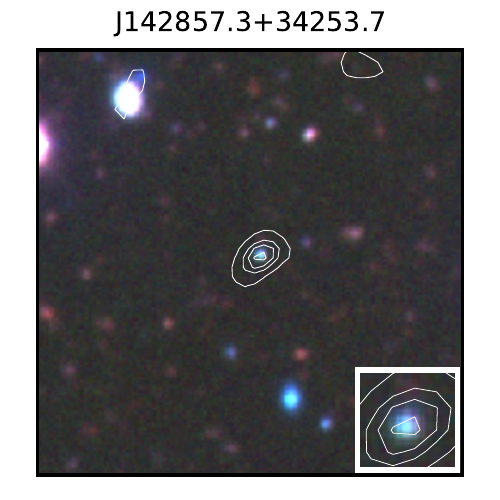}\includegraphics[clip,scale=0.92]{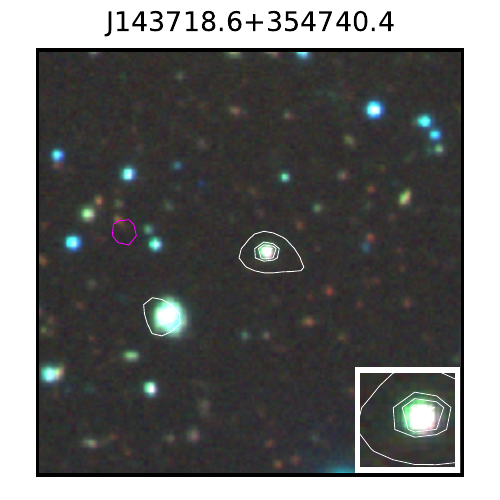}
\par\end{centering}
\begin{centering}
\includegraphics[clip,scale=0.92]{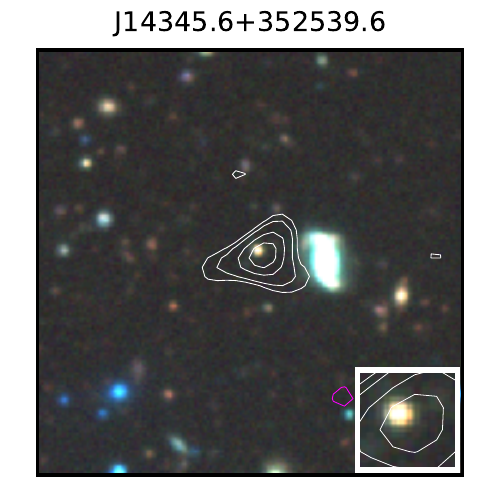}\includegraphics[clip,scale=0.92]{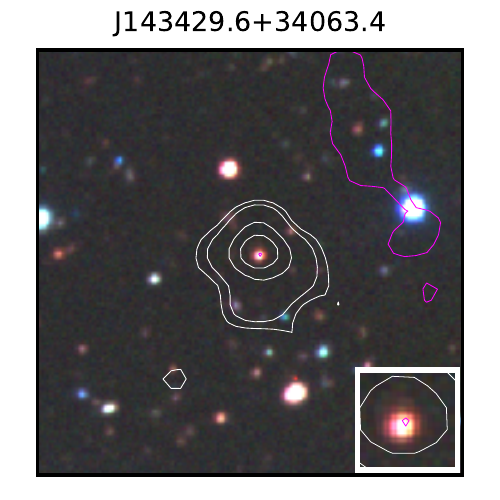}\includegraphics[clip,scale=0.92]{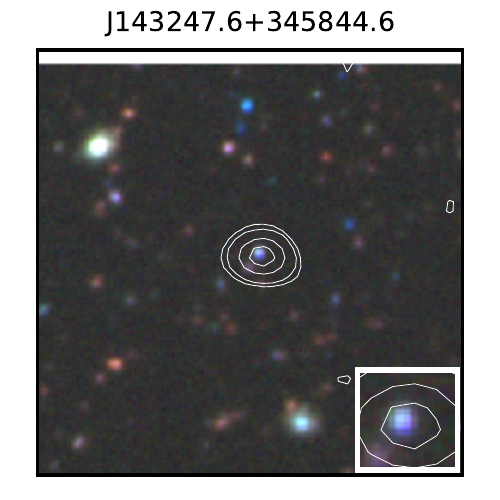}\includegraphics[clip,scale=0.92]{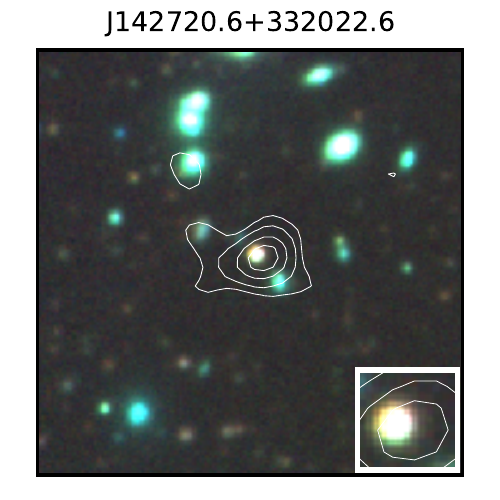}
\par\end{centering}
\centering{}\centering\caption{\label{fig:rgb_plots_photo_qso} A sample of false color RGB (R=$B_{W}$,
G=$R$, B=$I$) NDWFS images centered on photometric quasars. Each
image covers $70^{\prime\prime}\times70^{\prime\prime}$, and the
inset size is $7^{\prime\prime}\times7{}^{\prime\prime}$. The contours
are $[3,5,7,9,11,13]\times\sigma$ times the local noise level in
the LOFAR (white) and FIRST (purple) images. }
 
\end{figure*}

\end{appendix}

\bibliographystyle{aa}
\addcontentsline{toc}{section}{\refname}\bibliography{my_bib}

\begin{thebibliography}{241}
\expandafter\ifx\csname natexlab\endcsname\relax\def\natexlab#1{#1}\fi

\bibitem[{{Abolfathi} {et~al.}(2018){Abolfathi}, {Aguado}, {Aguilar}, {Allende
  Prieto}, {Almeida}, {Ananna}, {Anders}, {Anderson}, {Andrews}, {Anguiano}, \&
  et~al.}]{2018ApJS..235...42A}
{Abolfathi}, B., {Aguado}, D.~S., {Aguilar}, G., {et~al.} 2018, \apjs, 235, 42

\bibitem[{{Ahn} {et~al.}(2012){Ahn}, {Alexandroff}, {Allende Prieto},
  {Anderson}, {Anderton}, {Andrews}, {Aubourg}, {Bailey}, {Balbinot}, {Barnes},
  \& et~al.}]{2012ApJS..203...21A}
{Ahn}, C.~P., {Alexandroff}, R., {Allende Prieto}, C., {et~al.} 2012, \apjs,
  203, 21

\bibitem[{{Aird} {et~al.}(2015){Aird}, {Coil}, {Georgakakis}, {Nandra},
  {Barro}, \& {P{\'e}rez-Gonz{\'a}lez}}]{2015MNRAS.451.1892A}
{Aird}, J., {Coil}, A.~L., {Georgakakis}, A., {et~al.} 2015, \mnras, 451, 1892

\bibitem[{{Akiyama} {et~al.}(2018){Akiyama}, {He}, {Ikeda}, {Niida}, {Nagao},
  {Bosch}, {Coupon}, {Enoki}, {Imanishi}, {Kashikawa}, {Kawaguchi}, {Komiyama},
  {Lee}, {Matsuoka}, {Miyazaki}, {Nishizawa}, {Oguri}, {Ono}, {Onoue}, {Ouchi},
  {Schulze}, {Silverman}, {Tanaka}, {Tanaka}, {Terashima}, {Toba}, \&
  {Ueda}}]{2018PASJ...70S..34A}
{Akiyama}, M., {He}, W., {Ikeda}, H., {et~al.} 2018, \pasj, 70, S34

\bibitem[{{Alexander}(2000)}]{2000MNRAS.319....8A}
{Alexander}, P. 2000, \mnras, 319, 8

\bibitem[{Altman(1992)}]{doi:10.1080/00031305.1992.10475879}
Altman, N.~S. 1992, The American Statistician, 46, 175

\bibitem[{{Ashby} {et~al.}(2009){Ashby}, {Stern}, {Brodwin}, {Griffith},
  {Eisenhardt}, {Koz{\l}owski}, {Kochanek}, {Bock}, {Borys}, {Brand}, {Brown},
  {Cool}, {Cooray}, {Croft}, {Dey}, {Eisenstein}, {Gonzalez}, {Gorjian},
  {Grogin}, {Ivison}, {Jacob}, {Jannuzi}, {Mainzer}, {Moustakas},
  {R{\"o}ttgering}, {Seymour}, {Smith}, {Stanford}, {Stauffer}, {Sullivan},
  {van Breugel}, {Willner}, \& {Wright}}]{2009ApJ...701..428A}
{Ashby}, M.~L.~N., {Stern}, D., {Brodwin}, M., {et~al.} 2009, \apj, 701, 428

\bibitem[{{Assef} {et~al.}(2011){Assef}, {Kochanek}, {Ashby}, {Brodwin},
  {Brown}, {Cool}, {Forman}, {Gonzalez}, {Hickox}, {Jannuzi}, {Jones}, {Le
  Floc'h}, {Moustakas}, {Murray}, \& {Stern}}]{2011ApJ...728...56A}
{Assef}, R.~J., {Kochanek}, C.~S., {Ashby}, M.~L.~N., {et~al.} 2011, \apj, 728,
  56

\bibitem[{{Ba{\~n}ados} {et~al.}(2015){Ba{\~n}ados}, {Venemans}, {Morganson},
  {Hodge}, {Decarli}, {Walter}, {Stern}, {Schlafly}, {Farina}, {Greiner},
  {Chambers}, {Fan}, {Rix}, {Burgett}, {Draper}, {Flewelling}, {Kaiser},
  {Metcalfe}, {Morgan}, {Tonry}, \& {Wainscoat}}]{2015ApJ...804..118B}
{Ba{\~n}ados}, E., {Venemans}, B.~P., {Morganson}, E., {et~al.} 2015, \apj,
  804, 118

\bibitem[{{Baldwin}(1977)}]{1977ApJ...214..679B}
{Baldwin}, J.~A. 1977, \apj, 214, 679

\bibitem[{{Becker} {et~al.}(2000){Becker}, {White}, {Gregg}, {Brotherton},
  {Laurent-Muehleisen}, \& {Arav}}]{2000ApJ...538...72B}
{Becker}, R.~H., {White}, R.~L., {Gregg}, M.~D., {et~al.} 2000, \apj, 538, 72

\bibitem[{{Becker} {et~al.}(1995){Becker}, {White}, \&
  {Helfand}}]{1995ApJ...450..559B}
{Becker}, R.~H., {White}, R.~L., \& {Helfand}, D.~J. 1995, \apj, 450, 559

\bibitem[{{Bershady} {et~al.}(1999){Bershady}, {Charlton}, \&
  {Geoffroy}}]{1999ApJ...518..103B}
{Bershady}, M.~A., {Charlton}, J.~C., \& {Geoffroy}, J.~M. 1999, \apj, 518, 103

\bibitem[{{Bertin} \& {Arnouts}(1996)}]{1996AAS..117..393B}
{Bertin}, E. \& {Arnouts}, S. 1996, \aaps, 117, 393

\bibitem[{{Bian} {et~al.}(2013){Bian}, {Fan}, {Jiang}, {McGreer}, {Dey},
  {Green}, {Maiolino}, {Walter}, {Lee}, \& {Dav{\'e}}}]{2013ApJ...774...28B}
{Bian}, F., {Fan}, X., {Jiang}, L., {et~al.} 2013, \apj, 774, 28

\bibitem[{{Bongiorno} {et~al.}(2007){Bongiorno}, {Zamorani}, {Gavignaud},
  {Marano}, {Paltani}, {Mathez}, {M{\o}ller}, {Picat}, {Cirasuolo},
  {Lamareille}, {Bottini}, {Garilli}, {Le Brun}, {Le F{\`e}vre}, {Maccagni},
  {Scaramella}, {Scodeggio}, {Tresse}, {Vettolani}, {Zanichelli}, {Adami},
  {Arnouts}, {Bardelli}, {Bolzonella}, {Cappi}, {Charlot}, {Ciliegi},
  {Contini}, {Foucaud}, {Franzetti}, {Guzzo}, {Ilbert}, {Iovino}, {McCracken},
  {Marinoni}, {Mazure}, {Meneux}, {Merighi}, {Pell{\`o}}, {Pollo}, {Pozzetti},
  {Radovich}, {Zucca}, {Hatziminaoglou}, {Polletta}, {Bondi}, {Brinchmann},
  {Cucciati}, {de la Torre}, {Gregorini}, {Mellier}, {Merluzzi}, {Temporin},
  {Vergani}, \& {Walcher}}]{2007A&A...472..443B}
{Bongiorno}, A., {Zamorani}, G., {Gavignaud}, I., {et~al.} 2007, \aap, 472, 443

\bibitem[{{Bonzini} {et~al.}(2013){Bonzini}, {Padovani}, {Mainieri},
  {Kellermann}, {Miller}, {Rosati}, {Tozzi}, \&
  {Vattakunnel}}]{2013MNRAS.436.3759B}
{Bonzini}, M., {Padovani}, P., {Mainieri}, V., {et~al.} 2013, \mnras, 436, 3759

\bibitem[{{Boroson} \& {Green}(1992)}]{1992ApJS...80..109B}
{Boroson}, T.~A. \& {Green}, R.~F. 1992, \apjs, 80, 109

\bibitem[{{Bovy} {et~al.}(2011){Bovy}, {Hennawi}, {Hogg}, {Myers},
  {Kirkpatrick}, {Schlegel}, {Ross}, {Sheldon}, {McGreer}, {Schneider}, \&
  {Weaver}}]{2011ApJ...729..141B}
{Bovy}, J., {Hennawi}, J.~F., {Hogg}, D.~W., {et~al.} 2011, \apj, 729, 141

\bibitem[{{Boyle} {et~al.}(2000){Boyle}, {Shanks}, {Croom}, {Smith}, {Miller},
  {Loaring}, \& {Heymans}}]{2000MNRAS.317.1014B}
{Boyle}, B.~J., {Shanks}, T., {Croom}, S.~M., {et~al.} 2000, \mnras, 317, 1014

\bibitem[{{Brammer} {et~al.}(2008){Brammer}, {van Dokkum}, \&
  {Coppi}}]{2008ApJ...686.1503B}
{Brammer}, G.~B., {van Dokkum}, P.~G., \& {Coppi}, P. 2008, \apj, 686, 1503

\bibitem[{Breiman(1996)}]{Breiman1996}
Breiman, L. 1996, Machine Learning, 24, 123

\bibitem[{Breiman(2001)}]{Breiman2001}
Breiman, L. 2001, Machine Learning, 45, 5

\bibitem[{{Bridle} {et~al.}(1994){Bridle}, {Hough}, {Lonsdale}, {Burns}, \&
  {Laing}}]{1994AJ....108..766B}
{Bridle}, A.~H., {Hough}, D.~H., {Lonsdale}, C.~J., {Burns}, J.~O., \& {Laing},
  R.~A. 1994, \aj, 108, 766

\bibitem[{{Brotherton} \& {Francis}(1999)}]{1999ASPC..162..395B}
{Brotherton}, M.~S. \& {Francis}, P.~J. 1999, in Astronomical Society of the
  Pacific Conference Series, Vol. 162, Quasars and Cosmology, ed. G.~{Ferland}
  \& J.~{Baldwin}, 395

\bibitem[{{Brown} {et~al.}(2006){Brown}, {Brand}, {Dey}, {Jannuzi}, {Cool}, {Le
  Floc'h}, {Kochanek}, {Armus}, {Bian}, {Higdon}, {Higdon}, {Papovich},
  {Rieke}, {Rieke}, {Smith}, {Soifer}, \& {Weedman}}]{2006ApJ...638...88B}
{Brown}, M.~J.~I., {Brand}, K., {Dey}, A., {et~al.} 2006, \apj, 638, 88

\bibitem[{{Brown} {et~al.}(2007){Brown}, {Dey}, {Jannuzi}, {Brand}, {Benson},
  {Brodwin}, {Croton}, \& {Eisenhardt}}]{2007ApJ...654..858B}
{Brown}, M.~J.~I., {Dey}, A., {Jannuzi}, B.~T., {et~al.} 2007, \apj, 654, 858

\bibitem[{{Calzetti} {et~al.}(2000){Calzetti}, {Armus}, {Bohlin}, {Kinney},
  {Koornneef}, \& {Storchi-Bergmann}}]{2000ApJ...533..682C}
{Calzetti}, D., {Armus}, L., {Bohlin}, R.~C., {et~al.} 2000, \apj, 533, 682

\bibitem[{Campbell {et~al.}(2012)Campbell, Lo, \&
  MacKinlay}]{campbell2012econometrics}
Campbell, J., Lo, A., \& MacKinlay, A. 2012, The Econometrics of Financial
  Markets (Princeton University Press)

\bibitem[{{Carballo} {et~al.}(2006){Carballo}, {Gonz{\'a}lez-Serrano},
  {Montenegro-Montes}, {Benn}, {Mack}, {Pedani}, \&
  {Vigotti}}]{2006MNRAS.370.1034C}
{Carballo}, R., {Gonz{\'a}lez-Serrano}, J.~I., {Montenegro-Montes}, F.~M.,
  {et~al.} 2006, \mnras, 370, 1034

\bibitem[{{Carrasco} {et~al.}(2015){Carrasco}, {Barrientos}, {Pichara},
  {Anguita}, {Murphy}, {Gilbank}, {Gladders}, {Yee}, {Hsieh}, \&
  {L{\'o}pez}}]{2015A&A...584A..44C}
{Carrasco}, D., {Barrientos}, L.~F., {Pichara}, K., {et~al.} 2015, \aap, 584,
  A44

\bibitem[{{Chabrier} {et~al.}(2000){Chabrier}, {Baraffe}, {Allard}, \&
  {Hauschildt}}]{2000ApJ...542..464C}
{Chabrier}, G., {Baraffe}, I., {Allard}, F., \& {Hauschildt}, P. 2000, \apj,
  542, 464

\bibitem[{{Chambers} {et~al.}(2016){Chambers}, {Magnier}, {Metcalfe},
  {Flewelling}, {Huber}, {Waters}, {Denneau}, {Draper}, {Farrow}, {Finkbeiner},
  {Holmberg}, {Koppenhoefer}, {Price}, {Rest}, {Saglia}, {Schlafly}, {Smartt},
  {Sweeney}, {Wainscoat}, {Burgett}, {Chastel}, {Grav}, {Heasley}, {Hodapp},
  {Jedicke}, {Kaiser}, {Kudritzki}, {Luppino}, {Lupton}, {Monet}, {Morgan},
  {Onaka}, {Shiao}, {Stubbs}, {Tonry}, {White}, {Ba{\~n}ados}, {Bell},
  {Bender}, {Bernard}, {Boegner}, {Boffi}, {Botticella}, {Calamida},
  {Casertano}, {Chen}, {Chen}, {Cole}, {Deacon}, {Frenk}, {Fitzsimmons},
  {Gezari}, {Gibbs}, {Goessl}, {Goggia}, {Gourgue}, {Goldman}, {Grant},
  {Grebel}, {Hambly}, {Hasinger}, {Heavens}, {Heckman}, {Henderson}, {Henning},
  {Holman}, {Hopp}, {Ip}, {Isani}, {Jackson}, {Keyes}, {Koekemoer}, {Kotak},
  {Le}, {Liska}, {Long}, {Lucey}, {Liu}, {Martin}, {Masci}, {McLean}, {Mindel},
  {Misra}, {Morganson}, {Murphy}, {Obaika}, {Narayan}, {Nieto-Santisteban},
  {Norberg}, {Peacock}, {Pier}, {Postman}, {Primak}, {Rae}, {Rai}, {Riess},
  {Riffeser}, {Rix}, {R{\"o}ser}, {Russel}, {Rutz}, {Schilbach}, {Schultz},
  {Scolnic}, {Strolger}, {Szalay}, {Seitz}, {Small}, {Smith}, {Soderblom},
  {Taylor}, {Thomson}, {Taylor}, {Thakar}, {Thiel}, {Thilker}, {Unger},
  {Urata}, {Valenti}, {Wagner}, {Walder}, {Walter}, {Watters}, {Werner},
  {Wood-Vasey}, \& {Wyse}}]{2016arXiv161205560C}
{Chambers}, K.~C., {Magnier}, E.~A., {Metcalfe}, N., {et~al.} 2016, arXiv
  e-prints, arXiv:1612.05560

\bibitem[{{Chen} \& {Guestrin}(2016)}]{2016arXiv160302754C}
{Chen}, T. \& {Guestrin}, C. 2016, arXiv e-prints, arXiv:1603.02754

\bibitem[{{Cirasuolo} {et~al.}(2005){Cirasuolo}, {Magliocchetti}, \&
  {Celotti}}]{2005MNRAS.357.1267C}
{Cirasuolo}, M., {Magliocchetti}, M., \& {Celotti}, A. 2005, \mnras, 357, 1267

\bibitem[{{Cirasuolo} {et~al.}(2003){Cirasuolo}, {Magliocchetti}, {Celotti}, \&
  {Danese}}]{2003MNRAS.341..993C}
{Cirasuolo}, M., {Magliocchetti}, M., {Celotti}, A., \& {Danese}, L. 2003,
  \mnras, 341, 993

\bibitem[{{Cirasuolo} {et~al.}(2006){Cirasuolo}, {Magliocchetti}, {Gentile},
  {Celotti}, {Cristiani}, \& {Danese}}]{2006MNRAS.371..695C}
{Cirasuolo}, M., {Magliocchetti}, M., {Gentile}, G., {et~al.} 2006, \mnras,
  371, 695

\bibitem[{{Coatman} {et~al.}(2016){Coatman}, {Hewett}, {Banerji}, \&
  {Richards}}]{2016MNRAS.461..647C}
{Coatman}, L., {Hewett}, P.~C., {Banerji}, M., \& {Richards}, G.~T. 2016,
  \mnras, 461, 647

\bibitem[{{Collier} {et~al.}(2016){Collier}, {Norris}, {Filipovi{\'c}}, \&
  {Tothill}}]{2016AN....337...36C}
{Collier}, J.~D., {Norris}, R.~P., {Filipovi{\'c}}, M.~D., \& {Tothill},
  N.~F.~H. 2016, Astronomische Nachrichten, 337, 36

\bibitem[{{Condon} {et~al.}(2013){Condon}, {Kellermann}, {Kimball},
  {Ivezi{\'c}}, \& {Perley}}]{2013ApJ...768...37C}
{Condon}, J.~J., {Kellermann}, K.~I., {Kimball}, A.~E., {Ivezi{\'c}}, {\v Z}.,
  \& {Perley}, R.~A. 2013, \apj, 768, 37

\bibitem[{{Cool}(2007)}]{2007ApJS..169...21C}
{Cool}, R.~J. 2007, \apjs, 169, 21

\bibitem[{{Cool} {et~al.}(2006){Cool}, {Kochanek}, {Eisenstein}, {Stern},
  {Brand}, {Brown}, {Dey}, {Eisenhardt}, {Fan}, {Gonzalez}, {Green}, {Jannuzi},
  {McKenzie}, {Rieke}, {Rieke}, {Soifer}, {Spinrad}, \&
  {Elston}}]{2006AJ....132..823C}
{Cool}, R.~J., {Kochanek}, C.~S., {Eisenstein}, D.~J., {et~al.} 2006, \aj, 132,
  823

\bibitem[{Cortes \& Vapnik(1995)}]{Cortes1995}
Cortes, C. \& Vapnik, V. 1995, Machine Learning, 20, 273

\bibitem[{{Coziol} {et~al.}(2017){Coziol}, {Andernach}, {Torres-Papaqui},
  {Ortega-Minakata}, \& {Moreno del Rio}}]{2017MNRAS.466..921C}
{Coziol}, R., {Andernach}, H., {Torres-Papaqui}, J.~P., {Ortega-Minakata},
  R.~A., \& {Moreno del Rio}, F. 2017, \mnras, 466, 921

\bibitem[{{Cristiani} \& {Vio}(1990)}]{1990A&A...227..385C}
{Cristiani}, S. \& {Vio}, R. 1990, \aap, 227, 385

\bibitem[{{Croom} {et~al.}(2009{\natexlab{a}}){Croom}, {Richards}, {Shanks},
  {Boyle}, {Sharp}, {Bland-Hawthorn}, {Bridges}, {Brunner}, {Cannon}, {Carson},
  {Chiu}, {Colless}, {Couch}, {De Propris}, {Drinkwater}, {Edge}, {Fine},
  {Loveday}, {Miller}, {Myers}, {Nichol}, {Outram}, {Pimbblet}, {Roseboom},
  {Ross}, {Schneider}, {Smith}, {Stoughton}, {Strauss}, \&
  {Wake}}]{2009MNRAS.392...19C}
{Croom}, S.~M., {Richards}, G.~T., {Shanks}, T., {et~al.} 2009{\natexlab{a}},
  \mnras, 392, 19

\bibitem[{{Croom} {et~al.}(2009{\natexlab{b}}){Croom}, {Richards}, {Shanks},
  {Boyle}, {Strauss}, {Myers}, {Nichol}, {Pimbblet}, {Ross}, {Schneider},
  {Sharp}, \& {Wake}}]{2009MNRAS.399.1755C}
{Croom}, S.~M., {Richards}, G.~T., {Shanks}, T., {et~al.} 2009{\natexlab{b}},
  \mnras, 399, 1755

\bibitem[{{Croom} {et~al.}(2001){Croom}, {Smith}, {Boyle}, {Shanks}, {Loaring},
  {Miller}, \& {Lewis}}]{2001MNRAS.322L..29C}
{Croom}, S.~M., {Smith}, R.~J., {Boyle}, B.~J., {et~al.} 2001, \mnras, 322, L29

\bibitem[{{Croom} {et~al.}(2004){Croom}, {Smith}, {Boyle}, {Shanks}, {Miller},
  {Outram}, \& {Loaring}}]{2004MNRAS.349.1397C}
{Croom}, S.~M., {Smith}, R.~J., {Boyle}, B.~J., {et~al.} 2004, \mnras, 349,
  1397

\bibitem[{{Cutri}(2013)}]{2013yCat.2328....0C}
{Cutri}, R.~M. 2013, VizieR Online Data Catalog, 2328

\bibitem[{{Czerny} {et~al.}(2004){Czerny}, {Li}, {Loska}, \&
  {Szczerba}}]{2004MNRAS.348L..54C}
{Czerny}, B., {Li}, J., {Loska}, Z., \& {Szczerba}, R. 2004, \mnras, 348, L54

\bibitem[{{Dawson} {et~al.}(2016){Dawson}, {Kneib}, {Percival}, {Alam},
  {Albareti}, {Anderson}, {Armengaud}, {Aubourg}, {Bailey}, {Bautista},
  {Berlind}, {Bershady}, {Beutler}, {Bizyaev}, {Blanton}, {Blomqvist},
  {Bolton}, {Bovy}, {Brandt}, {Brinkmann}, {Brownstein}, {Burtin}, {Busca},
  {Cai}, {Chuang}, {Clerc}, {Comparat}, {Cope}, {Croft}, {Cruz-Gonzalez}, {da
  Costa}, {Cousinou}, {Darling}, {de la Macorra}, {de la Torre}, {Delubac}, {du
  Mas des Bourboux}, {Dwelly}, {Ealet}, {Eisenstein}, {Eracleous}, {Escoffier},
  {Fan}, {Finoguenov}, {Font-Ribera}, {Frinchaboy}, {Gaulme}, {Georgakakis},
  {Green}, {Guo}, {Guy}, {Ho}, {Holder}, {Huehnerhoff}, {Hutchinson}, {Jing},
  {Jullo}, {Kamble}, {Kinemuchi}, {Kirkby}, {Kitaura}, {Klaene}, {Laher},
  {Lang}, {Laurent}, {Le Goff}, {Li}, {Liang}, {Lima}, {Lin}, {Lin}, {Lin},
  {Long}, {Lundgren}, {MacDonald}, {Geimba Maia}, {Malanushenko},
  {Malanushenko}, {Mariappan}, {McBride}, {McGreer}, {M{\'e}nard}, {Merloni},
  {Meza}, {Montero-Dorta}, {Muna}, {Myers}, {Nandra}, {Naugle}, {Newman},
  {Noterdaeme}, {Nugent}, {Ogando}, {Olmstead}, {Oravetz}, {Oravetz},
  {Padmanabhan}, {Palanque-Delabrouille}, {Pan}, {Parejko}, {P{\^a}ris},
  {Peacock}, {Petitjean}, {Pieri}, {Pisani}, {Prada}, {Prakash}, {Raichoor},
  {Reid}, {Rich}, {Ridl}, {Rodriguez-Torres}, {Carnero Rosell}, {Ross},
  {Rossi}, {Ruan}, {Salvato}, {Sayres}, {Schneider}, {Schlegel}, {Seljak},
  {Seo}, {Sesar}, {Shandera}, {Shu}, {Slosar}, {Sobreira}, {Streblyanska},
  {Suzuki}, {Taylor}, {Tao}, {Tinker}, {Tojeiro}, {Vargas-Maga{\~n}a}, {Wang},
  {Weaver}, {Weinberg}, {White}, {Wood-Vasey}, {Yeche}, {Zhai}, {Zhao}, {Zhao},
  {Zheng}, {Ben Zhu}, \& {Zou}}]{2016AJ....151...44D}
{Dawson}, K.~S., {Kneib}, J.-P., {Percival}, W.~J., {et~al.} 2016, \aj, 151, 44

\bibitem[{{de Vries} {et~al.}(2006){de Vries}, {Becker}, \&
  {White}}]{2006AJ....131..666D}
{de Vries}, W.~H., {Becker}, R.~H., \& {White}, R.~L. 2006, \aj, 131, 666

\bibitem[{{de Vries} {et~al.}(2002){de Vries}, {Morganti}, {R{\"o}ttgering},
  {Vermeulen}, {van Breugel}, {Rengelink}, \& {Jarvis}}]{2002AJ....123.1784D}
{de Vries}, W.~H., {Morganti}, R., {R{\"o}ttgering}, H.~J.~A., {et~al.} 2002,
  \aj, 123, 1784

\bibitem[{{DESI Collaboration} {et~al.}(2016){DESI Collaboration}, {Aghamousa},
  {Aguilar}, {Ahlen}, {Alam}, {Allen}, {Allende Prieto}, {Annis}, {Bailey},
  {Balland}, \& et~al.}]{2016arXiv161100036D}
{DESI Collaboration}, {Aghamousa}, A., {Aguilar}, J., {et~al.} 2016, ArXiv
  e-prints [\eprint[arXiv]{1611.00036}]

\bibitem[{{DiPompeo} {et~al.}(2011){DiPompeo}, {Brotherton}, {De Breuck}, \&
  {Laurent-Muehleisen}}]{2011ApJ...743...71D}
{DiPompeo}, M.~A., {Brotherton}, M.~S., {De Breuck}, C., \&
  {Laurent-Muehleisen}, S. 2011, \apj, 743, 71

\bibitem[{{Donley} {et~al.}(2012){Donley}, {Koekemoer}, {Brusa}, {Capak},
  {Cardamone}, {Civano}, {Ilbert}, {Impey}, {Kartaltepe}, {Miyaji}, {Salvato},
  {Sanders}, {Trump}, \& {Zamorani}}]{2012ApJ...748..142D}
{Donley}, J.~L., {Koekemoer}, A.~M., {Brusa}, M., {et~al.} 2012, \apj, 748, 142

\bibitem[{{Fan}(1999)}]{1999AJ....117.2528F}
{Fan}, X. 1999, \aj, 117, 2528

\bibitem[{{Fan} {et~al.}(2001){Fan}, {Narayanan}, {Lupton}, {Strauss}, {Knapp},
  {Becker}, {White}, {Pentericci}, {Leggett}, {Haiman}, {Gunn}, {Ivezi{\'c}},
  {Schneider}, {Anderson}, {Brinkmann}, {Bahcall}, {Connolly}, {Csabai}, {Doi},
  {Fukugita}, {Geballe}, {Grebel}, {Harbeck}, {Hennessy}, {Lamb}, {Miknaitis},
  {Munn}, {Nichol}, {Okamura}, {Pier}, {Prada}, {Richards}, {Szalay}, \&
  {York}}]{2001AJ....122.2833F}
{Fan}, X., {Narayanan}, V.~K., {Lupton}, R.~H., {et~al.} 2001, \aj, 122, 2833

\bibitem[{{Fanaroff} \& {Riley}(1974)}]{1974MNRAS.167P..31F}
{Fanaroff}, B.~L. \& {Riley}, J.~M. 1974, \mnras, 167, 31P

\bibitem[{{Fanti} {et~al.}(1995){Fanti}, {Fanti}, {Dallacasa}, {Schilizzi},
  {Spencer}, \& {Stanghellini}}]{1995A&A...302..317F}
{Fanti}, C., {Fanti}, R., {Dallacasa}, D., {et~al.} 1995, \aap, 302, 317

\bibitem[{{Fanti} {et~al.}(1989){Fanti}, {Fanti}, {Parma}, {Venturi},
  {Schilizzi}, {Nan Rendong}, {Spencer}, {Muxlow}, \& {van
  Breugel}}]{1989A&A...217...44F}
{Fanti}, C., {Fanti}, R., {Parma}, P., {et~al.} 1989, \aap, 217, 44

\bibitem[{{Fanti} {et~al.}(1986){Fanti}, {Fanti}, {Schilizzi}, {Spencer}, \&
  {van Breugel}}]{1986A&A...170...10F}
{Fanti}, C., {Fanti}, R., {Schilizzi}, R.~T., {Spencer}, R.~E., \& {van
  Breugel}, W.~J.~M. 1986, \aap, 170, 10

\bibitem[{{Fanti} {et~al.}(1990){Fanti}, {Fanti}, {Schilizzi}, {Spencer}, {Nan
  Rendong}, {Parma}, {van Breugel}, \& {Venturi}}]{1990A&A...231..333F}
{Fanti}, R., {Fanti}, C., {Schilizzi}, R.~T., {et~al.} 1990, \aap, 231, 333

\bibitem[{{Flaugher}(2005)}]{2005IJMPA..20.3121F}
{Flaugher}, B. 2005, International Journal of Modern Physics A, 20, 3121

\bibitem[{{Flesch}(2015)}]{2015PASA...32...10F}
{Flesch}, E.~W. 2015, \pasa, 32, e010

\bibitem[{{Fontanot} {et~al.}(2007){Fontanot}, {Cristiani}, {Monaco}, {Nonino},
  {Vanzella}, {Brandt}, {Grazian}, \& {Mao}}]{2007AA...461...39F}
{Fontanot}, F., {Cristiani}, S., {Monaco}, P., {et~al.} 2007, \aap, 461, 39

\bibitem[{{Fukugita} {et~al.}(1996){Fukugita}, {Ichikawa}, {Gunn}, {Doi},
  {Shimasaku}, \& {Schneider}}]{1996AJ....111.1748F}
{Fukugita}, M., {Ichikawa}, T., {Gunn}, J.~E., {et~al.} 1996, \aj, 111, 1748

\bibitem[{{Gao} {et~al.}(2008){Gao}, {Zhang}, \& {Zhao}}]{2008MNRAS.386.1417G}
{Gao}, D., {Zhang}, Y.-X., \& {Zhao}, Y.-H. 2008, \mnras, 386, 1417

\bibitem[{{Gavignaud} {et~al.}(2006){Gavignaud}, {Bongiorno}, {Paltani},
  {Mathez}, {Zamorani}, {M{\o}ller}, {Picat}, {Le Brun}, {Marano}, \& {Le
  F{\`e}vre}}]{2006A&A...457...79G}
{Gavignaud}, I., {Bongiorno}, A., {Paltani}, S., {et~al.} 2006, \aap, 457, 79

\bibitem[{{Gehrels}(1986)}]{1986ApJ...303..336G}
{Gehrels}, N. 1986, \apj, 303, 336

\bibitem[{{Georgakakis} {et~al.}(2015){Georgakakis}, {Aird}, {Buchner},
  {Salvato}, {Menzel}, {Brandt}, {McGreer}, {Dwelly}, {Mountrichas}, {Koki},
  {Georgantopoulos}, {Hsu}, {Merloni}, {Liu}, {Nandra}, \&
  {Ross}}]{2015MNRAS.453.1946G}
{Georgakakis}, A., {Aird}, J., {Buchner}, J., {et~al.} 2015, \mnras, 453, 1946

\bibitem[{{Giallongo} {et~al.}(2015){Giallongo}, {Grazian}, {Fiore}, {Fontana},
  {Pentericci}, {Vanzella}, {Dickinson}, {Kocevski}, {Castellano}, {Cristiani},
  {Ferguson}, {Finkelstein}, {Grogin}, {Hathi}, {Koekemoer}, {Newman}, \&
  {Salvato}}]{2015AAA...578A..83G}
{Giallongo}, E., {Grazian}, A., {Fiore}, F., {et~al.} 2015, \aap, 578, A83

\bibitem[{{Giallongo} {et~al.}(2019){Giallongo}, {Grazian}, {Fiore}, {Kodra},
  {Urrutia}, {Castellano}, {Cristiani}, {Dickinson}, {Fontana}, {Menci},
  {Pentericci}, {Boutsia}, {Newman}, \& {Puccetti}}]{2019ApJ...884...19G}
{Giallongo}, E., {Grazian}, A., {Fiore}, F., {et~al.} 2019, \apj, 884, 19

\bibitem[{{Glikman} {et~al.}(2011){Glikman}, {Djorgovski}, {Stern}, {Dey},
  {Jannuzi}, \& {Lee}}]{2011ApJ...728L..26G}
{Glikman}, E., {Djorgovski}, S.~G., {Stern}, D., {et~al.} 2011, \apjl, 728, L26

\bibitem[{{Goldschmidt} {et~al.}(1999){Goldschmidt}, {Kukula}, {Miller}, \&
  {Dunlop}}]{1999ApJ...511..612G}
{Goldschmidt}, P., {Kukula}, M.~J., {Miller}, L., \& {Dunlop}, J.~S. 1999,
  \apj, 511, 612

\bibitem[{{Gunn} {et~al.}(2006){Gunn}, {Siegmund}, {Mannery}, {Owen}, {Hull},
  {Leger}, {Carey}, {Knapp}, {York}, {Boroski}, {Kent}, {Lupton}, {Rockosi},
  {Evans}, {Waddell}, {Anderson}, {Annis}, {Barentine}, {Bartoszek}, {Bastian},
  {Bracker}, {Brewington}, {Briegel}, {Brinkmann}, {Brown}, {Carr},
  {Czarapata}, {Drennan}, {Dombeck}, {Federwitz}, {Gillespie}, {Gonzales},
  {Hansen}, {Harvanek}, {Hayes}, {Jordan}, {Kinney}, {Klaene}, {Kleinman},
  {Kron}, {Kresinski}, {Lee}, {Limmongkol}, {Lindenmeyer}, {Long}, {Loomis},
  {McGehee}, {Mantsch}, {Neilsen}, {Neswold}, {Newman}, {Nitta}, {Peoples},
  {Pier}, {Prieto}, {Prosapio}, {Rivetta}, {Schneider}, {Snedden}, \&
  {Wang}}]{2006AJ....131.2332G}
{Gunn}, J.~E., {Siegmund}, W.~A., {Mannery}, E.~J., {et~al.} 2006, \aj, 131,
  2332

\bibitem[{{Gurkan}(2019)}]{2019Anda}
{Gurkan}, G. e.~a. 2019, \aap

\bibitem[{{Haardt} \& {Madau}(2012)}]{2012ApJ...746..125H}
{Haardt}, F. \& {Madau}, P. 2012, \apj, 746, 125

\bibitem[{H\"{a}rdle(1990)}]{hardle_1990}
H\"{a}rdle, W. 1990, Applied Nonparametric Regression, Econometric Society
  Monographs (Cambridge University Press)

\bibitem[{{Hasinger} {et~al.}(2005){Hasinger}, {Miyaji}, \&
  {Schmidt}}]{2005A&A...441..417H}
{Hasinger}, G., {Miyaji}, T., \& {Schmidt}, M. 2005, \aap, 441, 417

\bibitem[{{Hatziminaoglou} {et~al.}(2000){Hatziminaoglou}, {Mathez}, \&
  {Pell{\'o}}}]{2000A&A...359....9H}
{Hatziminaoglou}, E., {Mathez}, G., \& {Pell{\'o}}, R. 2000, \aap, 359, 9

\bibitem[{{Herrera Ruiz} {et~al.}(2016){Herrera Ruiz}, {Middelberg}, {Norris},
  \& {Maini}}]{2016A&A...589L...2H}
{Herrera Ruiz}, N., {Middelberg}, E., {Norris}, R.~P., \& {Maini}, A. 2016,
  \aap, 589, L2

\bibitem[{{Hertzsprung}(1909)}]{1909AN....179..373H}
{Hertzsprung}, E. 1909, Astronomische Nachrichten, 179, 373

\bibitem[{{Hodapp} {et~al.}(2004){Hodapp}, {Siegmund}, {Kaiser}, {Chambers},
  {Laux}, {Morgan}, \& {Mannery}}]{2004SPIE.5489..667H}
{Hodapp}, K.~W., {Siegmund}, W.~A., {Kaiser}, N., {et~al.} 2004, in Society of
  Photo-Optical Instrumentation Engineers (SPIE) Conference Series, Vol. 5489,
  Ground-based Telescopes, ed. J.~{Oschmann}, Jacobus~M., 667--678

\bibitem[{{Hook} {et~al.}(2002){Hook}, {McMahon}, {Shaver}, \&
  {Snellen}}]{2002A&A...391..509H}
{Hook}, I.~M., {McMahon}, R.~G., {Shaver}, P.~A., \& {Snellen}, I.~A.~G. 2002,
  \aap, 391, 509

\bibitem[{{Hook} {et~al.}(1998){Hook}, {Shaver}, \&
  {McMahon}}]{1998ASPC..146...17H}
{Hook}, I.~M., {Shaver}, P.~A., \& {McMahon}, R.~G. 1998, in Astronomical
  Society of the Pacific Conference Series, Vol. 146, The Young Universe:
  Galaxy Formation and Evolution at Intermediate and High Redshift, ed.
  S.~{D'Odorico}, A.~{Fontana}, \& E.~{Giallongo}, 17

\bibitem[{{Ilbert} {et~al.}(2006){Ilbert}, {Arnouts}, {McCracken},
  {Bolzonella}, {Bertin}, {Le F{\`e}vre}, {Mellier}, {Zamorani}, {Pell{\`o}},
  {Iovino}, {Tresse}, {Le Brun}, {Bottini}, {Garilli}, {Maccagni}, {Picat},
  {Scaramella}, {Scodeggio}, {Vettolani}, {Zanichelli}, {Adami}, {Bardelli},
  {Cappi}, {Charlot}, {Ciliegi}, {Contini}, {Cucciati}, {Foucaud}, {Franzetti},
  {Gavignaud}, {Guzzo}, {Marano}, {Marinoni}, {Mazure}, {Meneux}, {Merighi},
  {Paltani}, {Pollo}, {Pozzetti}, {Radovich}, {Zucca}, {Bondi}, {Bongiorno},
  {Busarello}, {de La Torre}, {Gregorini}, {Lamareille}, {Mathez}, {Merluzzi},
  {Ripepi}, {Rizzo}, \& {Vergani}}]{2006A&A...457..841I}
{Ilbert}, O., {Arnouts}, S., {McCracken}, H.~J., {et~al.} 2006, \aap, 457, 841

\bibitem[{{Ivezi{\'c}} {et~al.}(2014){Ivezi{\'c}}, {Brandt}, {Fan}, {MacLeod},
  {Richards}, \& {Yoachim}}]{2014IAUS..304...11I}
{Ivezi{\'c}}, {\v{Z}}., {Brandt}, W.~N., {Fan}, X., {et~al.} 2014, in IAU
  Symposium, Vol. 304, Multiwavelength AGN Surveys and Studies, ed. A.~M.
  {Mickaelian} \& D.~B. {Sanders}, 11--17

\bibitem[{{Ivezi{\'c}} {et~al.}(2019){Ivezi{\'c}}, {Kahn}, {Tyson}, {Abel},
  {Acosta}, {Allsman}, {Alonso}, {AlSayyad}, {Anderson}, {Andrew}, \&
  et~al.}]{2019ApJ...873..111I}
{Ivezi{\'c}}, {\v Z}., {Kahn}, S.~M., {Tyson}, J.~A., {et~al.} 2019, \apj, 873,
  111

\bibitem[{{Ivezi{\'c}} {et~al.}(2002){Ivezi{\'c}}, {Menou}, {Knapp}, {Strauss},
  {Lupton}, {Vanden Berk}, {Richards}, {Tremonti}, {Weinstein}, {Anderson},
  {Bahcall}, {Becker}, {Bernardi}, {Blanton}, {Eisenstein}, {Fan},
  {Finkbeiner}, {Finlator}, {Frieman}, {Gunn}, {Hall}, {Kim}, {Kinkhabwala},
  {Narayanan}, {Rockosi}, {Schlegel}, {Schneider}, {Strateva}, {SubbaRao},
  {Thakar}, {Voges}, {White}, {Yanny}, {Brinkmann}, {Doi}, {Fukugita},
  {Hennessy}, {Munn}, {Nichol}, \& {York}}]{2002AJ....124.2364I}
{Ivezi{\'c}}, {\v Z}., {Menou}, K., {Knapp}, G.~R., {et~al.} 2002, \aj, 124,
  2364

\bibitem[{{Jannuzi} {et~al.}(2010){Jannuzi}, {Weiner}, {Block}, {Borys},
  {Eisenstein}, {Kochanek}, {Rieke}, {Rieke}, {Armus}, {Brodwin}, {Brown},
  {Cool}, {Desai}, {Dey}, {Dickinson}, {Dole}, {Herrera}, {Le Floc'h},
  {Morrison}, {Papovich}, {P{\'e}rez-Gonz{\'a}lez}, {Stern}, {Rujopakarn}, \&
  {Zehavi}}]{2010AAS...21547001J}
{Jannuzi}, B., {Weiner}, B., {Block}, M., {et~al.} 2010, in Bulletin of the
  American Astronomical Society, Vol.~42, American Astronomical Society Meeting
  Abstracts \#215, 513

\bibitem[{{Jannuzi} \& {Dey}(1999)}]{1999ASPC..191..111J}
{Jannuzi}, B.~T. \& {Dey}, A. 1999, in Astronomical Society of the Pacific
  Conference Series, Vol. 191, Photometric Redshifts and the Detection of High
  Redshift Galaxies, ed. R.~{Weymann}, L.~{Storrie-Lombardi}, M.~{Sawicki}, \&
  R.~{Brunner}, 111

\bibitem[{{Jiang} {et~al.}(2007){Jiang}, {Fan}, {Ivezi{\'c}}, {Richards},
  {Schneider}, {Strauss}, \& {Kelly}}]{2007ApJ...656..680J}
{Jiang}, L., {Fan}, X., {Ivezi{\'c}}, {\v Z}., {et~al.} 2007, \apj, 656, 680

\bibitem[{{Jiang} {et~al.}(2016){Jiang}, {McGreer}, {Fan}, {Strauss},
  {Ba{\~n}ados}, {Becker}, {Bian}, {Farnsworth}, {Shen}, {Wang}, {Wang},
  {Wang}, {White}, {Wu}, {Wu}, {Yang}, \& {Yang}}]{2016ApJ...833..222J}
{Jiang}, L., {McGreer}, I.~D., {Fan}, X., {et~al.} 2016, \apj, 833, 222

\bibitem[{{Jin} {et~al.}(2019){Jin}, {Zhang}, {Zhang}, {Zhao}, {Wu}, \&
  {Fan}}]{2019MNRAS.485.4539J}
{Jin}, X., {Zhang}, Y., {Zhang}, J., {et~al.} 2019, \mnras, 485, 4539

\bibitem[{{Kashikawa} {et~al.}(2015){Kashikawa}, {Ishizaki}, {Willott},
  {Onoue}, {Im}, {Furusawa}, {Toshikawa}, {Ishikawa}, {Niino}, {Shimasaku},
  {Ouchi}, \& {Hibon}}]{2015ApJ...798...28K}
{Kashikawa}, N., {Ishizaki}, Y., {Willott}, C.~J., {et~al.} 2015, \apj, 798, 28

\bibitem[{{Kellermann} {et~al.}(1989){Kellermann}, {Sramek}, {Schmidt},
  {Shaffer}, \& {Green}}]{1989AJ.....98.1195K}
{Kellermann}, K.~I., {Sramek}, R., {Schmidt}, M., {Shaffer}, D.~B., \& {Green},
  R. 1989, \aj, 98, 1195

\bibitem[{{Kennefick} {et~al.}(1995){Kennefick}, {Djorgovski}, \& {de
  Carvalho}}]{1995AJ....110.2553K}
{Kennefick}, J.~D., {Djorgovski}, S.~G., \& {de Carvalho}, R.~R. 1995, \aj,
  110, 2553

\bibitem[{{Kenter} {et~al.}(2005){Kenter}, {Murray}, {Forman}, {Jones},
  {Green}, {Kochanek}, {Vikhlinin}, {Fabricant}, {Fazio}, {Brand}, {Brown},
  {Dey}, {Jannuzi}, {Najita}, {McNamara}, {Shields}, \&
  {Rieke}}]{2005ApJS..161....9K}
{Kenter}, A., {Murray}, S.~S., {Forman}, W.~R., {et~al.} 2005, \apjs, 161, 9

\bibitem[{{Kimball} {et~al.}(2011){Kimball}, {Kellermann}, {Condon},
  {Ivezi{\'c}}, \& {Perley}}]{2011ApJ...739L..29K}
{Kimball}, A.~E., {Kellermann}, K.~I., {Condon}, J.~J., {Ivezi{\'c}}, {\v Z}.,
  \& {Perley}, R.~A. 2011, \apjl, 739, L29

\bibitem[{{Kimball} {et~al.}(2009){Kimball}, {Knapp}, {Ivezi{\'c}}, {West},
  {Bochanski}, {Plotkin}, \& {Gordon}}]{2009ApJ...701..535K}
{Kimball}, A.~E., {Knapp}, G.~R., {Ivezi{\'c}}, {\v Z}., {et~al.} 2009, \apj,
  701, 535

\bibitem[{{Kirkpatrick} {et~al.}(2011){Kirkpatrick}, {Schlegel}, {Ross},
  {Myers}, {Hennawi}, {Sheldon}, {Schneider}, \&
  {Weaver}}]{2011ApJ...743..125K}
{Kirkpatrick}, J.~A., {Schlegel}, D.~J., {Ross}, N.~P., {et~al.} 2011, \apj,
  743, 125

\bibitem[{{Kochanek} {et~al.}(2012){Kochanek}, {Eisenstein}, {Cool},
  {Caldwell}, {Assef}, {Jannuzi}, {Jones}, {Murray}, {Forman}, {Dey}, {Brown},
  {Eisenhardt}, {Gonzalez}, {Green}, \& {Stern}}]{2012ApJS..200....8K}
{Kochanek}, C.~S., {Eisenstein}, D.~J., {Cool}, R.~J., {et~al.} 2012, \apjs,
  200, 8

\bibitem[{{Kratzer} \& {Richards}(2015)}]{2015AJ....149...61K}
{Kratzer}, R.~M. \& {Richards}, G.~T. 2015, \aj, 149, 61

\bibitem[{{Kunert-Bajraszewska} {et~al.}(2015){Kunert-Bajraszewska},
  {Ceg{\l}owski}, {Katarzy{\'n}ski}, \&
  {Roskowi{\'n}ski}}]{2015A&A...579A.109K}
{Kunert-Bajraszewska}, M., {Ceg{\l}owski}, M., {Katarzy{\'n}ski}, K., \&
  {Roskowi{\'n}ski}, C. 2015, \aap, 579, A109

\bibitem[{{La Franca} {et~al.}(1994){La Franca}, {Gregorini}, {Cristiani}, {de
  Ruiter}, \& {Owen}}]{1994AJ....108.1548L}
{La Franca}, F., {Gregorini}, L., {Cristiani}, S., {de Ruiter}, H., \& {Owen},
  F. 1994, \aj, 108, 1548

\bibitem[{{Lacy} {et~al.}(2007){Lacy}, {Petric}, {Sajina}, {Canalizo},
  {Storrie-Lombardi}, {Armus}, {Fadda}, \& {Marleau}}]{2007AJ....133..186L}
{Lacy}, M., {Petric}, A.~O., {Sajina}, A., {et~al.} 2007, \aj, 133, 186

\bibitem[{{Lacy} {et~al.}(2005){Lacy}, {Wilson}, {Masci}, {Storrie-Lombardi},
  {Appleton}, {Armus}, {Chapman}, {Choi}, {Fadda}, {Fang}, {Frayer},
  {Heinrichsen}, {Helou}, {Im}, {Laine}, {Marleau}, {Shupe}, {Soifer},
  {Squires}, {Surace}, {Teplitz}, \& {Yan}}]{2005ApJS..161...41L}
{Lacy}, M., {Wilson}, G., {Masci}, F., {et~al.} 2005, \apjs, 161, 41

\bibitem[{{Laureijs} {et~al.}(2011){Laureijs}, {Amiaux}, {Arduini},
  {Augu{\`e}res}, {Brinchmann}, {Cole}, {Cropper}, {Dabin}, {Duvet}, {Ealet},
  \& et~al.}]{2011arXiv1110.3193L}
{Laureijs}, R., {Amiaux}, J., {Arduini}, S., {et~al.} 2011, arXiv e-prints
  [\eprint[arXiv]{1110.3193}]

\bibitem[{{Le F{\`e}vre} {et~al.}(2013){Le F{\`e}vre}, {Cassata}, {Cucciati},
  {Garilli}, {Ilbert}, {Le Brun}, {Maccagni}, {Moreau}, {Scodeggio}, \&
  {Tresse}}]{2013A&A...559A..14L}
{Le F{\`e}vre}, O., {Cassata}, P., {Cucciati}, O., {et~al.} 2013, \aap, 559,
  A14

\bibitem[{Li \& Racine(2011)}]{li2011nonparametric}
Li, Q. \& Racine, J. 2011, Nonparametric Econometrics: Theory and Practice
  (Princeton University Press)

\bibitem[{{Liu} {et~al.}(2008){Liu}, {Jiang}, {Wang}, \&
  {Xie}}]{2008MNRAS.391..246L}
{Liu}, Y., {Jiang}, D.~R., {Wang}, T.~G., \& {Xie}, F.~G. 2008, \mnras, 391,
  246

\bibitem[{{Lonsdale} {et~al.}(2003){Lonsdale}, {Smith}, {Rowan-Robinson},
  {Surace}, {Shupe}, {Xu}, {Oliver}, {Padgett}, {Fang}, {Conrow},
  {Franceschini}, {Gautier}, {Griffin}, {Hacking}, {Masci}, {Morrison},
  {O'Linger}, {Owen}, {P{\'e}rez-Fournon}, {Pierre}, {Puetter}, {Stacey},
  {Castro}, {Polletta}, {Farrah}, {Jarrett}, {Frayer}, {Siana}, {Babbedge},
  {Dye}, {Fox}, {Gonzalez-Solares}, {Salaman}, {Berta}, {Condon}, {Dole}, \&
  {Serjeant}}]{2003PASP..115..897L}
{Lonsdale}, C.~J., {Smith}, H.~E., {Rowan-Robinson}, M., {et~al.} 2003, \pasp,
  115, 897

\bibitem[{{LSST Science Collaboration} {et~al.}(2009){LSST Science
  Collaboration}, {Abell}, {Allison}, {Anderson}, {Andrew}, {Angel}, {Armus},
  {Arnett}, {Asztalos}, {Axelrod}, \& et~al.}]{2009arXiv0912.0201L}
{LSST Science Collaboration}, {Abell}, P.~A., {Allison}, J., {et~al.} 2009,
  ArXiv e-prints [\eprint[arXiv]{0912.0201}]

\bibitem[{{Lu} {et~al.}(2007){Lu}, {Wang}, {Zhou}, \&
  {Wu}}]{2007AJ....133.1615L}
{Lu}, Y., {Wang}, T., {Zhou}, H., \& {Wu}, J. 2007, \aj, 133, 1615

\bibitem[{{Lupton} {et~al.}(1999){Lupton}, {Gunn}, \&
  {Szalay}}]{1999AJ....118.1406L}
{Lupton}, R.~H., {Gunn}, J.~E., \& {Szalay}, A.~S. 1999, \aj, 118, 1406

\bibitem[{{Maddox} {et~al.}(2012){Maddox}, {Hewett}, {P{\'e}roux}, {Nestor}, \&
  {Wisotzki}}]{2012MNRAS.424.2876M}
{Maddox}, N., {Hewett}, P.~C., {P{\'e}roux}, C., {Nestor}, D.~B., \&
  {Wisotzki}, L. 2012, \mnras, 424, 2876

\bibitem[{{Magnier} {et~al.}(2016){Magnier}, {Sweeney}, {Chambers},
  {Flewelling}, {Huber}, {Price}, {Waters}, {Denneau}, {Draper}, {Jedicke},
  {Hodapp}, {Kaiser}, {Kudritzki}, {Metcalfe}, {Stubbs}, \&
  {Wainscoast}}]{2016arXiv161205244M}
{Magnier}, E.~A., {Sweeney}, W.~E., {Chambers}, K.~C., {et~al.} 2016, arXiv
  e-prints, arXiv:1612.05244

\bibitem[{{Martin} {et~al.}(2005){Martin}, {Fanson}, {Schiminovich},
  {Morrissey}, {Friedman}, {Barlow}, {Conrow}, {Grange}, {Jelinsky},
  {Milliard}, {Siegmund}, {Bianchi}, {Byun}, {Donas}, {Forster}, {Heckman},
  {Lee}, {Madore}, {Malina}, {Neff}, {Rich}, {Small}, {Surber}, {Szalay},
  {Welsh}, \& {Wyder}}]{2005ApJ...619L...1M}
{Martin}, D.~C., {Fanson}, J., {Schiminovich}, D., {et~al.} 2005, \apjl, 619,
  L1

\bibitem[{{Masters} {et~al.}(2012){Masters}, {Capak}, {Salvato}, {Civano},
  {Mobasher}, {Siana}, {Hasinger}, {Impey}, {Nagao}, {Trump}, {Ikeda}, {Elvis},
  \& {Scoville}}]{2012ApJ...755..169M}
{Masters}, D., {Capak}, P., {Salvato}, M., {et~al.} 2012, \apj, 755, 169

\bibitem[{{Masters} {et~al.}(2015){Masters}, {Capak}, {Stern}, {Ilbert},
  {Salvato}, {Schmidt}, {Longo}, {Rhodes}, {Paltani}, {Mobasher}, {Hoekstra},
  {Hildebrandt}, {Coupon}, {Steinhardt}, {Speagle}, {Faisst}, {Kalinich},
  {Brodwin}, {Brescia}, \& {Cavuoti}}]{2015ApJ...813...53M}
{Masters}, D., {Capak}, P., {Stern}, D., {et~al.} 2015, \apj, 813, 53

\bibitem[{{Mauduit} {et~al.}(2012){Mauduit}, {Lacy}, {Farrah}, {Surace},
  {Jarvis}, {Oliver}, {Maraston}, {Vaccari}, {Marchetti}, {Zeimann},
  {Gonz{\'a}les-Solares}, {Pforr}, {Petric}, {Henriques}, {Thomas}, {Afonso},
  {Rettura}, {Wilson}, {Falder}, {Geach}, {Huynh}, {Norris}, {Seymour},
  {Richards}, {Stanford}, {Alexand er}, {Becker}, {Best}, {Bizzocchi},
  {Bonfield}, {Castro}, {Cava}, {Chapman}, {Christopher}, {Clements}, {Covone},
  {Dubois}, {Dunlop}, {Dyke}, {Edge}, {Ferguson}, {Foucaud}, {Franceschini},
  {Gal}, {Grant}, {Grossi}, {Hatziminaoglou}, {Hickey}, {Hodge}, {Huang},
  {Ivison}, {Kim}, {LeFevre}, {Lehnert}, {Lonsdale}, {Lubin}, {McLure},
  {Messias}, {Mart{\'\i}nez-Sansigre}, {Mortier}, {Nielsen}, {Ouchi}, {Parish},
  {Perez-Fournon}, {Pierre}, {Rawlings}, {Readhead}, {Ridgway}, {Rigopoulou},
  {Romer}, {Rosebloom}, {Rottgering}, {Rowan-Robinson}, {Sajina}, {Simpson},
  {Smail}, {Squires}, {Stevens}, {Taylor}, {Trichas}, {Urrutia}, {van Kampen},
  {Verma}, \& {Xu}}]{2012PASP..124.1135M}
{Mauduit}, J.~C., {Lacy}, M., {Farrah}, D., {et~al.} 2012, \pasp, 124, 1135

\bibitem[{{McGreer} {et~al.}(2006){McGreer}, {Becker}, {Helfand}, \&
  {White}}]{2006ApJ...652..157M}
{McGreer}, I.~D., {Becker}, R.~H., {Helfand}, D.~J., \& {White}, R.~L. 2006,
  \apj, 652, 157

\bibitem[{{McGreer} {et~al.}(2018){McGreer}, {Fan}, {Jiang}, \&
  {Cai}}]{2018AJ....155..131M}
{McGreer}, I.~D., {Fan}, X., {Jiang}, L., \& {Cai}, Z. 2018, \aj, 155, 131

\bibitem[{{McGreer} {et~al.}(2009){McGreer}, {Helfand}, \&
  {White}}]{2009AJ....138.1925M}
{McGreer}, I.~D., {Helfand}, D.~J., \& {White}, R.~L. 2009, \aj, 138, 1925

\bibitem[{{McGreer} {et~al.}(2013){McGreer}, {Jiang}, {Fan}, {Richards},
  {Strauss}, {Ross}, {White}, {Shen}, {Schneider}, {Myers}, {Brandt}, {DeGraf},
  {Glikman}, {Ge}, \& {Streblyanska}}]{2013ApJ...768..105M}
{McGreer}, I.~D., {Jiang}, L., {Fan}, X., {et~al.} 2013, \apj, 768, 105

\bibitem[{{Meisner} {et~al.}(2017){Meisner}, {Lang}, \&
  {Schlegel}}]{2017AJ....154..161M}
{Meisner}, A.~M., {Lang}, D., \& {Schlegel}, D.~J. 2017, \aj, 154, 161

\bibitem[{{Messias} {et~al.}(2012){Messias}, {Afonso}, {Salvato}, {Mobasher},
  \& {Hopkins}}]{2012ApJ...754..120M}
{Messias}, H., {Afonso}, J., {Salvato}, M., {Mobasher}, B., \& {Hopkins}, A.~M.
  2012, \apj, 754, 120

\bibitem[{{Miller} {et~al.}(1990){Miller}, {Peacock}, \&
  {Mead}}]{1990MNRAS.244..207M}
{Miller}, L., {Peacock}, J.~A., \& {Mead}, A.~R.~G. 1990, \mnras, 244, 207

\bibitem[{{Miyaji} {et~al.}(2015){Miyaji}, {Hasinger}, {Salvato}, {Brusa},
  {Cappelluti}, {Civano}, {Puccetti}, {Elvis}, {Brunner}, {Fotopoulou}, {Ueda},
  {Griffiths}, {Koekemoer}, {Akiyama}, {Comastri}, {Gilli}, {Lanzuisi},
  {Merloni}, \& {Vignali}}]{2015ApJ...804..104M}
{Miyaji}, T., {Hasinger}, G., {Salvato}, M., {et~al.} 2015, \apj, 804, 104

\bibitem[{{M{\o}ller} \& {Jakobsen}(1990)}]{1990AA...228..299M}
{M{\o}ller}, P. \& {Jakobsen}, P. 1990, \aap, 228, 299

\bibitem[{{Morabito} {et~al.}(2016){Morabito}, {Deller}, {R{\"o}ttgering},
  {Miley}, {Varenius}, {Shimwell}, {Mold{\'o}n}, {Jackson}, {Morganti}, {van
  Weeren}, \& {Oonk}}]{2016MNRAS.461.2676M}
{Morabito}, L.~K., {Deller}, A.~T., {R{\"o}ttgering}, H., {et~al.} 2016,
  \mnras, 461, 2676

\bibitem[{{Morabito} {et~al.}(2018){Morabito}, {Matthews}, {Best},
  {G{\"u}rkan}, {Jarvis}, {Prandoni}, {Duncan}, {Hardcastle},
  {Kunert-Bajraszewska}, {Mechev}, {Mooney}, {Sabater}, {R{\"o}ttgering},
  {Shimwell}, {Smith}, {Tasse}, \& {Williams}}]{2018arXiv181107931M}
{Morabito}, L.~K., {Matthews}, J.~H., {Best}, P.~N., {et~al.} 2018, ArXiv
  e-prints [\eprint[arXiv]{1811.07931}]

\bibitem[{{Mullin} {et~al.}(2008){Mullin}, {Riley}, \&
  {Hardcastle}}]{2008MNRAS.390..595M}
{Mullin}, L.~M., {Riley}, J.~M., \& {Hardcastle}, M.~J. 2008, \mnras, 390, 595

\bibitem[{{Myers} {et~al.}(2015){Myers}, {Palanque-Delabrouille}, {Prakash},
  {P{\^a}ris}, {Yeche}, {Dawson}, {Bovy}, {Lang}, {Schlegel}, {Newman},
  {Petitjean}, {Kneib}, {Laurent}, {Percival}, {Ross}, {Seo}, {Tinker},
  {Armengaud}, {Brownstein}, {Burtin}, {Cai}, {Comparat}, {Kasliwal},
  {Kulkarni}, {Laher}, {Levitan}, {McBride}, {McGreer}, {Miller}, {Nugent},
  {Ofek}, {Rossi}, {Ruan}, {Schneider}, {Sesar}, {Streblyanska}, \&
  {Surace}}]{2015ApJS..221...27M}
{Myers}, A.~D., {Palanque-Delabrouille}, N., {Prakash}, A., {et~al.} 2015,
  \apjs, 221, 27

\bibitem[{Nadaraya(1964)}]{doi:10.1137/1109020}
Nadaraya, E. 1964, Theory of Probability \& Its Applications, 9, 141

\bibitem[{{Nakoneczny} {et~al.}(2019){Nakoneczny}, {Bilicki}, {Solarz},
  {Pollo}, {Maddox}, {Spiniello}, {Brescia}, \&
  {Napolitano}}]{2019A&A...624A..13N}
{Nakoneczny}, S., {Bilicki}, M., {Solarz}, A., {et~al.} 2019, \aap, 624, A13

\bibitem[{{Niida} {et~al.}(2016){Niida}, {Nagao}, {Ikeda}, {Matsuoka},
  {Kobayashi}, {Toba}, \& {Taniguchi}}]{2016ApJ...832..208N}
{Niida}, M., {Nagao}, T., {Ikeda}, H., {et~al.} 2016, \apj, 832, 208

\bibitem[{{Norris}(2017)}]{2017NatAs...1..671N}
{Norris}, R.~P. 2017, Nature Astronomy, 1, 671

\bibitem[{{Norris} {et~al.}(2011){Norris}, {Hopkins}, {Afonso}, {Brown},
  {Condon}, {Dunne}, {Feain}, {Hollow}, {Jarvis}, {Johnston-Hollitt}, {Lenc},
  {Middelberg}, {Padovani}, {Prandoni}, {Rudnick}, {Seymour}, {Umana},
  {Andernach}, {Alexander}, {Appleton}, {Bacon}, {Banfield}, {Becker}, {Brown},
  {Ciliegi}, {Jackson}, {Eales}, {Edge}, {Gaensler}, {Giovannini}, {Hales},
  {Hancock}, {Huynh}, {Ibar}, {Ivison}, {Kennicutt}, {Kimball}, {Koekemoer},
  {Koribalski}, {L{\'o}pez-S{\'a}nchez}, {Mao}, {Murphy}, {Messias},
  {Pimbblet}, {Raccanelli}, {Randall}, {Reiprich}, {Roseboom},
  {R{\"o}ttgering}, {Saikia}, {Sharp}, {Slee}, {Smail}, {Thompson}, {Urquhart},
  {Wall}, \& {Zhao}}]{2011PASA...28..215N}
{Norris}, R.~P., {Hopkins}, A.~M., {Afonso}, J., {et~al.} 2011, \pasa, 28, 215

\bibitem[{O'Dea(1998)}]{10.1086/316162}
O'Dea, C. 1998, Publications of the Astronomical Society of the Pacific, 110,
  493

\bibitem[{{O'Dea}(1998)}]{1998PASP..110..493O}
{O'Dea}, C.~P. 1998, \pasp, 110, 493

\bibitem[{{Oke} \& {Gunn}(1983)}]{1983ApJ...266..713O}
{Oke}, J.~B. \& {Gunn}, J.~E. 1983, \apj, 266, 713

\bibitem[{{Orienti} {et~al.}(2007){Orienti}, {Dallacasa}, \&
  {Stanghellini}}]{2007A&A...461..923O}
{Orienti}, M., {Dallacasa}, D., \& {Stanghellini}, C. 2007, \aap, 461, 923

\bibitem[{{Padovani}(1993)}]{1993MNRAS.263..461P}
{Padovani}, P. 1993, \mnras, 263, 461

\bibitem[{{Padovani} {et~al.}(2011){Padovani}, {Miller}, {Kellermann},
  {Mainieri}, {Rosati}, \& {Tozzi}}]{2011ApJ...740...20P}
{Padovani}, P., {Miller}, N., {Kellermann}, K.~I., {et~al.} 2011, \apj, 740, 20

\bibitem[{{Palanque-Delabrouille} {et~al.}(2016){Palanque-Delabrouille},
  {Magneville}, {Y{\`e}che}, {P{\^a}ris}, {Petitjean}, {Burtin}, {Dawson},
  {McGreer}, {Myers}, {Rossi}, {Schlegel}, {Schneider}, {Streblyanska}, \&
  {Tinker}}]{2016A&A...587A..41P}
{Palanque-Delabrouille}, N., {Magneville}, C., {Y{\`e}che}, C., {et~al.} 2016,
  \aap, 587, A41

\bibitem[{{Papovich} {et~al.}(2006){Papovich}, {Cool}, {Eisenstein}, {Le
  Floc'h}, {Fan}, {Kennicutt}, {Smith}, {Rieke}, \&
  {Vestergaard}}]{2006AJ....132..231P}
{Papovich}, C., {Cool}, R., {Eisenstein}, D., {et~al.} 2006, \aj, 132, 231

\bibitem[{{Papovich} {et~al.}(2016){Papovich}, {Shipley}, {Mehrtens}, {Lanham},
  {Lacy}, {Ciardullo}, {Finkelstein}, {Bassett}, {Behroozi}, {Blanc}, {de
  Jong}, {DePoy}, {Drory}, {Gawiser}, {Gebhardt}, {Gronwall}, {Hill}, {Hopp},
  {Jogee}, {Kawinwanichakij}, {Marshall}, {McLinden}, {Mentuch Cooper},
  {Somerville}, {Steinmetz}, {Tran}, {Tuttle}, {Viero}, {Wechsler}, \&
  {Zeimann}}]{2016ApJS..224...28P}
{Papovich}, C., {Shipley}, H.~V., {Mehrtens}, N., {et~al.} 2016, \apjs, 224, 28

\bibitem[{{P{\^a}ris} {et~al.}(2018){P{\^a}ris}, {Petitjean}, {Aubourg},
  {Myers}, {Streblyanska}, {Lyke}, {Anderson}, {Armengaud}, {Bautista},
  {Blanton}, {Blomqvist}, {Brinkmann}, {Brownstein}, {Brandt}, {Burtin},
  {Dawson}, {de la Torre}, {Georgakakis}, {Gil-Mar{\'{\i}}n}, {Green}, {Hall},
  {Kneib}, {LaMassa}, {Le Goff}, {MacLeod}, {Mariappan}, {McGreer}, {Merloni},
  {Noterdaeme}, {Palanque-Delabrouille}, {Percival}, {Ross}, {Rossi},
  {Schneider}, {Seo}, {Tojeiro}, {Weaver}, {Weijmans}, {Y{\`e}che}, {Zarrouk},
  \& {Zhao}}]{2018A&A...613A..51P}
{P{\^a}ris}, I., {Petitjean}, P., {Aubourg}, {\'E}., {et~al.} 2018, \aap, 613,
  A51

\bibitem[{{Pasquet} {et~al.}(2019){Pasquet}, {Bertin}, {Treyer}, {Arnouts}, \&
  {Fouchez}}]{2019A&A...621A..26P}
{Pasquet}, J., {Bertin}, E., {Treyer}, M., {Arnouts}, S., \& {Fouchez}, D.
  2019, \aap, 621, A26

\bibitem[{{Pasquet-Itam} \& {Pasquet}(2018)}]{2018A&A...611A..97P}
{Pasquet-Itam}, J. \& {Pasquet}, J. 2018, \aap, 611, A97

\bibitem[{{Peacock} {et~al.}(1986){Peacock}, {Miller}, \&
  {Longair}}]{1986MNRAS.218..265P}
{Peacock}, J.~A., {Miller}, L., \& {Longair}, M.~S. 1986, \mnras, 218, 265

\bibitem[{{Peacock} \& {Wall}(1982)}]{1982MNRAS.198..843P}
{Peacock}, J.~A. \& {Wall}, J.~V. 1982, \mnras, 198, 843

\bibitem[{{Pei}(1992)}]{1992ApJ...395..130P}
{Pei}, Y.~C. 1992, \apj, 395, 130

\bibitem[{{Pei}(1995)}]{1995ApJ...438..623P}
{Pei}, Y.~C. 1995, \apj, 438, 623

\bibitem[{{Peng} {et~al.}(2012){Peng}, {Zhang}, {Zhao}, \&
  {Wu}}]{2012MNRAS.425.2599P}
{Peng}, N., {Zhang}, Y., {Zhao}, Y., \& {Wu}, X.-b. 2012, \mnras, 425, 2599

\bibitem[{{Peters} {et~al.}(2015){Peters}, {Richards}, {Myers}, {Strauss},
  {Schmidt}, {Ivezi{\'c}}, {Ross}, {MacLeod}, \&
  {Riegel}}]{2015ApJ...811...95P}
{Peters}, C.~M., {Richards}, G.~T., {Myers}, A.~D., {et~al.} 2015, \apj, 811,
  95

\bibitem[{{Prandoni} {et~al.}(2010){Prandoni}, {de Ruiter}, {Ricci}, {Parma},
  {Gregorini}, \& {Ekers}}]{2010A&A...510A..42P}
{Prandoni}, I., {de Ruiter}, H.~R., {Ricci}, R., {et~al.} 2010, \aap, 510, A42

\bibitem[{{Pu}(2013)}]{2013Ap&SS.345..355P}
{Pu}, X. 2013, \apss, 345, 355

\bibitem[{Qiu(2013)}]{qiu2013introduction}
Qiu, P. 2013, Introduction to Statistical Process Control, Chapman \& Hall/CRC
  Texts in Statistical Science (CRC Press)

\bibitem[{Retana-Montenegro \& R\"{o}ttgering(2018)}]{10.3389/fspas.2018.00005}
Retana-Montenegro, E. \& R\"{o}ttgering, H. 2018, Frontiers in Astronomy and
  Space Sciences, 5, 5

\bibitem[{{Retana-Montenegro} \& {R{\"o}ttgering}(2017)}]{2017A&A...600A..97R}
{Retana-Montenegro}, E. \& {R{\"o}ttgering}, H.~J.~A. 2017, \aap, 600, A97

\bibitem[{{Retana-Montenegro} {et~al.}(2018){Retana-Montenegro},
  {R{\"o}ttgering}, {Shimwell}, {van Weeren}, {Prandoni}, {Brunetti}, {Best},
  \& {Br{\"u}ggen}}]{2018arXiv180704878R}
{Retana-Montenegro}, E., {R{\"o}ttgering}, H.~J.~A., {Shimwell}, T.~W.,
  {et~al.} 2018, A\&A, 620, A74

\bibitem[{{Richards} {et~al.}(2002){Richards}, {Fan}, {Newberg}, {Strauss},
  {Vanden Berk}, {Schneider}, {Yanny}, {Boucher}, {Burles}, {Frieman}, {Gunn},
  {Hall}, {Ivezi{\'c}}, {Kent}, {Loveday}, {Lupton}, {Rockosi}, {Schlegel},
  {Stoughton}, {SubbaRao}, \& {York}}]{2002AJ....123.2945R}
{Richards}, G.~T., {Fan}, X., {Newberg}, H.~J., {et~al.} 2002, \aj, 123, 2945

\bibitem[{{Richards} {et~al.}(2011){Richards}, {Kruczek}, {Gallagher}, {Hall},
  {Hewett}, {Leighly}, {Deo}, {Kratzer}, \& {Shen}}]{2011AJ....141..167R}
{Richards}, G.~T., {Kruczek}, N.~E., {Gallagher}, S.~C., {et~al.} 2011, \aj,
  141, 167

\bibitem[{{Richards} {et~al.}(2009){Richards}, {Myers}, {Gray}, {Riegel},
  {Nichol}, {Brunner}, {Szalay}, {Schneider}, \&
  {Anderson}}]{2009ApJS..180...67R}
{Richards}, G.~T., {Myers}, A.~D., {Gray}, A.~G., {et~al.} 2009, \apjs, 180, 67

\bibitem[{{Richards} {et~al.}(2015){Richards}, {Myers}, {Peters}, {Krawczyk},
  {Chase}, {Ross}, {Fan}, {Jiang}, {Lacy}, {McGreer}, {Trump}, \&
  {Riegel}}]{2015ApJS..219...39R}
{Richards}, G.~T., {Myers}, A.~D., {Peters}, C.~M., {et~al.} 2015, \apjs, 219,
  39

\bibitem[{{Richards} {et~al.}(2006){Richards}, {Strauss}, {Fan}, {Hall},
  {Jester}, {Schneider}, {Vanden Berk}, {Stoughton}, {Anderson}, {Brunner},
  {Gray}, {Gunn}, {Ivezi{\'c}}, {Kirkland}, {Knapp}, {Loveday}, {Meiksin},
  {Pope}, {Szalay}, {Thakar}, {Yanny}, {York}, {Barentine}, {Brewington},
  {Brinkmann}, {Fukugita}, {Harvanek}, {Kent}, {Kleinman}, {Krzesi{\'n}ski},
  {Long}, {Lupton}, {Nash}, {Neilsen}, {Nitta}, {Schlegel}, \&
  {Snedden}}]{2006AJ....131.2766R}
{Richards}, G.~T., {Strauss}, M.~A., {Fan}, X., {et~al.} 2006, \aj, 131, 2766

\bibitem[{{Richards} {et~al.}(2001){Richards}, {Weinstein}, {Schneider}, {Fan},
  {Strauss}, {Vanden Berk}, {Annis}, {Burles}, {Laubacher}, {York}, {Frieman},
  {Johnston}, {Scranton}, {Gunn}, {Ivezi{\'c}}, {Nichol}, {Budav{\'a}ri},
  {Csabai}, {Szalay}, {Connolly}, {Szokoly}, {Bahcall}, {Ben{\'{\i}}tez},
  {Brinkmann}, {Brunner}, {Fukugita}, {Hall}, {Hennessy}, {Knapp}, {Kunszt},
  {Lamb}, {Munn}, {Newberg}, \& {Stoughton}}]{2001AJ....122.1151R}
{Richards}, G.~T., {Weinstein}, M.~A., {Schneider}, D.~P., {et~al.} 2001, \aj,
  122, 1151

\bibitem[{{Ross} {et~al.}(2013){Ross}, {McGreer}, {White}, {Richards}, {Myers},
  {Palanque-Delabrouille}, {Strauss}, {Anderson}, {Shen}, {Brandt},
  {Y{\`e}che}, {Swanson}, {Aubourg}, {Bailey}, {Bizyaev}, {Bovy}, {Brewington},
  {Brinkmann}, {DeGraf}, {Di Matteo}, {Ebelke}, {Fan}, {Ge}, {Malanushenko},
  {Malanushenko}, {Mandelbaum}, {Maraston}, {Muna}, {Oravetz}, {Pan},
  {P{\^a}ris}, {Petitjean}, {Schawinski}, {Schlegel}, {Schneider}, {Silverman},
  {Simmons}, {Snedden}, {Streblyanska}, {Suzuki}, {Weinberg}, \&
  {York}}]{2013ApJ...773...14R}
{Ross}, N.~P., {McGreer}, I.~D., {White}, M., {et~al.} 2013, \apj, 773, 14

\bibitem[{{R{\"o}ttgering} {et~al.}(2011){R{\"o}ttgering}, {Afonso}, {Barthel},
  {Batejat}, {Best}, {Bonafede}, {Br{\"u}ggen}, {Brunetti}, {Chy{\.z}y},
  {Conway}, {de Gasperin}, {Ferrari}, {Haverkorn}, {Heald}, {Hoeft}, {Jackson},
  {Jarvis}, {Ker}, {Lehnert}, {Macario}, {McKean}, {Miley}, {Morganti},
  {Oosterloo}, {Orr{\`u}}, {Pizzo}, {Rafferty}, {Shulevski}, {Tasse}, {van
  Bemmel}, {van der Tol}, {van Weeren}, {Verheijen}, {White}, \&
  {Wise}}]{2011JApA...32..557R}
{R{\"o}ttgering}, H., {Afonso}, J., {Barthel}, P., {et~al.} 2011, Journal of
  Astrophysics and Astronomy, 32, 557

\bibitem[{{Russell}(1914)}]{1914PA.....22..275R}
{Russell}, H.~N. 1914, Popular Astronomy, 22, 275

\bibitem[{{Salvato} {et~al.}(2009){Salvato}, {Hasinger}, {Ilbert}, {Zamorani},
  {Brusa}, {Scoville}, {Rau}, {Capak}, {Arnouts}, {Aussel}, {Bolzonella},
  {Buongiorno}, {Cappelluti}, {Caputi}, {Civano}, {Cook}, {Elvis}, {Gilli},
  {Jahnke}, {Kartaltepe}, {Impey}, {Lamareille}, {Le Floc'h}, {Lilly},
  {Mainieri}, {McCarthy}, {McCracken}, {Mignoli}, {Mobasher}, {Murayama},
  {Sasaki}, {Sanders}, {Schiminovich}, {Shioya}, {Shopbell}, {Silverman},
  {Smol{\v c}i{\'c}}, {Surace}, {Taniguchi}, {Thompson}, {Trump}, {Urry}, \&
  {Zamojski}}]{2009ApJ...690.1250S}
{Salvato}, M., {Hasinger}, G., {Ilbert}, O., {et~al.} 2009, \apj, 690, 1250

\bibitem[{{Sanders} {et~al.}(2007){Sanders}, {Salvato}, {Aussel}, {Ilbert},
  {Scoville}, {Surace}, {Frayer}, {Sheth}, {Helou}, {Brooke}, {Bhattacharya},
  {Yan}, {Kartaltepe}, {Barnes}, {Blain}, {Calzetti}, {Capak}, {Carilli},
  {Carollo}, {Comastri}, {Daddi}, {Ellis}, {Elvis}, {Fall}, {Franceschini},
  {Giavalisco}, {Hasinger}, {Impey}, {Koekemoer}, {Le F{\`e}vre}, {Lilly},
  {Liu}, {McCracken}, {Mobasher}, {Renzini}, {Rich}, {Schinnerer}, {Shopbell},
  {Taniguchi}, {Thompson}, {Urry}, \& {Williams}}]{2007ApJS..172...86S}
{Sanders}, D.~B., {Salvato}, M., {Aussel}, H., {et~al.} 2007, \apjs, 172, 86

\bibitem[{{Schindler} {et~al.}(2017){Schindler}, {Fan}, {McGreer}, {Yang},
  {Wu}, {Jiang}, \& {Green}}]{2017ApJ...851...13S}
{Schindler}, J.-T., {Fan}, X., {McGreer}, I.~D., {et~al.} 2017, \apj, 851, 13

\bibitem[{{Schlafly} \& {Finkbeiner}(2011)}]{2011ApJ...737..103S}
{Schlafly}, E.~F. \& {Finkbeiner}, D.~P. 2011, \apj, 737, 103

\bibitem[{{Schmidt}(1968)}]{1968ApJ...151..393S}
{Schmidt}, M. 1968, \apj, 151, 393

\bibitem[{{Schmidt} {et~al.}(1995){Schmidt}, {Schneider}, \&
  {Gunn}}]{1995AJ....110...68S}
{Schmidt}, M., {Schneider}, D.~P., \& {Gunn}, J.~E. 1995, \aj, 110, 68

\bibitem[{{Schneider} {et~al.}(2010){Schneider}, {Richards}, {Hall}, {Strauss},
  {Anderson}, {Boroson}, {Ross}, {Shen}, {Brandt}, {Fan}, {Inada}, {Jester},
  {Knapp}, {Krawczyk}, {Thakar}, {Vanden Berk}, {Voges}, {Yanny}, {York},
  {Bahcall}, {Bizyaev}, {Blanton}, {Brewington}, {Brinkmann}, {Eisenstein},
  {Frieman}, {Fukugita}, {Gray}, {Gunn}, {Hibon}, {Ivezi{\'c}}, {Kent}, {Kron},
  {Lee}, {Lupton}, {Malanushenko}, {Malanushenko}, {Oravetz}, {Pan}, {Pier},
  {Price}, {Saxe}, {Schlegel}, {Simmons}, {Snedden}, {SubbaRao}, {Szalay}, \&
  {Weinberg}}]{2010AJ....139.2360S}
{Schneider}, D.~P., {Richards}, G.~T., {Hall}, P.~B., {et~al.} 2010, \aj, 139,
  2360

\bibitem[{{Schulze} {et~al.}(2017){Schulze}, {Done}, {Lu}, {Zhang}, \&
  {Inoue}}]{2017ApJ...849....4S}
{Schulze}, A., {Done}, C., {Lu}, Y., {Zhang}, F., \& {Inoue}, Y. 2017, \apj,
  849, 4

\bibitem[{{Scranton} {et~al.}(2002){Scranton}, {Johnston}, {Dodelson},
  {Frieman}, {Connolly}, {Eisenstein}, {Gunn}, {Hui}, {Jain}, {Kent},
  {Loveday}, {Narayanan}, {Nichol}, {O'Connell}, {Scoccimarro}, {Sheth},
  {Stebbins}, {Strauss}, {Szalay}, {Szapudi}, {Tegmark}, {Vogeley}, {Zehavi},
  {Annis}, {Bahcall}, {Brinkman}, {Csabai}, {Hindsley}, {Ivezic}, {Kim},
  {Knapp}, {Lamb}, {Lee}, {Lupton}, {McKay}, {Munn}, {Peoples}, {Pier},
  {Richards}, {Rockosi}, {Schlegel}, {Schneider}, {Stoughton}, {Tucker},
  {Yanny}, \& {York}}]{2002ApJ...579...48S}
{Scranton}, R., {Johnston}, D., {Dodelson}, S., {et~al.} 2002, \apj, 579, 48

\bibitem[{{Shaver} {et~al.}(1996){Shaver}, {Wall}, {Kellermann}, {Jackson}, \&
  {Hawkins}}]{1996Natur.384..439S}
{Shaver}, P.~A., {Wall}, J.~V., {Kellermann}, K.~I., {Jackson}, C.~A., \&
  {Hawkins}, M.~R.~S. 1996, \nat, 384, 439

\bibitem[{{Shen} {et~al.}(2009){Shen}, {Strauss}, {Ross}, {Hall}, {Lin},
  {Richards}, {Schneider}, {Weinberg}, {Connolly}, {Fan}, {Hennawi}, {Shankar},
  {Vanden Berk}, {Bahcall}, \& {Brunner}}]{2009ApJ...697.1656S}
{Shen}, Y., {Strauss}, M.~A., {Ross}, N.~P., {et~al.} 2009, \apj, 697, 1656

\bibitem[{{Shimwell} {et~al.}(2017){Shimwell}, {R{\"o}ttgering}, {Best},
  {Williams}, {Dijkema}, {de Gasperin}, {Hardcastle}, {Heald}, {Hoang},
  {Horneffer}, {Intema}, {Mahony}, {Mandal}, {Mechev}, {Morabito}, {Oonk},
  {Rafferty}, {Retana-Montenegro}, {Sabater}, {Tasse}, {van Weeren},
  {Br{\"u}ggen}, {Brunetti}, {Chy{\.z}y}, {Conway}, {Haverkorn}, {Jackson},
  {Jarvis}, {McKean}, {Miley}, {Morganti}, {White}, {Wise}, {van Bemmel},
  {Beck}, {Brienza}, {Bonafede}, {Calistro Rivera}, {Cassano}, {Clarke},
  {Cseh}, {Deller}, {Drabent}, {van Driel}, {Engels}, {Falcke}, {Ferrari},
  {Fr{\"o}hlich}, {Garrett}, {Harwood}, {Heesen}, {Hoeft}, {Horellou},
  {Israel}, {Kapi{\'n}ska}, {Kunert-Bajraszewska}, {McKay}, {Mohan},
  {Orr{\'u}}, {Pizzo}, {Prandoni}, {Schwarz}, {Shulevski}, {Sipior}, {Smith},
  {Sridhar}, {Steinmetz}, {Stroe}, {Varenius}, {van der Werf}, {Zensus}, \&
  {Zwart}}]{2017A&A...598A.104S}
{Shimwell}, T.~W., {R{\"o}ttgering}, H.~J.~A., {Best}, P.~N., {et~al.} 2017,
  \aap, 598, A104

\bibitem[{{Shimwell} {et~al.}(2019){Shimwell}, {Tasse}, {Hardcastle}, {Mechev},
  {Williams}, {Best}, {R{\"o}ttgering}, {Callingham}, {Dijkema}, \& {de
  Gasperin}}]{2019A&A...622A...1S}
{Shimwell}, T.~W., {Tasse}, C., {Hardcastle}, M.~J., {et~al.} 2019, \aap, 622,
  A1

\bibitem[{{Siana} {et~al.}(2008){Siana}, {Polletta}, {Smith}, {Lonsdale},
  {Gonzalez-Solares}, {Farrah}, {Babbedge}, {Rowan-Robinson}, {Surace},
  {Shupe}, {Fang}, {Franceschini}, \& {Oliver}}]{2008ApJ...675...49S}
{Siana}, B., {Polletta}, M.~d.~C., {Smith}, H.~E., {et~al.} 2008, \apj, 675, 49

\bibitem[{{Sikora}(2009)}]{2009AN....330..291S}
{Sikora}, M. 2009, Astronomische Nachrichten, 330, 291

\bibitem[{{Sikora} {et~al.}(2007){Sikora}, {Stawarz}, \&
  {Lasota}}]{2007ApJ...658..815S}
{Sikora}, M., {Stawarz}, {\L}., \& {Lasota}, J.-P. 2007, \apj, 658, 815

\bibitem[{{Silverman} {et~al.}(2005){Silverman}, {Green}, {Barkhouse},
  {Cameron}, {Foltz}, {Jannuzi}, {Kim}, {Kim}, {Mossman}, {Tananbaum},
  {Wilkes}, {Smith}, {Smith}, \& {Smith}}]{2005ApJ...624..630S}
{Silverman}, J.~D., {Green}, P.~J., {Barkhouse}, W.~A., {et~al.} 2005, \apj,
  624, 630

\bibitem[{{Smith} {et~al.}(1993){Smith}, {Thompson}, \&
  {Djorgovski}}]{1993ASPC...43..189S}
{Smith}, J.~D., {Thompson}, D., \& {Djorgovski}, S. 1993, in Astronomical
  Society of the Pacific Conference Series, Vol.~43, Sky Surveys. Protostars to
  Protogalaxies, ed. B.~T. {Soifer}, 189

\bibitem[{{Snellen} {et~al.}(2000){Snellen}, {Schilizzi}, {Miley}, {de Bruyn},
  {Bremer}, \& {R{\"o}ttgering}}]{2000MNRAS.319..445S}
{Snellen}, I.~A.~G., {Schilizzi}, R.~T., {Miley}, G.~K., {et~al.} 2000, \mnras,
  319, 445

\bibitem[{{Somerville} {et~al.}(2008){Somerville}, {Hopkins}, {Cox},
  {Robertson}, \& {Hernquist}}]{2008MNRAS.391..481S}
{Somerville}, R.~S., {Hopkins}, P.~F., {Cox}, T.~J., {Robertson}, B.~E., \&
  {Hernquist}, L. 2008, \mnras, 391, 481

\bibitem[{{Spergel} {et~al.}(2015){Spergel}, {Gehrels}, {Baltay}, {Bennett},
  {Breckinridge}, {Donahue}, {Dressler}, {Gaudi}, {Greene}, {Guyon}, {Hirata},
  {Kalirai}, {Kasdin}, {Macintosh}, {Moos}, {Perlmutter}, {Postman},
  {Rauscher}, {Rhodes}, {Wang}, {Weinberg}, {Benford}, {Hudson}, {Jeong},
  {Mellier}, {Traub}, {Yamada}, {Capak}, {Colbert}, {Masters}, {Penny},
  {Savransky}, {Stern}, {Zimmerman}, {Barry}, {Bartusek}, {Carpenter}, {Cheng},
  {Content}, {Dekens}, {Demers}, {Grady}, {Jackson}, {Kuan}, {Kruk}, {Melton},
  {Nemati}, {Parvin}, {Poberezhskiy}, {Peddie}, {Ruffa}, {Wallace}, {Whipple},
  {Wollack}, \& {Zhao}}]{2015arXiv150303757S}
{Spergel}, D., {Gehrels}, N., {Baltay}, C., {et~al.} 2015, arXiv e-prints
  [\eprint[arXiv]{1503.03757}]

\bibitem[{{Spergel} {et~al.}(2013){Spergel}, {Gehrels}, {Breckinridge},
  {Donahue}, {Dressler}, {Gaudi}, {Greene}, {Guyon}, {Hirata}, {Kalirai},
  {Kasdin}, {Moos}, {Perlmutter}, {Postman}, {Rauscher}, {Rhodes}, {Wang},
  {Weinberg}, {Centrella}, {Traub}, {Baltay}, {Colbert}, {Bennett},
  {Kiessling}, {Macintosh}, {Merten}, {Mortonson}, {Penny}, {Rozo},
  {Savransky}, {Stapelfeldt}, {Zu}, {Baker}, {Cheng}, {Content}, {Dooley},
  {Foote}, {Goullioud}, {Grady}, {Jackson}, {Kruk}, {Levine}, {Melton},
  {Peddie}, {Ruffa}, \& {Shaklan}}]{2013arXiv1305.5422S}
{Spergel}, D., {Gehrels}, N., {Breckinridge}, J., {et~al.} 2013, ArXiv e-prints
  [\eprint[arXiv]{1305.5422}]

\bibitem[{{Stern} {et~al.}(2000){Stern}, {Djorgovski}, {Perley}, {de Carvalho},
  \& {Wall}}]{2000AJ....119.1526S}
{Stern}, D., {Djorgovski}, S.~G., {Perley}, R.~A., {de Carvalho}, R.~R., \&
  {Wall}, J.~V. 2000, \aj, 119, 1526

\bibitem[{{Stern} {et~al.}(2005){Stern}, {Eisenhardt}, {Gorjian}, {Kochanek},
  {Caldwell}, {Eisenstein}, {Brodwin}, {Brown}, {Cool}, {Dey}, {Green},
  {Jannuzi}, {Murray}, {Pahre}, \& {Willner}}]{2005ApJ...631..163S}
{Stern}, D., {Eisenhardt}, P., {Gorjian}, V., {et~al.} 2005, \apj, 631, 163

\bibitem[{{Stocke} {et~al.}(1992){Stocke}, {Morris}, {Weymann}, \&
  {Foltz}}]{1992ApJ...396..487S}
{Stocke}, J.~T., {Morris}, S.~L., {Weymann}, R.~J., \& {Foltz}, C.~B. 1992,
  \apj, 396, 487

\bibitem[{{Sulentic} \& {Marziani}(2015)}]{2015FrASS...2....6S}
{Sulentic}, J. \& {Marziani}, P. 2015, Frontiers in Astronomy and Space
  Sciences, 2, 6

\bibitem[{{Sulentic} {et~al.}(2007){Sulentic}, {Bachev}, {Marziani}, {Negrete},
  \& {Dultzin}}]{2007ApJ...666..757S}
{Sulentic}, J.~W., {Bachev}, R., {Marziani}, P., {Negrete}, C.~A., \&
  {Dultzin}, D. 2007, \apj, 666, 757

\bibitem[{{Sulentic} {et~al.}(2000{\natexlab{a}}){Sulentic}, {Marziani}, \&
  {Dultzin-Hacyan}}]{2000ARA&A..38..521S}
{Sulentic}, J.~W., {Marziani}, P., \& {Dultzin-Hacyan}, D. 2000{\natexlab{a}},
  Annual Review of Astronomy and Astrophysics, 38, 521

\bibitem[{{Sulentic} {et~al.}(2003){Sulentic}, {Zamfir}, {Marziani}, {Bachev},
  {Calvani}, \& {Dultzin-Hacyan}}]{2003ApJ...597L..17S}
{Sulentic}, J.~W., {Zamfir}, S., {Marziani}, P., {et~al.} 2003, \apjl, 597, L17

\bibitem[{{Sulentic} {et~al.}(2000{\natexlab{b}}){Sulentic}, {Zwitter},
  {Marziani}, \& {Dultzin-Hacyan}}]{2000ApJ...536L...5S}
{Sulentic}, J.~W., {Zwitter}, T., {Marziani}, P., \& {Dultzin-Hacyan}, D.
  2000{\natexlab{b}}, \apjl, 536, L5

\bibitem[{{Telfer} {et~al.}(2002){Telfer}, {Zheng}, {Kriss}, \&
  {Davidsen}}]{2002ApJ...565..773T}
{Telfer}, R.~C., {Zheng}, W., {Kriss}, G.~A., \& {Davidsen}, A.~F. 2002, \apj,
  565, 773

\bibitem[{{Timlin} {et~al.}(2016){Timlin}, {Ross}, {Richards}, {Lacy}, {Ryan},
  {Stone}, {Bauer}, {Brandt}, {Fan}, {Glikman}, {Haggard}, {Jiang}, {LaMassa},
  {Lin}, {Makler}, {McGehee}, {Myers}, {Schneider}, {Urry}, {Wollack}, \&
  {Zakamska}}]{2016ApJS..225....1T}
{Timlin}, J.~D., {Ross}, N.~P., {Richards}, G.~T., {et~al.} 2016, \apjs, 225, 1

\bibitem[{{Timlin} {et~al.}(2018){Timlin}, {Ross}, {Richards}, {Myers},
  {Pellegrino}, {Bauer}, {Lacy}, {Schneider}, {Wollack}, \&
  {Zakamska}}]{2018ApJ...859...20T}
{Timlin}, J.~D., {Ross}, N.~P., {Richards}, G.~T., {et~al.} 2018, \apj, 859, 20

\bibitem[{{Tonry} {et~al.}(2012){Tonry}, {Stubbs}, {Lykke}, {Doherty},
  {Shivvers}, {Burgett}, {Chambers}, {Hodapp}, {Kaiser}, \&
  {Kudritzki}}]{2012ApJ...750...99T}
{Tonry}, J.~L., {Stubbs}, C.~W., {Lykke}, K.~R., {et~al.} 2012, \apj, 750, 99

\bibitem[{{Trump} {et~al.}(2009){Trump}, {Impey}, {Elvis}, {McCarthy},
  {Huchra}, {Brusa}, {Salvato}, {Capak}, {Cappelluti}, {Civano}, {Comastri},
  {Gabor}, {Hao}, {Hasinger}, {Jahnke}, {Kelly}, {Lilly}, {Schinnerer},
  {Scoville}, \& {Smol{\v{c}}i{\'c}}}]{2009ApJ...696.1195T}
{Trump}, J.~R., {Impey}, C.~D., {Elvis}, M., {et~al.} 2009, \apj, 696, 1195

\bibitem[{{Tuccillo} {et~al.}(2015){Tuccillo}, {Gonz{\'a}lez-Serrano}, \&
  {Benn}}]{2015MNRAS.449.2818T}
{Tuccillo}, D., {Gonz{\'a}lez-Serrano}, J.~I., \& {Benn}, C.~R. 2015, \mnras,
  449, 2818

\bibitem[{{Tyson}(2002)}]{2002SPIE.4836...10T}
{Tyson}, J.~A. 2002, in \procspie, Vol. 4836, Survey and Other Telescope
  Technologies and Discoveries, ed. J.~A. {Tyson} \& S.~{Wolff}, 10--20

\bibitem[{{Ueda} {et~al.}(2003){Ueda}, {Akiyama}, {Ohta}, \&
  {Miyaji}}]{2003ApJ...598..886U}
{Ueda}, Y., {Akiyama}, M., {Ohta}, K., \& {Miyaji}, T. 2003, \apj, 598, 886

\bibitem[{{van Breugel} {et~al.}(1984){van Breugel}, {Miley}, \&
  {Heckman}}]{1984AJ.....89....5V}
{van Breugel}, W., {Miley}, G., \& {Heckman}, T. 1984, \aj, 89, 5

\bibitem[{{van Haarlem} {et~al.}(2013){van Haarlem}, {Wise}, {Gunst}, {Heald},
  {McKean}, {Hessels}, {de Bruyn}, {Nijboer}, {Swinbank}, {Fallows},
  {Brentjens}, {Nelles}, {Beck}, {Falcke}, {Fender}, {H{\"o}randel},
  {Koopmans}, {Mann}, {Miley}, {R{\"o}ttgering}, {Stappers}, {Wijers},
  {Zaroubi}, {van den Akker}, {Alexov}, {Anderson}, {Anderson}, {van Ardenne},
  {Arts}, {Asgekar}, {Avruch}, {Batejat}, {B{\"a}hren}, {Bell}, {Bell}, {van
  Bemmel}, {Bennema}, {Bentum}, {Bernardi}, {Best}, {B{\^\i}rzan}, {Bonafede},
  {Boonstra}, {Braun}, {Bregman}, {Breitling}, {van de Brink}, {Broderick},
  {Broekema}, {Brouw}, {Br{\"u}ggen}, {Butcher}, {van Cappellen}, {Ciardi},
  {Coenen}, {Conway}, {Coolen}, {Corstanje}, {Damstra}, {Davies}, {Deller},
  {Dettmar}, {van Diepen}, {Dijkstra}, {Donker}, {Doorduin}, {Dromer}, {Drost},
  {van Duin}, {Eisl{\"o}ffel}, {van Enst}, {Ferrari}, {Frieswijk}, {Gankema},
  {Garrett}, {de Gasperin}, {Gerbers}, {de Geus}, {Grie{\ss}meier}, {Grit},
  {Gruppen}, {Hamaker}, {Hassall}, {Hoeft}, {Holties}, {Horneffer}, {van der
  Horst}, {van Houwelingen}, {Huijgen}, {Iacobelli}, {Intema}, {Jackson},
  {Jelic}, {de Jong}, {Juette}, {Kant}, {Karastergiou}, {Koers}, {Kollen},
  {Kondratiev}, {Kooistra}, {Koopman}, {Koster}, {Kuniyoshi}, {Kramer},
  {Kuper}, {Lambropoulos}, {Law}, {van Leeuwen}, {Lemaitre}, {Loose}, {Maat},
  {Macario}, {Markoff}, {Masters}, {McFadden}, {McKay-Bukowski}, {Meijering},
  {Meulman}, {Mevius}, {Middelberg}, {Millenaar}, {Miller-Jones}, {Mohan},
  {Mol}, {Morawietz}, {Morganti}, {Mulcahy}, {Mulder}, {Munk}, {Nieuwenhuis},
  {van Nieuwpoort}, {Noordam}, {Norden}, {Noutsos}, {Offringa}, {Olofsson},
  {Omar}, {Orr{\'u}}, {Overeem}, {Paas}, {Pand ey-Pommier}, {Pandey}, {Pizzo},
  {Polatidis}, {Rafferty}, {Rawlings}, {Reich}, {de Reijer}, {Reitsma},
  {Renting}, {Riemers}, {Rol}, {Romein}, {Roosjen}, {Ruiter}, {Scaife}, {van
  der Schaaf}, {Scheers}, {Schellart}, {Schoenmakers}, {Schoonderbeek},
  {Serylak}, {Shulevski}, {Sluman}, {Smirnov}, {Sobey}, {Spreeuw}, {Steinmetz},
  {Sterks}, {Stiepel}, {Stuurwold}, {Tagger}, {Tang}, {Tasse}, {Thomas},
  {Thoudam}, {Toribio}, {van der Tol}, {Usov}, {van Veelen}, {van der Veen},
  {ter Veen}, {Verbiest}, {Vermeulen}, {Vermaas}, {Vocks}, {Vogt}, {de Vos},
  {van der Wal}, {van Weeren}, {Weggemans}, {Weltevrede}, {White}, {Wijnholds},
  {Wilhelmsson}, {Wucknitz}, {Yatawatta}, {Zarka}, {Zensus}, \& {van
  Zwieten}}]{2013A&A...556A...2V}
{van Haarlem}, M.~P., {Wise}, M.~W., {Gunst}, A.~W., {et~al.} 2013, \aap, 556,
  A2

\bibitem[{{Vanden Berk} {et~al.}(2001){Vanden Berk}, {Richards}, {Bauer},
  {Strauss}, {Schneider}, {Heckman}, {York}, {Hall}, {Fan}, {Knapp},
  {Anderson}, {Annis}, {Bahcall}, {Bernardi}, {Briggs}, {Brinkmann}, {Brunner},
  {Burles}, {Carey}, {Castander}, {Connolly}, {Crocker}, {Csabai}, {Doi},
  {Finkbeiner}, {Friedman}, {Frieman}, {Fukugita}, {Gunn}, {Hennessy},
  {Ivezi{\'c}}, {Kent}, {Kunszt}, {Lamb}, {Leger}, {Long}, {Loveday}, {Lupton},
  {Meiksin}, {Merelli}, {Munn}, {Newberg}, {Newcomb}, {Nichol}, {Owen}, {Pier},
  {Pope}, {Rockosi}, {Schlegel}, {Siegmund}, {Smee}, {Snir}, {Stoughton},
  {Stubbs}, {SubbaRao}, {Szalay}, {Szokoly}, {Tremonti}, {Uomoto}, {Waddell},
  {Yanny}, \& {Zheng}}]{2001AJ....122..549V}
{Vanden Berk}, D.~E., {Richards}, G.~T., {Bauer}, A., {et~al.} 2001, \aj, 122,
  549

\bibitem[{{Varenius} {et~al.}(2015){Varenius}, {Conway}, {Mart{\'{\i}}-Vidal},
  {Beswick}, {Deller}, {Wucknitz}, {Jackson}, {Adebahr}, {P{\'e}rez-Torres},
  {Chy{\.z}y}, {Carozzi}, {Mold{\'o}n}, {Aalto}, {Beck}, {Best}, {Dettmar},
  {van Driel}, {Brunetti}, {Br{\"u}ggen}, {Haverkorn}, {Heald}, {Horellou},
  {Jarvis}, {Morabito}, {Miley}, {R{\"o}ttgering}, {Toribio}, \&
  {White}}]{2015A&A...574A.114V}
{Varenius}, E., {Conway}, J.~E., {Mart{\'{\i}}-Vidal}, I., {et~al.} 2015, \aap,
  574, A114

\bibitem[{{Vasconcellos} {et~al.}(2011){Vasconcellos}, {de Carvalho}, {Gal},
  {LaBarbera}, {Capelato}, {Frago Campos Velho}, {Trevisan}, \&
  {Ruiz}}]{2011AJ....141..189V}
{Vasconcellos}, E.~C., {de Carvalho}, R.~R., {Gal}, R.~R., {et~al.} 2011, \aj,
  141, 189

\bibitem[{{Vigotti} {et~al.}(2003){Vigotti}, {Carballo}, {Benn}, {De Zotti},
  {Fanti}, {Gonzalez Serrano}, {Mack}, \& {Holt}}]{2003ApJ...591...43V}
{Vigotti}, M., {Carballo}, R., {Benn}, C.~R., {et~al.} 2003, \apj, 591, 43

\bibitem[{{Visnovsky} {et~al.}(1992){Visnovsky}, {Impey}, {Foltz}, {Hewett},
  {Weymann}, \& {Morris}}]{1992ApJ...391..560V}
{Visnovsky}, K.~L., {Impey}, C.~D., {Foltz}, C.~B., {et~al.} 1992, \apj, 391,
  560

\bibitem[{{Vito} {et~al.}(2014){Vito}, {Gilli}, {Vignali}, {Comastri}, {Brusa},
  {Cappelluti}, \& {Iwasawa}}]{2014MNRAS.445.3557V}
{Vito}, F., {Gilli}, R., {Vignali}, C., {et~al.} 2014, \mnras, 445, 3557

\bibitem[{{Volonteri} \& {Rees}(2005)}]{2005ApJ...633..624V}
{Volonteri}, M. \& {Rees}, M.~J. 2005, \apj, 633, 624

\bibitem[{{Wang} {et~al.}(2007){Wang}, {Zhang}, {Liu}, \&
  {Zhao}}]{2007MNRAS.382.1601W}
{Wang}, D., {Zhang}, Y.~X., {Liu}, C., \& {Zhao}, Y.~H. 2007, \mnras, 382, 1601

\bibitem[{{Warren} {et~al.}(1994){Warren}, {Hewett}, \&
  {Osmer}}]{1994ApJ...421..412W}
{Warren}, S.~J., {Hewett}, P.~C., \& {Osmer}, P.~S. 1994, \apj, 421, 412

\bibitem[{Watson(1964)}]{10.2307/25049340}
Watson, G.~S. 1964, Sankhya: The Indian Journal of Statistics, Series A
  (1961-2002), 26, 359

\bibitem[{{Weinstein} {et~al.}(2004){Weinstein}, {Richards}, {Schneider},
  {Younger}, {Strauss}, {Hall}, {Budav{\'a}ri}, {Gunn}, {York}, \&
  {Brinkmann}}]{2004ApJS..155..243W}
{Weinstein}, M.~A., {Richards}, G.~T., {Schneider}, D.~P., {et~al.} 2004,
  \apjs, 155, 243

\bibitem[{{Welling} {et~al.}(2014){Welling}, {Miller}, {Brandt}, {Capellupo},
  \& {Gibson}}]{2014MNRAS.440.2474W}
{Welling}, C.~A., {Miller}, B.~P., {Brandt}, W.~N., {Capellupo}, D.~M., \&
  {Gibson}, R.~R. 2014, \mnras, 440, 2474

\bibitem[{{Williams} {et~al.}(2013){Williams}, {Intema}, \&
  {R{\"o}ttgering}}]{2013AA...549A..55W}
{Williams}, W.~L., {Intema}, H.~T., \& {R{\"o}ttgering}, H.~J.~A. 2013, \aap,
  549, A55

\bibitem[{{Williams} {et~al.}(2016){Williams}, {van Weeren}, {R{\"o}ttgering},
  {Best}, {Dijkema}, {de Gasperin}, {Hardcastle}, {Heald}, {Prandoni},
  {Sabater}, {Shimwell}, {Tasse}, {van Bemmel}, {Br{\"u}ggen}, {Brunetti},
  {Conway}, {En{\ss}lin}, {Engels}, {Falcke}, {Ferrari}, {Haverkorn},
  {Jackson}, {Jarvis}, {Kapi{\'n}ska}, {Mahony}, {Miley}, {Morabito},
  {Morganti}, {Orr{\'u}}, {Retana-Montenegro}, {Sridhar}, {Toribio}, {White},
  {Wise}, \& {Zwart}}]{2016MNRAS.460.2385W}
{Williams}, W.~L., {van Weeren}, R.~J., {R{\"o}ttgering}, H.~J.~A., {et~al.}
  2016, \mnras, 460, 2385

\bibitem[{{Willott} {et~al.}(2010){Willott}, {Delorme}, {Reyl{\'e}}, {Albert},
  {Bergeron}, {Crampton}, {Delfosse}, {Forveille}, {Hutchings}, {McLure},
  {Omont}, \& {Schade}}]{2010AJ....139..906W}
{Willott}, C.~J., {Delorme}, P., {Reyl{\'e}}, C., {et~al.} 2010, \aj, 139, 906

\bibitem[{{Wolf} {et~al.}(2003){Wolf}, {Wisotzki}, {Borch}, {Dye},
  {Kleinheinrich}, \& {Meisenheimer}}]{2003A&A...408..499W}
{Wolf}, C., {Wisotzki}, L., {Borch}, A., {et~al.} 2003, \aap, 408, 499

\bibitem[{{Wright} {et~al.}(2010){Wright}, {Eisenhardt}, {Mainzer}, {Ressler},
  {Cutri}, {Jarrett}, {Kirkpatrick}, {Padgett}, {McMillan}, {Skrutskie},
  {Stanford}, {Cohen}, {Walker}, {Mather}, {Leisawitz}, {Gautier}, {McLean},
  {Benford}, {Lonsdale}, {Blain}, {Mendez}, {Irace}, {Duval}, {Liu}, {Royer},
  {Heinrichsen}, {Howard}, {Shannon}, {Kendall}, {Walsh}, {Larsen}, {Cardon},
  {Schick}, {Schwalm}, {Abid}, {Fabinsky}, {Naes}, \&
  {Tsai}}]{2010AJ....140.1868W}
{Wright}, E.~L., {Eisenhardt}, P.~R.~M., {Mainzer}, A.~K., {et~al.} 2010, \aj,
  140, 1868

\bibitem[{{Wu} \&
  {Zhang}(2006)}]{RePEc:bes:jnlasa:v:102:y:2007:m:june:p:761-761}
{Wu}, H. \& {Zhang}, J. 2006 (Wiley Press)

\bibitem[{{Yang} {et~al.}(2018){Yang}, {Wu}, {Liu}, {Fan}, {Yang}, {Wang},
  {McGreer}, {Fan}, {Yuan}, \& {Shan}}]{2018AJ....155..110Y}
{Yang}, J., {Wu}, X.-B., {Liu}, D., {et~al.} 2018, \aj, 155, 110

\bibitem[{{Yang} {et~al.}(2017){Yang}, {Wu}, {Fan}, {Jiang}, {McGreer},
  {Green}, {Yang}, {Schindler}, {Wang}, {Zuo}, \& {Fu}}]{2017AJ....154..269Y}
{Yang}, Q., {Wu}, X.-B., {Fan}, X., {et~al.} 2017, \aj, 154, 269

\bibitem[{{Yao} {et~al.}(2019){Yao}, {Wu}, {Ai}, {Yang}, {Yang}, {Dong},
  {Joshi}, {Wang}, {Feng}, {Fu}, {Hou}, {Luo}, {Kong}, {Liu}, {Zhao}, {Zhang},
  {Yuan}, \& {Shen}}]{2019ApJS..240....6Y}
{Yao}, S., {Wu}, X.-B., {Ai}, Y.~L., {et~al.} 2019, \apjs, 240, 6

\bibitem[{{Y{\`e}che} {et~al.}(2010){Y{\`e}che}, {Petitjean}, {Rich},
  {Aubourg}, {Busca}, {Hamilton}, {Le Goff}, {Paris}, {Peirani}, {Pichon},
  {Rollinde}, \& {Vargas-Maga{\~n}a}}]{2010A&A...523A..14Y}
{Y{\`e}che}, C., {Petitjean}, P., {Rich}, J., {et~al.} 2010, \aap, 523, A14

\bibitem[{{York} {et~al.}(2000){York}, {Adelman}, {Anderson}, {Anderson},
  {Annis}, {Bahcall}, {Bakken}, {Barkhouser}, {Bastian}, {Berman}, {Boroski},
  {Bracker}, {Briegel}, {Briggs}, {Brinkmann}, {Brunner}, {Burles}, {Carey},
  {Carr}, {Castander}, {Chen}, {Colestock}, {Connolly}, {Crocker}, {Csabai},
  {Czarapata}, {Davis}, {Doi}, {Dombeck}, {Eisenstein}, {Ellman}, {Elms},
  {Evans}, {Fan}, {Federwitz}, {Fiscelli}, {Friedman}, {Frieman}, {Fukugita},
  {Gillespie}, {Gunn}, {Gurbani}, {de Haas}, {Haldeman}, {Harris}, {Hayes},
  {Heckman}, {Hennessy}, {Hindsley}, {Holm}, {Holmgren}, {Huang}, {Hull},
  {Husby}, {Ichikawa}, {Ichikawa}, {Ivezi{\'c}}, {Kent}, {Kim}, {Kinney},
  {Klaene}, {Kleinman}, {Kleinman}, {Knapp}, {Korienek}, {Kron}, {Kunszt},
  {Lamb}, {Lee}, {Leger}, {Limmongkol}, {Lindenmeyer}, {Long}, {Loomis},
  {Loveday}, {Lucinio}, {Lupton}, {MacKinnon}, {Mannery}, {Mantsch}, {Margon},
  {McGehee}, {McKay}, {Meiksin}, {Merelli}, {Monet}, {Munn}, {Narayanan},
  {Nash}, {Neilsen}, {Neswold}, {Newberg}, {Nichol}, {Nicinski}, {Nonino},
  {Okada}, {Okamura}, {Ostriker}, {Owen}, {Pauls}, {Peoples}, {Peterson},
  {Petravick}, {Pier}, {Pope}, {Pordes}, {Prosapio}, {Rechenmacher}, {Quinn},
  {Richards}, {Richmond}, {Rivetta}, {Rockosi}, {Ruthmansdorfer}, {Sandford},
  {Schlegel}, {Schneider}, {Sekiguchi}, {Sergey}, {Shimasaku}, {Siegmund},
  {Smee}, {Smith}, {Snedden}, {Stone}, {Stoughton}, {Strauss}, {Stubbs},
  {SubbaRao}, {Szalay}, {Szapudi}, {Szokoly}, {Thakar}, {Tremonti}, {Tucker},
  {Uomoto}, {Vanden Berk}, {Vogeley}, {Waddell}, {Wang}, {Watanabe},
  {Weinberg}, {Yanny}, {Yasuda}, \& {SDSS Collaboration}}]{2000AJ....120.1579Y}
{York}, D.~G., {Adelman}, J., {Anderson}, Jr., J.~E., {et~al.} 2000, \aj, 120,
  1579

\bibitem[{{Zakamska} \& {Greene}(2014)}]{2014MNRAS.442..784Z}
{Zakamska}, N.~L. \& {Greene}, J.~E. 2014, \mnras, 442, 784

\bibitem[{{Zakamska} {et~al.}(2016){Zakamska}, {Lampayan}, {Petric}, {Dicken},
  {Greene}, {Heckman}, {Hickox}, {Ho}, {Krolik}, {Nesvadba}, {Strauss},
  {Geach}, {Oguri}, \& {Strateva}}]{2016MNRAS.455.4191Z}
{Zakamska}, N.~L., {Lampayan}, K., {Petric}, A., {et~al.} 2016, \mnras, 455,
  4191

\bibitem[{{Zamfir} {et~al.}(2008){Zamfir}, {Sulentic}, \&
  {Marziani}}]{2008MNRAS.387..856Z}
{Zamfir}, S., {Sulentic}, J.~W., \& {Marziani}, P. 2008, \mnras, 387, 856

\bibitem[{{Zeimann} {et~al.}(2011){Zeimann}, {White}, {Becker}, {Hodge},
  {Stanford}, \& {Richards}}]{2011ApJ...736...57Z}
{Zeimann}, G.~R., {White}, R.~L., {Becker}, R.~H., {et~al.} 2011, \apj, 736, 57

\end{thebibliography}

\end{document}